\documentclass[aps,prl,reprint,superscriptaddress,longbibliography]{revtex4-2}
\usepackage{amsmath,amssymb,bm,graphicx}
\usepackage{mathrsfs}
\usepackage[colorlinks=true,allcolors=blue]{hyperref}
\hypersetup{pdftitle={Lie-Poisson Formulation of Dissipative Spin Hydrodynamics},pdfauthor={Zi-Hao Liu, Mei Huang}}
\begin{document}
\title{Lie--Poisson Formulation of Dissipative Spin Hydrodynamics}
\author{Zi-Hao Liu}
\email{zihaomology@gmail.com}
\affiliation{School of Physical Sciences, University of Chinese Academy of Sciences, Beijing 101408, China}
\author{Mei Huang}
\email{huangmei@ucas.ac.cn}
\affiliation{School of Nuclear Science and Technology, University of Chinese Academy of Sciences, Beijing 100049, China}
\date{\today}
\begin{abstract}
Covariance and the second law cannot fix how fluid flow couples to spin, since reversible couplings produce no entropy. We show that the spin commutation relations and the Poincar\'e algebra fix these couplings. Reversibly, a flow exerts no torque on the spin: its magnitude per particle is conserved, its direction is Fermi--Walker transported, and spin--shear couplings reflect a deforming triad. The magnitude changes only through higher multipoles, as in spinor condensates, or dissipation, whose first-order spin--shear coupling reorients the spin away from the vorticity above a critical shear rate.
\end{abstract}
\maketitle

\textit{Introduction.}---The quark--gluon plasma produced in heavy-ion collisions is the most vortical fluid observed: its vorticity, of order $10^{22}\,\mathrm{s}^{-1}$, polarizes the emitted hadrons~\cite{STAR2017}. The polarization along the beam direction~\cite{STAR2019} has an azimuthal dependence opposite in sign to that predicted from the thermal vorticity alone, and calculations reproduce it only when shear-induced polarization is included~\cite{Fu2021,Becattini2021}. Both calculations assume the local-equilibrium spin set by the hydrodynamic gradients at freeze-out. If spin relaxes slowly, its density is instead an independent hydrodynamic variable~\cite{Hattori2019,Hongo2021,Gallegos2021}, as in molecular fluids~\cite{Stark2005} and spinor condensates~\cite{Kawaguchi2012}, and the polarization depends on how the acceleration, rotation, and shear of the flow act on the spin density throughout the evolution.

\begin{figure}[t]
 \includegraphics[width=\columnwidth]{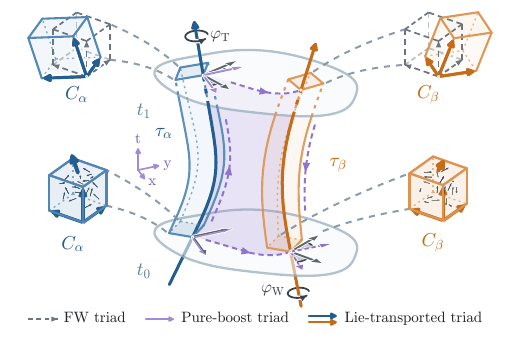}
 \caption{Spin transport along fluid worldtubes (schematic; angles exaggerated). Cells $C_\alpha$ and $C_\beta$ evolve from $t_0$ to $t_1$ with proper times $\tau_\alpha$ and $\tau_\beta$. Insets: at $t_0$ the constituent spins sum to the cell spin; at $t_1$ a Lie-transported triad (solid) has deformed with the cell, while the Fermi--Walker (FW) triad (dashed) remains orthonormal. Without torque the spin has constant components in the FW triad but, by \eqref{eq:triad}, not in the Lie-transported one. Shaded surface: the pure-boost triad of $C_\alpha$ (solid purple) reaches $C_\beta(t_0)$ by a direct boost (dashed purple) or by FW transport along $\tau_\alpha$, a boost at $t_1$, and FW transport back along $\tau_\beta$ (gray); noncommuting boosts make the two differ by the Wigner rotation $\varphi_{\rm W}$. Along $\tau_\alpha$ the FW triad rotates relative to the pure-boost triad by the Thomas angle $\varphi_{\rm T}$ [rate~\eqref{eq:thomas-rate}].}
 \label{fig:transport}
\end{figure}

The bottom-up approach that goes back to Eckart~\cite{Eckart1940} lists the covariant terms formed from the acceleration, shear, vorticity, and expansion of the flow and retains those the second law allows. For spin, this fixes the relaxation toward the thermal vorticity~\cite{Hattori2019,Fukushima2021,Li2021}, which produces entropy, but not the reversible couplings, which produce none: spin equations with and without reversible couplings to shear and expansion~\cite{Fang2026} satisfy the second law equally well. Nor can the second law distinguish a torque from a change of variables: if the triad that defines the spin components deforms with the fluid, shear enters the spin equation without any torque. Whether shear exerts a torque on the spin is thus left undetermined at first order in gradients. The reversible couplings of Ref.~\cite{Fang2026} persist even without a spin--orbit interaction: in the nonrelativistic limit, where spin-independent interactions conserve spin separately, they would exchange angular momentum between spin and orbital motion~\cite{Supplement}.

Reversible hydrodynamics is, however, Hamiltonian: its couplings are constrained by a Lie--Poisson bracket that is inherited from the commutators of the microscopic observables before any equation of state is specified, as in superfluids, magnets, and liquid crystals~\cite{Dzyaloshinskii1980,Stark2005}. We follow this top-down approach for a relativistic fluid with spin: tracking comoving fluid elements (the Lagrangian picture, Fig.~\ref{fig:transport}), we derive the Eulerian Lie--Poisson bracket of the local densities, require it to realize the Poincar\'e algebra, and add dissipation through a symmetric bracket with the same conservation laws.

We find that, reversibly, a flow exerts no torque on the spin: the spin algebra makes the spin magnitude per particle a Casimir invariant and relates orbital circulation to the Berry phase of the spin, and the Poincar\'e algebra enforces Fermi--Walker (FW) transport, so that reversible shear couplings arise only from a deforming triad. Only spin currents driven by higher internal multipoles, as in a spinor condensate, change the magnitude reversibly. Irreversibly, the spin relaxes toward the vorticity; when this relaxation is fast, eliminating the spin leaves an odd (Hall) viscosity. At finite spin density, dissipation admits a unique first-order shear torque, which competes with relaxation and reorients the spin away from the vorticity above a critical shear rate.

\textit{Lie--Poisson bracket.}---A comoving cell is characterized by its densities on an equal-time hypersurface: particle number $n$, entropy $s$, momentum $\pi_i$, and spin $s^A$ ($A,B,C=1,2,3$). Their Poisson brackets follow from the correspondence $\{\langle\widehat A\rangle,\langle\widehat B\rangle\}=\langle[\widehat A,\widehat B]\rangle/i\hbar$ for expectation values of Lie-algebraic observables~\cite{Bonet2016}. Because the spins of distinct particles commute and each obeys $[\widehat S_A,\widehat S_B]=i\hbar\epsilon_{ABC}\widehat S_C$, the coarse-grained spin density obeys the $\mathfrak{su}(2)$ algebra locally, while the momentum density generates the spatial diffeomorphisms that advect the spin with the cell. In the continuum limit~\cite{Supplement},
\begin{equation}
\begin{aligned}
 \{s^A(\bm x),s^B(\bm y)\}&=\epsilon_{ABC}\,s^C(\bm x)\,\delta(\bm x-\bm y),\\
 \{\pi_i(\bm x),s^A(\bm y)\}&=s^A(\bm x)\,\partial_{y^i}\delta(\bm x-\bm y),
\end{aligned}\label{eq:bracket}
\end{equation}
which combine with the standard ideal-fluid brackets into a semidirect-product Lie--Poisson structure~\cite{Morrison1980,Marsden1984,Holm1998} independent of $\hbar$ and particle statistics. If the total angular momentum generates simultaneous rotations of positions, momenta, and spins, the mixed bracket $\{\pi_i,s^A\}$ in \eqref{eq:bracket} is the only one that is local, linear in the fields, first order in gradients, and consistent with the Jacobi identity.

For any Hamiltonian $H[n,s,\pi_i,s^A]$, with fluid velocity $v^i=\delta H/\delta\pi_i$ and spin potential $2\omega_A=\delta H/\delta s^A$, the evolution $\partial_tF=\{F,H\}$ yields
\begin{equation}
 \partial_ts^A+\partial_i(v^is^A)=2\epsilon_{ABC}\,\omega_Bs^C .
 \label{eq:spin-eom}
\end{equation}
Because the precession term is orthogonal to $s^A$ and $n$ is advected by the same velocity $v^i$, the spin magnitude per particle is a local Casimir invariant of the bracket:
\begin{equation}
 (\partial_t+v^i\partial_i)\frac{s^As^A}{n^2}=0 .
 \label{eq:invariant}
\end{equation}
For any $H$, however strongly the flow deforms a cell, reversible dynamics only advects and rotates its spin.

The same bracket couples orbital circulation to the orientation of the spin. In a spinless fluid, Kelvin's theorem conserves the circulation $\oint_C(\pi_i/n)\,dx^i$ around a comoving loop $C$ of uniform entropy per particle $s/n$. With spin, gradients of the spin potential along $C$ exert a Stern--Gerlach-like force through $\{\pi_i,s^A\}$ and induce differential spin precession through $\{s^A,s^B\}$; for any $H$ and any comoving loop on which $s/n$ and $|\bm s|/n$ are uniform, the two combine into the exact invariant
\begin{equation}
 \frac{d}{dt}\biggl[\oint_C\frac{\pi_i}{n}\,dx^i-\frac{|\bm s|}{n}\,\Phi_C\biggr]=0 ,
 \label{eq:circulation}
\end{equation}
where $\Phi_C$ is the solid angle enclosed on the unit sphere by the path of the spin direction $\hat{\bm s}=\bm s/|\bm s|$ along $C$~\cite{Supplement}. In units of $\hbar$, the second term is the Berry phase~\cite{Berry1984} of a spin of magnitude $|\bm s|/n$ taken around $C$: inhomogeneous spin precession generates orbital vorticity. The invariant holds for mixed states far below spin saturation; analogous invariants are known for pure spin states~\cite{Mahajan2011,Foskett2024}, and in polarized superfluids it becomes the quantized Mermin--Ho relation~\cite{MerminHo1976,Lamacraft2008}.

\textit{Spinor condensates and higher multipoles.}---Reversible dynamics can change $|\bm s|/n$ only if a fluid element has internal multipoles beyond the spin dipole. A spin-$F$ Bose--Einstein condensate of atoms of mass $M$ is such a fluid. Its wavefunction $\psi=\sqrt n\,e^{i\varphi}z$, with phase $\varphi$ and normalized $(2F+1)$-component spinor $z$, gives each element the rank-1 density matrix $\rho=zz^\dagger$, and the spin density $s^A=\hbar n\operatorname{tr}(F_A\rho)$, with $[F_A,F_B]=i\epsilon_{ABC}F_C$, is only its dipole moment. For Hermitian generators $A$ and $B$, the canonical bracket of $\psi$ reduces to the $\mathfrak u(2F+1)$ Lie--Poisson bracket~\cite{Supplement}
\begin{equation}
 \{(\psi^\dagger A\psi)(\bm x),(\psi^\dagger B\psi)(\bm y)\}=\frac{n}{i\hbar}\operatorname{tr}(\rho[A,B])\,\delta(\bm x-\bm y),
 \label{eq:condensate}
\end{equation}
which reduces to \eqref{eq:bracket} for $A=\hbar F_A$ and $B=\hbar F_B$. With the mean-field Gross--Pitaevskii functional~\cite{Ho1998,Ohmi1998} for $F=1$, \eqref{eq:condensate} yields the coupled hydrodynamics of the spin and nematic (quadrupolar) densities $N_{AB}=\tfrac12n\operatorname{tr}[(F_AF_B+F_BF_A)\rho]$~\cite{Yukawa2012}, with the reversible spin current
\begin{equation}
 J_i^A=s^Av_i-\frac{\epsilon_{ABC}s^B\partial_is^C}{4Mn}-\frac{\hbar^2\epsilon_{ABC}N_{BD}\partial_iN_{CD}}{Mn}.
 \label{eq:current}
\end{equation}
The nematic term of $J_i^A$ persists even when $\bm s=0$, and its divergence modifies $|\bm s|/n$ reversibly: the full $\mathfrak u(2F+1)$ bracket preserves the purity $\operatorname{tr}(\rho^2)=1$ rather than the dipole magnitude alone.

\textit{Boost invariance and Fermi--Walker transport.}---For a relativistic fluid with only a spin dipole, Lorentz invariance requires that $H$, the momentum $P_i=\int\pi_i$, the angular momentum $J_i=\int(\epsilon_{ijk}x_j\pi_k+s_i)$, and the boost generators $K_i=tP_i-\int(x_i\mathcal H-\Sigma^{00i})$ close the Poincar\'e algebra under the bracket~\eqref{eq:bracket}, in particular $\{K_i,H\}=-P_i$ and $\{K_i,K_j\}=-\epsilon_{ijk}J_k$. Here $\mathcal H$ is the energy density and $\Sigma^{00i}$ the time--space component of the spin density $\Sigma^{0\mu\nu}$. For a pressure gradient $\nabla p\neq0$, gradient-free $\mathcal H$ and $\Sigma^{00i}$ cannot close the algebra at linear order in $\bm s$~\cite{Supplement}, where closure to all orders in gradients requires
\begin{equation}
 \Sigma^{00i}=\frac{\gamma}{\gamma+1}\,(\bm s\times\bm v)^i,\qquad
 \mathcal H=\mathcal H_0+\frac{\bm s\cdot(\bm v\times\nabla p)}{(\varepsilon+p)(\gamma+1)} ,
 \label{eq:thomas}
\end{equation}
where $\mathcal H_0$ is the spinless energy density, $\varepsilon+p$ the rest-frame enthalpy density, and $\gamma=(1-\bm v^2)^{-1/2}$. Physically, $\Sigma^{00i}$ generates the momentum-dependent Wigner rotation~\cite{Wigner1939} of the spin under boosts. The canonical densities obeying~\eqref{eq:bracket} are densities of Newton--Wigner positions~\cite{NewtonWigner1949,Pryce1948}, whereas the pressure acts on the covariant centroids (rest-frame centers of mass), displaced from them by $\bm\Delta=\bm s\times\bm v/[(\varepsilon+p)(\gamma+1)]$; the gradient term in $\mathcal H$ is the work $\bm\Delta\cdot\nabla p$ of the pressure on this displacement.

Through \eqref{eq:spin-eom}, the gradient term in $\mathcal H$ induces a precession $\bm\Omega_{\rm T}\times\bm s$ at the Thomas frequency~\cite{Thomas1926}
\begin{equation}
 \bm\Omega_{\rm T}=\frac{\bm v\times\nabla p}{(\varepsilon+p)(\gamma+1)}=\frac{\gamma^2}{\gamma+1}\,\dot{\bm v}\times\bm v ,
 \label{eq:thomas-rate}
\end{equation}
where $\dot{\bm v}=(\partial_t+\bm v\cdot\nabla)\bm v$. The canonical components $s^A$ refer to the pure-boost triad $e_A{}^\mu$ (laboratory axes boosted to the local rest frame without rotation), and the proper densities $s^A/\gamma$ form the spin four-vector $s^\mu=e_A{}^\mu s^A/\gamma$. Along an accelerated worldline this triad rotates at $-\bm\Omega_{\rm T}$ relative to the nonrotating FW triad (Fig.~\ref{fig:transport}), canceling the Thomas precession of the components: that precession reflects the rotation of the triad, and no torque acts. With four-velocity $u^\mu$, metric signature $(+,-,-,-)$, and $\varepsilon^{0123}=+1$, the spin four-vector satisfies
\begin{equation}
 u^\lambda\nabla_\lambda s^\mu+\theta s^\mu=-u^\mu a_\nu s^\nu
 -2\varepsilon^{\mu\nu\rho\sigma}u_\nu\omega_\rho s_\sigma ,
 \label{eq:fw}
\end{equation}
where $\theta=\nabla_\mu u^\mu$ is the expansion scalar, $a^\mu=u^\lambda\nabla_\lambda u^\mu$ the four-acceleration, and $\omega^\mu$ the covariant spin potential, which reduces to $\omega_A$ in the rest frame. The first term on the right enforces FW transport: without the internal precession about $\omega^\mu$, the spin does not rotate in the local rest frame and has no reversible coupling to shear, expansion, or vorticity.

At second order in spin and gradients the Poincar\'e algebra still closes without further degrees of freedom, once the energy density is evaluated at the covariant centroid and supplemented by the hidden-momentum term implied by the Frenkel condition $S^{\mu\nu}u_\nu=0$ on the spin tensor of a Weyssenhoff fluid~\cite{Frenkel1926,Weyssenhoff1947}, as we verified for three equations of state~\cite{Supplement}. Poincar\'e invariance also permits a Lorentz-scalar spin--vorticity energy $-\alpha\,s_\mu\Omega^\mu$ [$\alpha\,\bm s\cdot\bm\Omega$ in the rest frame, where $\bm\Omega=\tfrac12\nabla\times\bm v$]; it is absent in a Weyssenhoff fluid, and we set $\alpha=0$.

\textit{Choice of triad and pseudogauge.}---To distinguish torques from the motion of the triad, we expand $s^\mu=e_A{}^\mu s^A$ in an arbitrary spatial triad $e_A{}^\mu$ orthogonal to $u^\mu$ (taking $s^A$ here as proper densities). Applying the FW derivative $(\mathcal D_{\rm FW}s)^\mu=u^\lambda\nabla_\lambda s^\mu+u^\mu a_\nu s^\nu$, defined likewise for each $e_A{}^\mu$, gives
\begin{equation}
 (\mathcal D_{\rm FW}s)^\mu+\theta s^\mu=e_A{}^\mu(u^\lambda\partial_\lambda+\theta)s^A+s^A(\mathcal D_{\rm FW}e_A)^\mu .
 \label{eq:triad}
\end{equation}
The first term on the right is the Hamiltonian evolution of the components $s^A$; the second accounts for the rotation and deformation of the triad, which leaves the Casimir invariant~\eqref{eq:invariant} unchanged. It vanishes for the FW triad, cancels the Thomas precession for the pure-boost triad, and equals the velocity-gradient coupling $s^\nu\nabla_\nu u^\mu$ for a triad Lie-transported by the flow (Fig.~\ref{fig:transport}). In the FW triad the only reversible term is the internal precession $2\bm\omega\times\bm s$, which vanishes for an isotropic equation of state $\varepsilon(n,s,\bm s^2)$.

A different spin supplementary condition shifts the relativistic centroid and induces a pseudogauge transformation of the conserved currents. Like any pseudogauge transformation built from the hydrodynamic fields, it is a change of variables: the brackets and the form of $K_i$ in terms of the fields change, but not the Poincar\'e charges, the total entropy and its production, or the FW transport of the rest-frame spin defined by the Pauli--Luba\'nski vector of a fluid element. The pseudogauge transformation from the convective spin current $u^\lambda S^{\mu\nu}$ to the totally antisymmetric one of Ref.~\cite{Fang2026} adds $\nabla_\lambda(u^\nu S^{\mu\lambda})$ to the energy--momentum tensor~\cite{Daher2023}; the energy--momentum tensor of that work reproduces this term only at global equilibrium, so its reversible couplings are not a pseudogauge effect~\cite{Supplement}.

\begin{figure*}[t]
 \includegraphics[width=\textwidth]{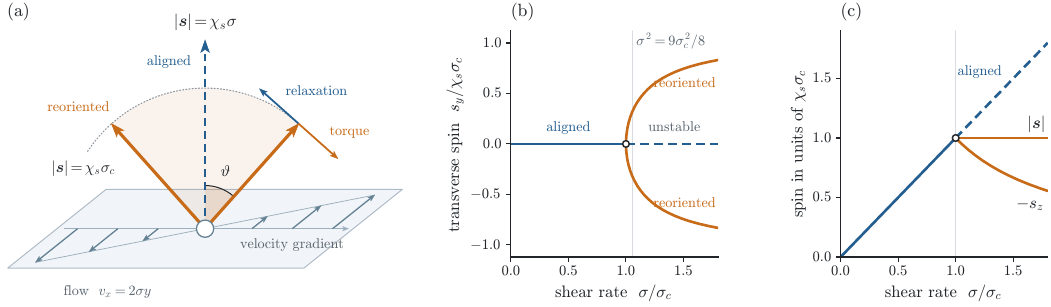}
 \caption{Shear-driven spin bifurcation. (a) Simple shear (flow plane shaded) at $\sigma=1.34\,\sigma_c$: relaxation balances the shear torque in the two stationary states, reoriented by $\vartheta$ toward the velocity gradient. (b) Supercritical pitchfork bifurcation of \eqref{eq:ode}: the aligned branch (blue) is unstable beyond $\sigma_c$ (dashed); perturbations of the reoriented branches (orange) decay monotonically up to the thin line and as damped oscillations beyond. (c) Stationary spin magnitude $|\bm s|$ and component $-s_z$ along $\bm\Omega$, in units of $\chi_s\sigma_c$.}
 \label{fig:bifurcation}
\end{figure*}

\textit{Dissipative dynamics.}---The flow thus exerts no reversible torque on the spin; any torque must be dissipative. The general equation for the nonequilibrium reversible--irreversible coupling (GENERIC)~\cite{Grmela1997,Ottinger1997} reads $\partial_tF=\{F,H\}+(F,\mathcal S)$, where $\mathcal S=\int s$ is the total entropy and $(\cdot,\cdot)$ a symmetric, positive-semidefinite bracket required to conserve $H$, $P_i$, and $J_i$. Its Onsager matrix relates the fluxes to the thermodynamic forces: the difference between twice the spin potential and the vorticity, ${2\omega_A-\Omega_A}$, the shear tensor $\sigma_{ij}=\tfrac12(\partial_iv_j+\partial_jv_i)-\tfrac13\delta_{ij}\partial_kv_k$, and the expansion scalar $\theta$. An isotropic Onsager matrix gives the standard spin relaxation $\partial_ts^A=-\lambda{(2\omega_A-\Omega_A)}$ with rotational viscosity $\lambda$ and an antisymmetric stress that returns the angular momentum to the orbital motion~\cite{Hattori2019,Hongo2021}, but no coupling of spin to shear.

When the relaxation rate $\Gamma_s=\lambda/\chi_s$ (for spin energy $\bm s^2/2\chi_s$ with susceptibility $\chi_s$) exceeds the hydrodynamic rates, the spin is in equilibrium with the vorticity, $\bm s=\chi_s\bm\Omega$, and the theory reduces to second-order hydrodynamics with a symmetric stress tensor~\cite{Li2021,Hongo2021}. Eliminating the spin within the Lie--Poisson structure leaves a nondissipative transport coefficient: because the Belinfante momentum density $\bm\pi+\tfrac12\nabla\times\bm s$ contains the spin, which the commutator of two shear deformations rotates, the rest-frame momentum flux acquires the contribution~\cite{Supplement}
\begin{equation}
 \Pi^{\rm odd}_{ij}=\tfrac12\left(\epsilon_{ikl}\sigma_{kj}+\epsilon_{jkl}\sigma_{ki}\right)s_l,\qquad\bm s=\chi_s\bm\Omega ,
 \label{eq:odd}
\end{equation}
an odd viscosity~\cite{Banerjee2017,Markovich2021} equal to $\tfrac12|\bm s|$, implicitly present in the pseudogauge-transformed stress tensor of Ref.~\cite{Li2021}. Because vortex stretching changes $|\bm\Omega|/n$ while the Casimir invariant~\eqref{eq:invariant} fixes $|\bm s|/n$, Hamiltonian dynamics alone cannot maintain $\bm s=\chi_s\bm\Omega$: it is a nonholonomic constraint enforced by fast dissipation, like rolling without slipping.

At finite spin density, however, the local state is anisotropic and the Onsager matrix may depend on $\bm s$. Because $\bm s$ and the forces $({2\omega_A-\Omega_A},\sigma_{ij},\theta)$ are all odd under time reversal, Onsager--Casimir reciprocity~\cite{Onsager1931,Casimir1945} makes the symmetric (dissipative) part of the matrix even in $\bm s$ and the antisymmetric (reversible) part odd, leaving eight dissipative and six reversible tensor structures~\cite{Supplement}. The Lie--Poisson bracket~\eqref{eq:bracket} and an isotropic equation of state fix the six reversible coefficients to zero; one of them would change $|\bm s|/n$ through shear, violating the Casimir invariant~\eqref{eq:invariant}. Of the eight dissipative structures, three are the isotropic shear, bulk, and rotational viscosities, four make them anisotropic, and a single tensor couples spin relaxation directly to shear:
\begin{equation}
 \Xi_{Aij}=\xi(\bm s^2)\left(\epsilon_{iAk}s_ks_j+\epsilon_{jAk}s_ks_i\right).
 \label{eq:A}
\end{equation}
Keeping the isotropic viscosities and $\xi$, with $u^\mu$ fixed by $u_\mu\Theta^{\mu\nu}=\varepsilon u^\nu$ (zero rest-frame momentum density), the energy--momentum tensor and the spin current read, to first order in gradients and all orders in spin,
\begin{equation}
\begin{aligned}
 \Theta^{\mu\nu}={}&\varepsilon\,u^\mu u^\nu-p\Delta^{\mu\nu}+S^{\mu\rho}a_\rho u^\nu\\
 &+2\eta\sigma^{\mu\nu}+\zeta\theta\Delta^{\mu\nu}+\xi\,(Y^\mu s^\nu+s^\mu Y^\nu)\\
 &-\tfrac12\varepsilon^{\mu\nu\rho\sigma}u_\rho[\lambda(2\omega_\sigma-\Omega_\sigma)+Z_\sigma],\\
 \Sigma^{\lambda\mu\nu}={}&u^\lambda S^{\mu\nu}-\varepsilon^{\mu\nu\rho\sigma}u_\rho\,j^\lambda_{{\rm s}\sigma},
\end{aligned}\label{eq:cross}
\end{equation}
where $\eta$ and $\zeta$ are the shear and bulk viscosities, $\Delta^{\mu\nu}=g^{\mu\nu}-u^\mu u^\nu$, $S^{\mu\nu}=-\varepsilon^{\mu\nu\rho\sigma}u_\rho s_\sigma$ is the spin tensor, $\sigma^{\mu\nu}=\Delta^{\mu\alpha}\Delta^{\nu\beta}\nabla_{(\alpha}u_{\beta)}-\tfrac13\theta\Delta^{\mu\nu}$, and $j_{\rm s}^{\lambda\mu}$ is the spin diffusion current. Here $Y^\mu=\varepsilon^{\mu\nu\rho\sigma}u_\nu{(2\omega_\rho-\Omega_\rho)}s_\sigma$ and $Z^\mu=-2\xi\,\varepsilon^{\mu\nu\rho\sigma}u_\nu s_\rho\sigma_{\sigma\lambda}s^\lambda$, with the kinematic vorticity $\Omega^\mu=\tfrac12\varepsilon^{\mu\nu\rho\sigma}u_\nu\nabla_\rho u_\sigma$. Combined with particle-number conservation, $\nabla_\mu\Theta^{\mu\nu}=0$ and $\nabla_\lambda\Sigma^{\lambda\mu\nu}=-2\Theta^{[\mu\nu]}$ are the equations of motion. In this frame the relativistic hidden momentum required by angular-momentum conservation appears as the energy flux $S^{\mu\rho}a_\rho$ ($\bm s\times\bm a$ in the rest frame). The antisymmetric stress $\Theta^{[\mu\nu]}$ contains the isotropic relaxation $\lambda{(2\omega^\mu-\Omega^\mu)}$ and the dissipative shear torque $-Z^\mu$ [$-2\xi\,\bm s\times(\bm\sigma\bm s)$ in the rest frame], which rotates $\bm s$ at fixed magnitude and exchanges angular momentum with the orbital motion; $\xi(Y^\mu s^\nu+s^\mu Y^\nu)$ is the Onsager-reciprocal symmetric shear stress.

Positivity of entropy production requires $\xi^2(\bm s^2)^2\le\lambda\eta$, which bounds $|\xi|$ but not its sign. The coefficient $\xi$ cannot be removed by field redefinitions and is fixed microscopically by a Green--Kubo formula, the time integral of the correlator of the antisymmetric stress (torque) with the symmetric shear stress at finite spin density. The isotropic dissipative relations of Ref.~\cite{Fang2026} omit $\xi$, although its power counting, $s^\mu=O(1)$ and $\omega^\mu=O(\partial)$, places $\xi$ at the same order~\cite{Supplement}; GENERIC supplies it systematically through the $\bm s$-dependent Onsager matrix.

\textit{Shear-driven spin reorientation.}---The torque $-2\xi\,\bm s\times(\bm\sigma\bm s)$ is quadratic in the spin density on which it acts, unlike momentum-dependent shear-induced polarization~\cite{Fu2021,Becattini2021}, which is linear in the shear and generates no net spin density, and unlike the corresponding couplings of polar active fluids, built on a vector order parameter distinct from the angular momentum~\cite{Markovich2019,MarkovichPRL2019}. It shares the algebraic structure of the Euler equations of a rigid body~\cite{LandauMechanics}, making alignment with the intermediate eigenvector of $\bm\sigma$ a saddle point of the rotation it generates.

In homogeneous simple shear, $v_x=2\sigma y$ [Fig.~\ref{fig:bifurcation}(a)], $\sigma_{xy}=\sigma=(\tfrac12\sigma_{ij}\sigma_{ij})^{1/2}$ and the vorticity $\bm\Omega=-\sigma\hat{\bm z}$ lies along the intermediate eigenvector of $\bm\sigma$, with eigenvalue $0$ between $\pm\sigma$. With $\omega_A=s_A/2\chi_s$, the rest-frame spin obeys
\begin{equation}
 \frac{d\bm s}{dt}=-\Gamma_s(\bm s-\chi_s\bm\Omega)-2\xi\,\bm s\times(\bm\sigma\bm s) .
 \label{eq:ode}
\end{equation}
The shear torque vanishes in the vorticity-aligned state $\bm s=\chi_s\bm\Omega$, but this state sits at the saddle point. For $\xi>0$ the torque amplifies perturbations of the spin along the velocity-gradient ($\hat{\bm y}$) axis at rate $2\xi\chi_s\sigma^2$ and damps those along the flow ($\hat{\bm x}$) axis, while relaxation restores the aligned state at rate $\Gamma_s$. When the amplification rate exceeds $\Gamma_s$,
\begin{equation}
 2\xi\chi_s\sigma^2>\Gamma_s\quad\Longleftrightarrow\quad\sigma>\sigma_c=\frac{\Gamma_s}{\sqrt{2\xi\lambda}} ,
 \label{eq:R}
\end{equation}
the aligned state loses stability in a supercritical pitchfork bifurcation to two stationary states [Fig.~\ref{fig:bifurcation}(b)] that break the symmetry under rotation by $\pi$ about $\bm\Omega$. For $\sigma>\sigma_c$ the spin has the shear-independent magnitude $|\bm s|=\chi_s\sigma_c=\sqrt{\lambda/2\xi}$ and is reoriented from $\bm\Omega$ toward $\pm\hat{\bm y}$ by $\vartheta=\arccos(\sigma_c/\sigma)$ [Fig.~\ref{fig:bifurcation}(c)]. For $\xi<0$ it reorients toward the flow axis $\pm\hat{\bm x}$, with $\xi\to|\xi|$ in \eqref{eq:R}. For the aligned spin, $|\bm s|=\chi_s\sigma$, the threshold becomes $2|\xi|\bm s^2=\lambda$, which does not involve $\chi_s$ and holds for arbitrary functions $\chi_s(\bm s^2)$, $\lambda(\bm s^2)$, and $\xi(\bm s^2)$ whenever $\bm\Omega$ lies along the intermediate eigenvector of $\bm\sigma$, as in any incompressible planar flow~\cite{Supplement}.

In three-dimensional flows, where $\bm\Omega$ is in general not an eigenvector of $\bm\sigma$, the torque deflects the stationary spin from the vorticity even for $\sigma\ll\sigma_c$, by $\delta\bm s=-(2\xi\chi_s^2/\Gamma_s)\,\bm\Omega\times(\bm\sigma\bm\Omega)$ at leading order in $\xi$, a direct signature of the dissipative spin--shear coupling.

\textit{Conclusions.}---Without dissipation, a flow exerts no torque on the spin: the commutation relations and the Poincar\'e algebra allow reversible shear couplings only as the effect of a deforming triad. All torques are therefore dissipative: relaxation toward the vorticity, whose fast limit leaves the odd viscosity $\tfrac12|\bm s|$, and a single first-order shear torque, with coefficient $\xi$, which reorients the spin above $\sigma_c$. Evaluating the Green--Kubo formula for $\xi$ will show whether the quark--gluon plasma reaches this regime.

\begin{acknowledgments}
\textit{Acknowledgments.}---We thank Jin Hu and Yi Yin for fruitful discussions, and Giorgio Torrieri for insightful questions. Large language models (OpenAI GPT and Anthropic Claude) assisted with symbolic cross-checks, numerical code, figure preparation, and text editing. The authors take full responsibility for the content of this work. This work was supported in part by the National Natural Science Foundation of China (NSFC) under Grant Nos. 12235016 and 12221005.
\end{acknowledgments}
\nocite{Baier2008,Becattini2024,BecattiniFlorkowskiSperanza2019,Belinfante1940,Bloch2005,Braginskii1965,Chudo2014,Chudo2015,Dixon1970,FangHattoriHu2025,Foskett2019,GayBalmaz2009,GayBalmaz2011,Haehl2015,Hu2023,Jantzen1992,Kibble1979,Kirillov1962,Kostant1970,Kudo2010,Leanhardt2002,Maugin1976,Monteiro2023,Montenegro2017,Obukhov1987,Olmsted2008,Rosenfeld1940,Schwinger1963,Simon1983,Son2026,Souriau1967,Tulczyjew1959,YangYan2024}
\bibliography{references}
\end{document}


\maketitle

This Supplemental Material contains the derivations, comparisons and numerical checks behind the Letter:
\begin{itemize}
\item \textbf{Section~\ref{sec:operators}}: the Lie--Poisson bracket~\Lbracket{} follows from the commutators of the constituents and is the only one of its class; for every Hamiltonian it gives the spin equation~\Lspineom{} and the Casimir invariant~\Linvariant{}.
\item \textbf{Section~\ref{sec:circulation}}: the circulation of a comoving loop changes only together with the Berry phase of the spins on the loop, \Lcirculation{}.
\item \textbf{Section~\ref{sec:gp-algebra}}: in a spinor condensate, described by the Gross--Pitaevskii (GP) equation, a fluid element retains multipoles beyond the spin dipole; the bracket~\Lcondensate{} and the GP energy give the spin current~\Lcurrent{}, which changes the spin magnitude per particle without dissipation, and the condensate quantizes what the circulation law conserves.
\item \textbf{Section~\ref{sec:poincare}}: the Poincar\'e algebra fixes the time--space component of the spin density and requires the Thomas term in the energy, \Lthomas{}; the canonical components then precess at the Thomas rate while their pure-boost triad rotates at the opposite rate, so the spin four-vector is Fermi--Walker (FW) transported, \Lfw{}. The section ends with the completion at second order in the spin.
\item \textbf{Section~\ref{sec:frames}}: by the triad law~\Ltriad{}, a term of a spin equation is a torque only if it remains in the FW triad. A momentum redefinition of Belinfante type and a change of spin supplementary condition, which acts on the currents as a pseudogauge transformation, are changes of variables: the charges, the entropy and its production, and the transport of the rest-frame spin do not change.
\item \textbf{Section~\ref{sec:generic}}: in the general equation for the nonequilibrium reversible--irreversible coupling (GENERIC) the spin relaxes toward the vorticity, and at finite spin density the Onsager matrix contains a single dissipative coupling between spin relaxation and shear, \LA{}, with its Onsager-reciprocal stress, positivity bound, Green--Kubo formula and power counting.
\item \textbf{Section~\ref{sec:rel-diss}}: the covariant constitutive relations~\Lcross{}, checked against the Ward identity and the second law, and the covariant circulation law.
\item \textbf{Section~\ref{sec:fast-relaxation}}: when relaxation is fast, friction holds the spin at $\chi_s\bm\Omega$, a nonholonomic constraint; eliminating the spin leaves the odd (Hall) viscosity~\Lodd{}, which a fluid with the same spin density inserted into an action lacks.
\item \textbf{Section~\ref{sec:fang}}: the construction of Ref.~\cite{Fang2026} uses the same spin density. Its constitutive relations differ from \Lcross{} by the term that a pseudogauge transformation to its totally antisymmetric spin current adds to the energy--momentum tensor, by reversible couplings odd in the spin, which no change of variables generates and which persist even without a spin--orbit interaction, and by the absence of~\LA{}.
\item \textbf{Section~\ref{sec:bifurcation}}: the shear-driven reorientation, \Lode{} and~\LR{}, for general coefficient functions and flows, its stresses and its stability at finite wave number, and the modulated and banded states.
\end{itemize}
Equations of the Letter are quoted by their numbers there, \Lbracket{}--\LR{}; equations of this Supplemental Material carry the prefix S.

Throughout, the reversible evolution of a functional $F$ of the fields is $\partial_tF=\{F,H\}$ with the Hamiltonian $H$, the Levi-Civita symbol has $\epsilon_{123}=+1$, the speed of light and the Boltzmann constant are $c=k_B=1$, and $\hbar$ is kept in microscopic and condensate expressions. Spatial Cartesian indices are contracted with $\delta_{ij}$, and Minkowski spacetime has signature $(+,-,-,-)$. Internal spin indices are $A,B,C$, laboratory indices $i,j,k$ and spacetime indices $\mu,\nu,\rho,\sigma$; $s$ without an index is the entropy density, while $s^A$, $\bm s$ and $s^\mu$ denote the spin density. Functional derivatives are written out in full. Bold symbols denote three-vectors and, in index-free products, the $3\times3$ matrices that act on them, such as the shear tensor $\bm\sigma$ with components $\sigma_{ij}$, the velocity gradient $\nabla\bm v$ and the unit matrix $\mathbf 1$; components carry indices and are lightface, and $\mathsf T$ denotes the transpose. Lorentz transformations and the matrices of the internal space of a condensate are lightface; $\bm s^2$ is the squared norm of the spin density and $|\bm s|$ its magnitude. Parentheses and square brackets around indices denote symmetrization and antisymmetrization with the factor $\tfrac12$, as in $\partial_{(i}v_{j)}=\tfrac12(\partial_iv_j+\partial_jv_i)$, and angle brackets the transverse, symmetric and traceless part. The derivations were checked symbolically or numerically with independent Python and Mathematica scripts.

\section{Lie--Poisson Bracket of a Fluid with Spin}
\label{sec:operators}

The bracket~\Lbracket{} of the hydrodynamic fields is inherited from the commutators of the microscopic observables of which the fields are coarse-grained averages (Sec.~\ref{sec:coarse}); within its class it admits no deformation (Sec.~\ref{sec:uniqueness}), and for every Hamiltonian it gives \Lspineom{} and \Linvariant{} (Sec.~\ref{sec:jacobi}).

\subsection{Coarse graining of the microscopic spin algebra}
\label{sec:coarse}
The starting point is the algebra of the positions $\widehat X_{ai}$, the momenta $\widehat P_{ai}$ and the spins $\widehat S_{aA}$ of the particles,
\begin{equation}
 [\widehat X_{ai},\widehat P_{bj}]=i\hbar\delta_{ab}\delta_{ij},\quad
 [\widehat S_{aA},\widehat S_{bB}]=i\hbar\delta_{ab}\epsilon_{ABC}\widehat S_{aC},\quad
 [\widehat S_{aA},\widehat X_{bi}]=[\widehat S_{aA},\widehat P_{bi}]=0 .
 \label{eq:micro}
\end{equation}
Here $a,b$ label particles, $A,B,C=1,2,3$ spin components and $i,j,k=1,2,3$ Cartesian coordinates; positions commute among themselves, and so do momenta. Simultaneous rotations of space and spin are generated by the total angular momentum $\sum_a(\widehat{\bm X}_a\times\widehat{\bm P}_a+\widehat{\bm S}_a)$. The spin commutator is the Lie algebra $\mathfrak{su}(2)\simeq\mathfrak{so}(3)$ itself, so neither the value of the spin nor the statistics enters the bracket: integer and half-integer spins differ only globally, the latter acquiring a factor $-1$ under a rotation by $2\pi$, and the statistics enters through the admissible states and the energy. Section~\ref{sec:uniqueness} shows which requirements lead to the factorized form~\eqref{eq:micro}.

The hydrodynamic variable is the spin of a cell divided by its volume. Let $\chi_\alpha(\bm x)$ be the indicator function of the cell $C_\alpha$ of volume $V_\alpha$, with $\chi_\alpha\chi_\beta=\delta_{\alpha\beta}\chi_\alpha$ for disjoint cells. The spin-density operator of the cell and its commutator are
\begin{equation}
\widehat s_\alpha^A=\frac1{V_\alpha}\sum_a
 \chi_\alpha(\widehat{\bm X}_a)\widehat S_{aA},\qquad [\widehat s_\alpha^A,\widehat s_\beta^B]
 =\frac{i\hbar\epsilon_{ABC}}{V_\alpha V_\beta}
 \sum_a\chi_\alpha(\widehat{\bm X}_a)\chi_\beta(\widehat{\bm X}_a)\widehat S_{aC}
 =\frac{i\hbar\delta_{\alpha\beta}}{V_\alpha}\epsilon_{ABC}\widehat s_\alpha^C .
 \label{eq:cell-derivation}
\end{equation}
Only terms with the same particle contribute. Because $\widehat s_\alpha^A$ is a density, one factor $1/V_\alpha$ remains on the right-hand side; it becomes the delta function of the continuum bracket. The product $\chi_\alpha(\widehat{\bm X}_a)\widehat S_{aA}$ preserves the correlation between the position and the spin of each particle.

Expectation values inherit a bracket from the commutator. For a density operator $\widehat\rho$ the Lie--Poisson bracket of two functions of the state is~\cite{Marsden1984,Bonet2016}
\begin{equation}
 \{F,G\}_{\widehat\rho}
 =\frac{1}{i\hbar}\operatorname{Tr}\!\left(\widehat\rho
 \left[\frac{\delta F}{\delta\widehat\rho},
       \frac{\delta G}{\delta\widehat\rho}\right]\right),
 \label{eq:state-bracket}
\end{equation}
where $\operatorname{Tr}$ is the trace over the many-body Hilbert space ($\tr$ is reserved for traces over the internal spin space). For functions of the cell expectations $s_\alpha^A=\operatorname{Tr}(\widehat\rho\widehat s_\alpha^A)$, the chain rule and \eqref{eq:cell-derivation} give
\begin{equation}
 \frac{\delta F}{\delta\widehat\rho}
 =\sum_{\alpha,A}\frac{\partial F}{\partial s_\alpha^A}\widehat s_\alpha^A,
 \qquad
 \{F,G\}=\sum_\alpha\frac{\bm s_\alpha}{V_\alpha}\cdot
 \left(\frac{\partial F}{\partial\bm s_\alpha}\times
       \frac{\partial G}{\partial\bm s_\alpha}\right).
 \label{eq:expectation}
\end{equation}
The factor $\hbar$ of the commutator cancels against the prefactor of the bracket. With an energy that is a function of the retained variables, the bracket~\eqref{eq:expectation} generates their reversible evolution.

At finite resolution the spin field is piecewise constant, $s^A(\bm x)=\sum_\alpha\chi_\alpha(\bm x)s_\alpha^A$, and the points of one cell share one variable:
\begin{equation}
 \{s^A(\bm x),s^B(\bm y)\}
 =\epsilon_{ABC}\sum_\alpha
 \frac{\chi_\alpha(\bm x)\chi_\alpha(\bm y)}{V_\alpha}s_\alpha^C.
 \label{eq:finite-kernel}
\end{equation}
The continuum limit applies when the cells are small compared with the scale on which the fields vary and large enough to average over microscopic fluctuations. For a smooth functional $F[s^A]$, $\partial F/\partial s_\alpha^A\simeq V_\alpha\,\delta F/\delta s^A(\bm x_\alpha)$; in \eqref{eq:expectation} the two derivatives contribute $V_\alpha^2$, which with the prefactor $1/V_\alpha$ leaves the volume element of a Riemann sum, $\sum_\alpha V_\alpha\to\int d^3x$. Equivalently, $\sum_\alpha\chi_\alpha(\bm x)\chi_\alpha(\bm y)/V_\alpha\to\delta^{(3)}(\bm x-\bm y)$, and
\begin{equation}
 \{s^A(\bm x),s^B(\bm y)\}
 =\epsilon_{ABC}s^C(\bm x)\delta^{(3)}(\bm x-\bm y).
 \label{eq:continuum-spin}
\end{equation}
This is the first line of \Lbracket{} of the Letter. The limit is an approximation in spatial resolution and leaves the spin algebra unchanged. Whether the spin density is a hydrodynamic variable at all is a separate, dynamical question: its relaxation must be slow compared with the microscopic rates.

\begin{figure}[t]
 \centering
 \includegraphics[width=0.36\textwidth]{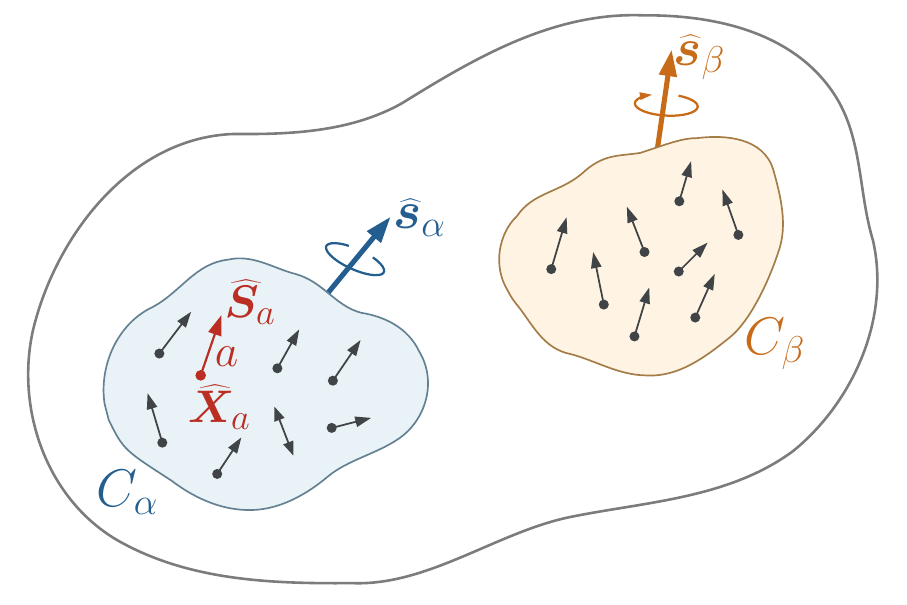}
 \caption{Coarse graining of microscopic spins (schematic). Summing the particle spins within a fluid cell gives the total cell spin, and dividing by the cell volume defines the local spin density. The volume normalization yields the inverse-volume factor in the finite-cell bracket, which becomes a delta function in the continuum limit.}
 \label{fig:cells}
\end{figure}

The second line of \Lbracket{} expresses that the momentum density generates the transport of the other densities. To derive it we smear the microscopic operators with smooth test functions $\phi(\bm x)$, $\theta_A(\bm x)$ and $w^i(\bm x)$:
\begin{equation}
 \widehat N[\phi]=\sum_b\phi(\widehat{\bm X}_b),\quad
 \widehat S[\bm\theta]=\sum_b\theta_A(\widehat{\bm X}_b)\widehat S_{bA},\quad
 \widehat P[\bm w]=\frac12\sum_b
 \bigl(w^i(\widehat{\bm X}_b)\widehat P_{bi}+\widehat P_{bi}w^i(\widehat{\bm X}_b)\bigr).
 \label{eq:smears}
\end{equation}
In the position representation the symmetrized momentum operator of one particle acts as $-i\hbar[w^i\partial_i+\tfrac12(\partial_i w^i)]$, and the commutators are
\begin{equation}
\begin{gathered}
\frac{[\widehat P[\bm w],\widehat P[\bm w']]}{i\hbar}=-\widehat P[[\bm w,\bm w']],\qquad \frac{[\widehat P[\bm w],\widehat N[\phi]]}{i\hbar}=-\widehat N[\bm w\cdot\nabla\phi],\\
\frac{[\widehat P[\bm w],\widehat S[\bm\theta]]}{i\hbar}=-\widehat S[\bm w\cdot\nabla\bm\theta],\qquad \frac{[\widehat S[\bm\theta],\widehat S[\bm\theta']]}{i\hbar}=\widehat S[\bm\theta\times\bm\theta'],
\end{gathered}
 \label{eq:smeared-algebra}
\end{equation}
where $[\bm w,\bm w']^i=w^j\partial_j w'^i-w'^j\partial_j w^i$ is the Lie bracket of vector fields. The expectation values are the functionals $P[\bm w]=\int d^3x\,w^i\pi_i$, $N[\phi]=\int d^3x\,\phi n$ and $S[\bm\theta]=\int d^3x\,\theta_A s^A$ of the momentum density $\pi_i$, the particle density $n$ and the spin density $s^A$. The entropy density $s$ is included as a further density advected by the momentum. Removing the test functions gives the remaining nonvanishing brackets,
\begin{equation}
\begin{gathered}
 \{\pi_i(\bm x),\pi_j(\bm y)\}
 =-\pi_j(\bm x)\partial_{x^i}\delta^{(3)}(\bm x-\bm y)
   +\pi_i(\bm y)\partial_{y^j}\delta^{(3)}(\bm x-\bm y),\qquad \{\pi_i(\bm x),n(\bm y)\}=n(\bm x)\partial_{y^i}\delta^{(3)}(\bm x-\bm y),\\
 \{\pi_i(\bm x),s(\bm y)\}=s(\bm x)\partial_{y^i}\delta^{(3)}(\bm x-\bm y),\qquad \{\pi_i(\bm x),s^A(\bm y)\}=s^A(\bm x)\partial_{y^i}\delta^{(3)}(\bm x-\bm y).
\end{gathered}
 \label{eq:local-brackets}
\end{equation}
The densities $n$ and $s$ commute with $s^A$ and with each other, and the brackets with exchanged arguments follow by antisymmetry. Boundary terms are assumed to vanish, by periodicity or by decay of the fields. Figure~\ref{fig:cells} illustrates the construction.

\subsection{Representation conditions and uniqueness}
\label{sec:uniqueness}
We prove that \Lbracket{} is the only local bracket, linear in the fields and of first order in gradients, that satisfies the Jacobi identity if the total angular momentum generates simultaneous rotations of the positions, momenta and spins of the constituents. The algebra of the total angular momentum alone does not fix the commutators of the spin with positions and momenta, so we first state the conditions that do; the Jacobi identity then excludes every deformation of the resulting hydrodynamic bracket.

\emph{Representation conditions.}
Let $\widehat J^{ij}=\widehat L^{ij}+\sum_a\widehat S_a^{ij}$ be the total angular momentum, with the orbital part $\widehat L^{ij}=\sum_a(\widehat X_a^i\widehat P_a^j-\widehat X_a^j\widehat P_a^i)$ and the spin of particle $a$ written as the antisymmetric tensor $\widehat S_a^{ij}=\epsilon_{ijk}\widehat S_{ak}$. For any unitary $\widehat U$, $\widehat J'=\widehat U\widehat J\widehat U^{-1}$ obeys the same $\mathfrak{so}(3)$ relations, and so does the ``spin'' $\widehat S'=\widehat J'-\widehat L$ in the sense of closure. For one spin-$\tfrac12$ particle with position $(\widehat X,\widehat Y,\widehat Z)$, the Pauli matrix $\sigma_z$, a constant $\alpha_0$ and $\widehat U=\exp(i\alpha_0\widehat Z\sigma_z/2)$, however, $[\widehat J'_x,\widehat P_z]+i\hbar\widehat P_y=O(\alpha_0)\neq0$: $\widehat J'$ closes but does not rotate the momentum as a vector. The content is therefore in the following conditions:
(R1) $\widehat J^{ij}$ rotates $\widehat{\bm X}_a$ and $\widehat{\bm P}_a$ as vectors;
(R2) $\widehat J^{ij}$ rotates the laboratory spin $\widehat S_a^{kl}$ as a tensor;
(L) spin operators of different particles commute, and $\widehat{\bm S}_a$ commutes with $\widehat{\bm X}_b,\widehat{\bm P}_b$ for $b\neq a$.
Since the orbital part already satisfies (R1), that condition reads $[\sum_b\widehat S_b^{ij},\widehat X_a^k]=0=[\sum_b\widehat S_b^{ij},\widehat P_a^k]$, and (L) restricts the sum to $b=a$. Then $[\widehat L^{ij},\widehat S_a^{kl}]=0$, and (R2) becomes the $\mathfrak{so}(3)$ algebra of $\widehat{\bm S}_a$. Hence (R1), (R2) and (L) imply \eqref{eq:micro} for the laboratory spin. If the spin is resolved in a position-dependent orthonormal triad, $\widehat S_a^A=e^A{}_i(\widehat{\bm X}_a)\widehat S_a^i$, the only change is
\begin{equation}
 [\widehat S_a^A,\widehat P_{bk}]=-i\hbar\,\delta_{ab}\,\Gamma_k{}^A{}_B(\widehat{\bm X}_a)\widehat S_a^B,\qquad
 \Gamma_k{}^A{}_B=e^A{}_j\,\partial_ke^j{}_B ,
 \label{eq:frame-connection}
\end{equation}
a flat connection, $\partial_k\bm\Gamma_l-\partial_l\bm\Gamma_k+[\bm\Gamma_k,\bm\Gamma_l]=0$ for the matrices $(\bm\Gamma_k)^A{}_B=\Gamma_k{}^A{}_B$, which adds $\Gamma_i{}^A{}_Bs^B\delta(\bm x-\bm y)$ to $\{\pi_i(\bm x),s^A(\bm y)\}$. It is the connection~\eqref{eq:mixed-frame} of a position-dependent change of triad (Sec.~\ref{sec:frames}).

\emph{Jacobi identity.}
Write $P[\bm w]=\int d^3x\,w^i\pi_i$ and $S[\bm\theta]=\int d^3x\,\theta_As^A$ for smooth test fields $\bm w(\bm x)$ and $\bm\theta(\bm x)$. Keep the ideal-fluid brackets, $\{s^A,s^B\}=\epsilon_{ABC}s^C\delta$ and $\{n,s^A\}=0$, and allow the most general mixed bracket that is local, linear in the fields, of first order in gradients, and relates spatial and internal indices only through the triad:
\begin{equation}
 \{P[\bm w],S[\bm\theta]\}=-S\bigl[w^k\partial_k\bm\theta+w^k\bm\Lambda_k\bm\theta+c_{\rm r}\,(\nabla\times\bm w)\times\bm\theta+\bm{\mathcal M}(\nabla\bm w)\,\bm\theta\bigr],
 \label{eq:general-mixed}
\end{equation}
with a matrix field $\bm\Lambda_k$, a constant $c_{\rm r}$, and a linear map $\bm{\mathcal M}$ from the symmetric part of $\nabla\bm w$ to $3\times3$ matrices. The Jacobi identity for $(S,S,P)$ requires $P[\bm w]$ to act on $(\mathbb R^3,\times)$ as a derivation. The derivations of the cross product form the Lie algebra $\mathfrak{so}(3)$ of infinitesimal rotations, so $\bm\Lambda_k$ and $\bm{\mathcal M}$ take values in it; an isotropic linear map from symmetric tensors to vectors vanishes, hence $\bm{\mathcal M}=0$. The identity for $(P,P,S)$ requires the action of $P[\bm w]$ on $\bm\theta$ to represent the Lie algebra of vector fields. For constant $\bm w,\bm w'$ the failure of the identity is the curvature of $\bm\Lambda$, so $\bm\Lambda_k$ is flat and can be removed by a position-dependent rotation of the spin axes; for linear fields $\bm w$ and $\bm w'$ with constant symmetric gradients the failure is $-c_{\rm r}\,\nabla\times[\bm w,\bm w']\times\bm\theta$, which vanishes only for $c_{\rm r}=0$. Therefore the bracket of Sec.~\ref{sec:coarse} is the only one in this class: neither corotational transport, the rotation of the spin with $\tfrac12\nabla\times\bm w$ ($c_{\rm r}=-\tfrac12$), nor Lie transport of the spin as a comoving line element (corotation with $\bm{\mathcal M}$ equal to the symmetric part of $\nabla\bm w$) is Hamiltonian for a variable with the $\mathfrak{so}(3)$ bracket, and no reversible bracket of this kind relaxes the conservation of $s^As^A/n^2$ (Sec.~\ref{sec:jacobi}). Both transports arise only in other variables: the momentum density $\pi'_i$ of Sec.~\ref{sec:hep-comparison} transports the spin corotationally but lies outside this class, because its bracket with itself, \eqref{eq:improved-self-bracket}, is not that of the generators of spatial diffeomorphisms; and components with the $\mathfrak{so}(3)$ bracket assigned to a Lie-transported triad give a Lie-transported vector whose own bracket depends on the triad (Sec.~\ref{sec:frames}). A reversible change of $|\bm s|/n$ requires non-canonical positions, as in the projected Dirac spin, $[\widehat s_i,\widehat s_j]=i\hbar\epsilon_{ijk}(\widehat s_k-\widehat P_k\,\widehat{\bm P}\cdot\widehat{\bm s}/\widehat E^2)$ with the momentum $\widehat{\bm P}$ and the energy $\widehat E$, or additional internal variables (Sec.~\ref{sec:gp-algebra}).

\subsection{Equations of motion and Casimir invariants}
\label{sec:jacobi}
For arbitrary functionals $F$ and $G$, the brackets \eqref{eq:continuum-spin} and~\eqref{eq:local-brackets} combine into
\begin{equation}
\begin{aligned}
 \{F,G\}=\int d^3x\,\Bigg[&
 \pi_i\left(\frac{\delta G}{\delta\pi_j}\partial_j\frac{\delta F}{\delta\pi_i}
           -\frac{\delta F}{\delta\pi_j}\partial_j\frac{\delta G}{\delta\pi_i}\right)+n\left(\frac{\delta G}{\delta\pi_j}\partial_j\frac{\delta F}{\delta n}
           -\frac{\delta F}{\delta\pi_j}\partial_j\frac{\delta G}{\delta n}\right)\\
 &+s\left(\frac{\delta G}{\delta\pi_j}\partial_j\frac{\delta F}{\delta s}
           -\frac{\delta F}{\delta\pi_j}\partial_j\frac{\delta G}{\delta s}\right)+s^A\left(\frac{\delta G}{\delta\pi_j}\partial_j\frac{\delta F}{\delta s^A}
           -\frac{\delta F}{\delta\pi_j}\partial_j\frac{\delta G}{\delta s^A}\right)
  +\epsilon_{ABC}s^C\frac{\delta F}{\delta s^A}\frac{\delta G}{\delta s^B}\Bigg].
\end{aligned}
 \label{eq:functional}
\end{equation}
This bracket satisfies the Jacobi identity because it is the Lie--Poisson bracket of a semidirect-product Lie algebra~\cite{Marsden1984,Holm1998}: the momentum and spin generators inherit the identity from the commutators~\eqref{eq:smeared-algebra}, and the entropy density transforms as $n$ does and has no internal bracket. The mixed identity for momentum and spin is the Leibniz rule for the Lie derivative of the cross product. For nonlinear functionals the terms with second variations cancel by symmetry, so that the identity reduces to that of the linear generators. Without the spin, the bracket~\eqref{eq:functional} is that of an ideal fluid in Ref.~\cite{Morrison1980}, with the particle density in place of the mass density.

No Hamiltonian has been specified so far; for any $H[n,s,\pi_i,s^A]$ define the conjugates of the fields, the velocity $v^i=\delta H/\delta\pi_i$, the chemical potential $\mu=\delta H/\delta n$, the temperature $T=\delta H/\delta s$ and the spin potential $\omega_A$ with $2\omega_A=\delta H/\delta s^A$, each at fixed remaining fields. Then $\partial_tF=\{F,H\}$ gives
\begin{equation}
\begin{gathered}
 \partial_t n=-\partial_i(nv^i),\quad \partial_t s=-\partial_i(sv^i),\qquad \partial_t s^A=-\partial_i(s^Av^i)+2\epsilon_{ABC}\omega_Bs^C,\\
 \partial_t\pi_i=-\partial_j(v^j\pi_i)-\pi_j\partial_i v^j
 -n\partial_i\mu-s\partial_iT-2s^A\partial_i\omega_A.
\end{gathered}
 \label{eq:eom-derived}
\end{equation}
The spin equation, \Lspineom{} of the Letter, follows from $\{S[\bm\theta],H\}=\int d^3x\,[s^Av^i\partial_i\theta_A+2\theta_A\epsilon_{ABC}\omega_Bs^C]$ after an integration by parts. The same mixed bracket $\{\pi_i,s^A\}$ produces the force $-2s^A\partial_i\omega_A$ in the momentum equation, the analogue of the Stern--Gerlach force.

With the comoving derivative $D_t=\partial_t+v^i\partial_i$, the densities $n$, $s$ and $s^A$ change with the volume of the fluid element at the same relative rate $-\partial_iv^i$, and the precession $2\epsilon_{ABC}\omega_Bs^C$ is perpendicular to the spin. For $n>0$, therefore,
\begin{equation}
 D_t\frac{s}{n}=0,\qquad
 D_t\frac{s^As^A}{n^2}=0,\qquad
 C_f=\int d^3x\,n f\!\left(\frac{s}{n},\frac{s^As^A}{n^2}\right),
 \quad \{C_f,H\}=0.
 \label{eq:casimirs}
\end{equation}
The second relation is \Linvariant{} of the Letter. These invariants follow from the bracket alone: every $C_f$ with smooth $f$ is a Casimir invariant of \eqref{eq:functional}, conserved for any $H$. A fluid element thus conserves its entropy per particle and the magnitude of its spin per particle, $|\bm s|/n$. The spheres of constant $|\bm s|/n$ are the coadjoint orbits of $\mathfrak{so}(3)^*$~\cite{Kirillov1962,Kostant1970,Souriau1967}; precession moves the spin on its sphere, and compression rescales $n$ and $s^A$ in proportion. The integral $\int d^3x\,\bm s^2$ is not conserved in a compressible flow, although $|\bm s|/n$ is.

\subsection{Local thermodynamics}
\label{sec:thermo}
Consider a local energy without gradient terms or external fields, $H=\int d^3x\,\varepsilon(s,n,\pi_i,s^A)$, with arbitrary dependence on $\pi_i$ and $s^A$. Here $\varepsilon$ is the energy density in the frame of the equal-time variables; in Sec.~\ref{sec:poincare}, where laboratory and rest-frame quantities occur together, this density is written $\mathcal H$ and $\varepsilon$ denotes its value in the local rest frame. The first law is
\begin{equation}
 T=\frac{\partial\varepsilon}{\partial s},\qquad
 \mu=\frac{\partial\varepsilon}{\partial n},\qquad
 v^i=\frac{\partial\varepsilon}{\partial\pi_i},\qquad
 2\omega_A=\frac{\partial\varepsilon}{\partial s^A},\qquad
 d\varepsilon=T\,ds+\mu\,dn+v^i\,d\pi_i+2\omega_A\,ds^A.
 \label{eq:first-law}
\end{equation}
The pressure is defined by the change of the energy under a uniform change of volume at fixed total entropy, particle number, momentum and spin, $p=Ts+\mu n+v^i\pi_i+2\omega_As^A-\varepsilon$. It obeys the Gibbs--Duhem relation
\begin{equation}
 dp=s\,dT+n\,d\mu+\pi_i\,dv^i+2s^A\,d\omega_A.
 \label{eq:gibbs-duhem}
\end{equation}
With \eqref{eq:gibbs-duhem} the momentum equation in \eqref{eq:eom-derived} becomes $D_t\pi_i=-\pi_i\partial_jv^j-\partial_i p$, and the chain rule for $\varepsilon(s,n,\pi_i,s^A)$ gives the conservation of energy,
\begin{equation}
\begin{aligned}
 D_t\varepsilon
 &=-(Ts+\mu n+v^i\pi_i+2\omega_As^A)\partial_jv^j
   -v^i\partial_i p+4\omega_A\epsilon_{ABC}\omega_Bs^C\\
 &=-(\varepsilon+p)\partial_jv^j-v^i\partial_i p,\qquad
 \partial_t\varepsilon+\partial_i[(\varepsilon+p)v^i]=0,
\end{aligned}\label{eq:energy}
\end{equation}
where the precession term $4\omega_A\epsilon_{ABC}\omega_Bs^C$ vanishes by antisymmetry. With gradient or nonlocal energies the total energy is still conserved, by the antisymmetry of the bracket, and the fluxes of momentum and energy acquire the corresponding stresses.

\section{Circulation Law and the Berry Phase of the Spin}
\label{sec:circulation}

The bracket also couples the momentum of the fluid to the orientation of its spins. Without spin, the circulation of the momentum per particle around a comoving loop is conserved where the entropy per particle is uniform (Kelvin's theorem); with spin, for every energy $H$, it changes only together with the Berry phase of the spins on the loop, \Lcirculation{} of the Letter. We derive the law in the canonical variables of one frame; its covariant form is derived in Sec.~\ref{sec:kelvin-cov}, and its quantized limit in a condensate in Sec.~\ref{sec:gp-bracket}.

\subsection{Circulation of a comoving loop}
\label{sec:kelvin}
Dividing the momentum equation in \eqref{eq:eom-derived} by $n$ and using $D_tn=-n\partial_jv^j$ gives the equation of motion of the momentum per particle,
\begin{equation}
 D_t\frac{\pi_i}{n}+\frac{\pi_j}{n}\,\partial_iv^j
 =-\partial_i\mu-\frac sn\,\partial_iT-\frac{s^A}{n}\,\partial_i(2\omega_A) .
 \label{eq:kelvin-momentum}
\end{equation}
A line element that moves with the fluid changes at the rate $D_t(dx^i)=dx^j\partial_jv^i$, so that the left-hand side, contracted with $dx^i$, is the rate of change of $(\pi_i/n)\,dx^i$ on such an element. For a closed loop $C$ that moves with the fluid the gradient of $\mu$ integrates to zero, and
\begin{equation}
 \frac{d}{dt}\oint_C\frac{\pi_i}{n}\,dx^i=-\oint_C\frac sn\,dT-\oint_C\frac{s^A}{n}\,d(2\omega_A) .
 \label{eq:kelvin-loop}
\end{equation}
The first term on the right is the baroclinic source of an ordinary fluid. The second is the work done around the loop, per particle, by the force $-2s^A\partial_i\omega_A$ of \eqref{eq:eom-derived}, with which a spin potential that varies in space acts on the fluid.

The same spin potential makes the spins precess. Write $s^A=|\bm s|\,\hat s_A$ with the unit vector $\hat{\bm s}$; by \eqref{eq:eom-derived} and~\eqref{eq:casimirs}, $D_t\hat{\bm s}=2\bm\omega\times\hat{\bm s}$. As the loop is traversed, $\hat{\bm s}$ traces a closed curve on the unit sphere. Let $\Phi_C$ be the solid angle that this curve encloses, defined modulo $4\pi$ and counted positive when the curve runs counterclockwise around the enclosed region as seen from outside the sphere. When the points of a closed curve on the unit sphere are displaced by $\delta\hat{\bm s}$, the enclosed solid angle changes by $\oint\hat{\bm s}\cdot(\delta\hat{\bm s}\times d\hat{\bm s})$. For $\delta\hat{\bm s}=2\bm\omega\times\hat{\bm s}\,\delta t$ the integrand is $2\omega_A\,d\hat s_A\,\delta t$, because $\hat{\bm s}\cdot d\hat{\bm s}=0$, and therefore
\begin{equation}
 \frac{d\Phi_C}{dt}=\oint_C2\omega_A\,d\hat s_A .
 \label{eq:kelvin-solid-angle}
\end{equation}
Let $|\bm s|/n$ be uniform on the loop; by \eqref{eq:casimirs} it remains so. The last term of \eqref{eq:kelvin-loop} is then $-(|\bm s|/n)\oint_C\hat s_A\,d(2\omega_A)=(|\bm s|/n)\oint_C2\omega_A\,d\hat s_A$, and with \eqref{eq:kelvin-solid-angle}
\begin{equation}
 \frac{d}{dt}\left[\oint_C\frac{\pi_i}{n}\,dx^i-\frac{|\bm s|}{n}\,\Phi_C\right]=-\oint_C\frac sn\,dT .
 \label{eq:kelvin-law}
\end{equation}
For uniform $s/n$ the right-hand side vanishes, which is \Lcirculation{} of the Letter. Only \eqref{eq:eom-derived} was used, so that the law holds for every $H$, including energies with gradient terms and with applied fields.

The quantity $(|\bm s|/n)\,\Phi_C$ is, in units of $\hbar$ and up to the sign convention for the phase, the geometric phase of Berry~\cite{Berry1984} for a spin of magnitude $|\bm s|/n$ whose direction is taken around the curve traced by $\hat{\bm s}$. The force that changes the circulation and the precession that changes the phase are generated by the same spin potential through the same bracket, $\{\pi_i,s^A\}$ and $\{s^A,s^B\}$ in \Lbracket{} of the Letter, and the invariance of $|\bm s|/n$ makes the two changes equal. Rotating the spins on a loop by angles that differ along it therefore sets the fluid on the loop into circulation.

\subsection{Local form and the Berry curvature}
\label{sec:kelvin-local}
For a surface $S$ that spans $C$ and on which $\hat{\bm s}$ is smooth, $\Phi_C$ is the area that the image of $S$ covers on the unit sphere. Where $|\bm s|/n$ is uniform on $S$, $(|\bm s|/n)\,\Phi_C$ is therefore the flux through $S$ of
\begin{equation}
 \mathcal B_i=\frac{|\bm s|}{2n}\,\epsilon_{ijk}\,\hat{\bm s}\cdot(\partial_j\hat{\bm s}\times\partial_k\hat{\bm s}) ,
 \label{eq:kelvin-berry}
\end{equation}
the Berry curvature of the spin per particle. The local form of \eqref{eq:kelvin-law}, without any assumption on $s/n$ and $|\bm s|/n$, is the equation of the vector $\bm W=\nabla\times(\bm\pi/n)-\bm{\mathcal B}$,
\begin{equation}
 \partial_t\bm W-\nabla\times(\bm v\times\bm W)+\bm v\,\nabla\cdot\bm W
 =-\nabla\frac sn\times\nabla T-\nabla\frac{|\bm s|}{n}\times\bigl(\hat s_A\nabla(2\omega_A)\bigr) .
 \label{eq:kelvin-local}
\end{equation}
The left-hand side is the rate of change of the flux of $\bm W$ through a surface element that moves with the fluid. Where $s/n$ and $|\bm s|/n$ are uniform, this flux is frozen into the fluid, as that of the vorticity is in an ideal fluid without spin, and $\nabla\cdot\bm W=-\nabla\cdot\bm{\mathcal B}=0$; gradients of the two invariants act as baroclinic sources. Equation~\eqref{eq:kelvin-local} follows from the curl of \eqref{eq:kelvin-momentum}, the invariance of $|\bm s|/n$ and the precession $D_t\hat{\bm s}=2\bm\omega\times\hat{\bm s}$; it holds only with both the Berry curvature in $\bm W$ and the source term in $\nabla(|\bm s|/n)$.

No purity of the spin state is assumed: $|\bm s|/n$ is whatever the state of the element gives, in a normal fluid a thermal average far below saturation, and it enters \eqref{eq:kelvin-law} only through its invariance. The circulation and the phase exchange only where the precession rate varies along the loop. For an energy that depends on the spin through $\bm s^2$ alone, $2\bm\omega$ is parallel to $\bm s$, every spin direction is constant, and the two are conserved separately; they exchange in an applied or exchange field, for an anisotropic or gradient energy, and through the Thomas precession of Sec.~\ref{sec:poincare} wherever the acceleration varies along the loop. For example, a field that rotates the spins on a loop around an axis from parallel to antiparallel, each along a meridian whose azimuth winds once around the loop, changes $\Phi_C$ by $4\pi$ in magnitude and imprints a circulation $4\pi|\bm s|/n$, an orbital angular momentum $2|\bm s|/n$ per particle about the axis. In a condensate with $|\bm s|/n=\hbar|m_F|$ this is the vortex of winding number $2|m_F|$ imprinted by adiabatically inverting the axial field of a magnetic trap~\cite{Leanhardt2002}; in a normal fluid the imprinted circulation is not quantized.

\subsection{Relation to earlier work on circulation}
\label{sec:kelvin-earlier}
The Kelvin--Noether theorems for fluids with advected and internal variables~\cite{Holm1998,GayBalmaz2009} state the balance \eqref{eq:kelvin-loop} with the source kept. Its absorption into a phase of the spin texture is known for pure states: for the spin-$\tfrac12$ fluid of the Pauli equation and for the spin quantum plasma the circulation law contains the curvature of the spin direction~\cite{Mahajan2011,Foskett2024}, and the Mermin--Ho relation~\cite{MerminHo1976}, in the generalized form~\eqref{eq:gp-circulation}, is the corresponding statement for a condensate. Equation~\eqref{eq:kelvin-law} shows that purity is not needed, only the invariant~\eqref{eq:casimirs}. A classical counterpart exists for nematic liquid crystals whose director $\bm d$ has a rotational inertia: in ideal incompressible flow, and for any free energy of the director, the conserved circulation is that of the velocity plus the angular momentum of the director contracted with $\bm d\times\nabla\bm d$~\cite{GayBalmaz2011}. There the internal variable has a configuration and a conjugate angular momentum. A spin density has no such pair, and the solid angle $\Phi_C$ takes the place of their product; it is a line integral along $C$ only for a vector potential on the unit sphere that is singular at some point. The relativistic results of Refs.~\cite{Maugin1976,Son2026} are compared with the covariant form of the law in Sec.~\ref{sec:kelvin-cov}.

\section{Spinor Condensates and Higher Multipoles}
\label{sec:gp-algebra}

The invariant $|\bm s|/n$ rests on the premise that the spin density is the only internal variable of a fluid element. In a spin-$F$ condensate of atoms of mass $M$ at zero temperature, described by the Gross--Pitaevskii (GP) equation with $\psi=\sqrt n\,e^{i\varphi}z$, phase $\varphi$ and normalized spinor $z$, $z^\dagger z=1$, an element retains the whole internal state $\rho=zz^\dagger$, of which the spin density $s_A=\hbar n\tr(F_A\rho)$, with the spin-$F$ matrices $F_A$, is only the dipole component. The algebra of the constituents is that of Sec.~\ref{sec:operators}; only the retained moments differ. Results that are standard for spinor condensates~\cite{Kawaguchi2012,Kudo2010,Yukawa2012,Foskett2019} are quoted rather than derived.

\subsection{Bracket of the internal state}
\label{sec:gp-bracket}
The GP action $\int dt\,[\int d^3x\,\tfrac{i\hbar}{2}(\psi^\dagger\partial_t\psi-\partial_t\psi^\dagger\psi)-H]$ of the $2F+1$ components $\psi_m$, $m=-F,\ldots,F$, defines the canonical bracket of the field,
\begin{equation}
 \{\psi_m(\bm x),\psi_{m'}^*(\bm y)\}=\frac{\delta_{mm'}}{i\hbar}\delta^{(3)}(\bm x-\bm y),
 \label{eq:gp-canonical-bracket}
\end{equation}
with which $\partial_t F=\{F,H\}$ is the GP equation $i\hbar\partial_t\psi_m=\delta H/\delta\psi_m^*$. The observables of an element are the bilinear densities $\psi^\dagger A\psi=n\tr(A\rho)$ with constant Hermitian matrices $A$. Since $\delta(\psi^\dagger A\psi)/\delta\psi_m=(\psi^\dagger A)_m$ and $\delta(\psi^\dagger A\psi)/\delta\psi_m^*=(A\psi)_m$, the bracket~\eqref{eq:gp-canonical-bracket} gives \Lcondensate{} of the Letter,
\begin{equation}
\{(\psi^\dagger A\psi)(\bm x),(\psi^\dagger B\psi)(\bm y)\}
 =\frac{(\psi^\dagger[A,B]\psi)(\bm x)}{i\hbar}\delta^{(3)}(\bm x-\bm y)
 =\frac{n(\bm x)}{i\hbar}\tr\!\bigl(\rho(\bm x)[A,B]\bigr)\delta^{(3)}(\bm x-\bm y),
 \label{eq:gp-observable-bracket}
\end{equation}
with the matrix commutator $[A,B]=AB-BA$. The spin matrices obey $[F_A,F_B]=i\epsilon_{ABC}F_C$, and $s_A=s^A$ in Cartesian components; the index is raised where it is the free index of a balance equation or of a current. For $A=\hbar F_A$ and $B=\hbar F_B$, the bracket~\eqref{eq:gp-observable-bracket} is the spin bracket~\eqref{eq:continuum-spin}: the bracket of Sec.~\ref{sec:operators} is the restriction of the condensate bracket to the spin-dipole component. The components of the density matrix $n\rho=\psi\psi^\dagger$ thus obey the Lie--Poisson bracket of $\mathfrak u(2F+1)$, and the conditions $\rho^\dagger=\rho$, $\rho^2=\rho$ and $\tr\rho=1$ restrict the state to the coadjoint orbit of pure states, $\mathbb{CP}^{2F}$, of real dimension $4F$~\cite{Bonet2016,Foskett2019}. For $F\geq1$ this orbit contains quadrupole (nematic) and higher multipole components besides the spin dipole; the fully polarized, spin-coherent states form the submanifold $\mathbb{CP}^1\simeq S^2$.

In the variables $(n,\varphi,\rho)$ the canonical one-form of the action is $-\hbar\int d^3x\,n\,(d\varphi-\mathcal A)$, up to an exact differential, with the Berry connection $\mathcal A=iz^\dagger dz$ on the manifold of internal states~\cite{Berry1984,Simon1983}. A change of the representative spinor, $z\mapsto e^{i\chi}z$ with $\varphi\mapsto\varphi-\chi$, shifts $\mathcal A$ by $-d\chi$ and leaves $d\varphi-\mathcal A$ invariant, so that the projector $\rho$ is a gauge-invariant coordinate on $\mathbb{CP}^{2F}$~\cite{Kibble1979}. At fixed $n$ the symplectic form on $\mathbb{CP}^{2F}$ is $\hbar n\,d\mathcal A$, with $d\mathcal A=i\tr(\rho\,d\rho\wedge d\rho)$, which is the origin of the factor $1/(\hbar n)$ in the bracket of $\rho$ with itself in \eqref{eq:gp-fluid-bracket}. The connection belongs to the exact phase-space structure of the GP action; no adiabatic approximation is involved. Transport is generated, as in Sec.~\ref{sec:operators}, by the momentum density, here $\pi_i=\hbar\operatorname{Im}(\psi^\dagger\partial_i\psi)$: under $P[\bm w]=\int d^3x\,w^i\pi_i$, $n$ is transported as a density and $\rho$ and $\varphi$ as scalars. For functionals of $(n,\pi_i,\rho)$ the brackets combine into the Lie--Poisson bracket of the semidirect product of the fluid algebra with the unitary algebra~\cite{Foskett2019},
\begin{equation}
\begin{aligned}
 \{F,G\}_{\rm LP}=\int d^3x\,\Bigg[&
 \pi_j\left(\frac{\delta G}{\delta\pi_i}\partial_i\frac{\delta F}{\delta\pi_j}
            -\frac{\delta F}{\delta\pi_i}\partial_i\frac{\delta G}{\delta\pi_j}\right)
 +n\left(\frac{\delta G}{\delta\pi_i}\partial_i\frac{\delta F}{\delta n}
          -\frac{\delta F}{\delta\pi_i}\partial_i\frac{\delta G}{\delta n}\right)\\
 &+\tr\!\left[\left(\frac{\delta F}{\delta\pi_i}\frac{\delta G}{\delta\rho}
                    -\frac{\delta G}{\delta\pi_i}\frac{\delta F}{\delta\rho}\right)\partial_i\rho\right]
 +\frac1{i\hbar n}\tr\!\left(\rho\left[\frac{\delta F}{\delta\rho},\frac{\delta G}{\delta\rho}\right]\right)\Bigg],
\end{aligned}
 \label{eq:gp-fluid-bracket}
\end{equation}
where the derivatives with respect to $\rho$ are defined by the trace pairing, $\delta F=\int d^3x\,[(\delta F/\delta n)\delta n+(\delta F/\delta\pi_i)\delta\pi_i+\tr((\delta F/\delta\rho)\delta\rho)]$. This is the counterpart of \eqref{eq:functional} with the projector in place of the spin density. A condensate occupies the submanifold of this phase space on which the momentum density is fixed by the phase and the Berry connection,
\begin{equation}
 \mathcal A_\mu=iz^\dagger\partial_\mu z,\qquad
 v_i=\frac{\hbar}{M}(\partial_i\varphi-\mathcal A_i),\qquad \pi_i=Mn v_i,
 \label{eq:berry-velocity}
\end{equation}
where $\mu$ is $t$ or a spatial index; the same gauge-invariant combination appears in the time component, $-\hbar n(\partial_t\varphi-\mathcal A_t)$. A condensate is thus specified by $4F+2$ real fields, $n$, $\varphi$ and $\rho\in\mathbb{CP}^{2F}$. The curl of the velocity gives the generalized Mermin--Ho relation~\cite{MerminHo1976,Lamacraft2008} and the quantization of the circulation,
\begin{equation}
 \partial_iv_j-\partial_jv_i=-\frac{i\hbar}{M}\tr\bigl(\rho[\partial_i\rho,\partial_j\rho]\bigr),\qquad
 \oint\left(\frac{M v_i}{\hbar}+\mathcal A_i\right)dx^i=2\pi\ell,\quad \ell\in\mathbb Z.
 \label{eq:gp-circulation}
\end{equation}
Smooth variations of the internal state thus produce vorticity without singular vortices. On the spin-coherent states, where $|\bm s|/n=\hbar F$ and, by \eqref{eq:berry-velocity}, $\pi_i/n=\hbar(\partial_i\varphi-\mathcal A_i)$, the curvature of the Berry connection is that of the spin direction~\cite{Berry1984}, $\hbar(\partial_i\mathcal A_j-\partial_j\mathcal A_i)=-\epsilon_{ijk}\mathcal B_k$ with $\bm{\mathcal B}$ of \eqref{eq:kelvin-berry}, so that the vector $\bm W$ of \eqref{eq:kelvin-local} vanishes wherever $\varphi$ is smooth. The law~\eqref{eq:kelvin-law} has then become a constraint, the Mermin--Ho relation~\eqref{eq:gp-circulation}, and its conserved quantity is quantized,
\begin{equation}
 \oint_C\frac{\pi_i}{n}\,dx^i-\frac{|\bm s|}{n}\,\Phi_C=2\pi\hbar\,\ell,\qquad\ell\in\mathbb Z .
 \label{eq:kelvin-quantized}
\end{equation}
A normal fluid with spin conserves what the condensate quantizes, as an ideal fluid conserves the circulation that a scalar superfluid quantizes; for internal states that are not spin-coherent the Berry curvature of $\rho$ in \eqref{eq:gp-circulation} takes the place of $\bm{\mathcal B}$.

\subsection{Equations of motion, purity and the spin current}
\label{sec:gp-dynamics}
The energy of a dilute spin-$F$ Bose gas is~\cite{Ho1998,Ohmi1998,Kawaguchi2012}
\begin{equation}
 H=\int d^3x\left[\frac{\hbar^2}{2M}\partial_i\psi^\dagger\partial_i\psi
 +Un-b_As_A+q_{AB}N_{AB}+\varepsilon_c\right],\qquad
 N_{AB}=n\tr\!\left(\frac{\{F_A,F_B\}}2\rho\right).
 \label{eq:gp-energy}
\end{equation}
Here $U(\bm x)$ is the trap potential, $-b_As_A$ the linear Zeeman energy with Larmor frequency $b_A$, and $q_{AB}N_{AB}$ the quadratic Zeeman coupling to the nematic tensor $N_{AB}$, with $\{F_A,F_B\}=F_AF_B+F_BF_A$ and $q_{AB}$ symmetric and traceless. The contact energy $\varepsilon_c$ is the sum over the $s$-wave channels of even total spin of a pair; it is proportional to $n^2$ at fixed $\rho$ and invariant under spin rotations~\cite{Ho1998,Kawaguchi2012}.

With $|\partial_i\psi|^2=(\partial_i\sqrt n)^2+n(\partial_i\varphi-\mathcal A_i)^2+\tfrac n2\tr(\partial_i\rho\,\partial_i\rho)$ the energy becomes a functional of the hydrodynamic variables,
\begin{equation}
 H=\int d^3x\left[\frac{\bm\pi^2}{2Mn}+\frac{\hbar^2(\partial_i n)^2}{8Mn}
 +\frac{\hbar^2 n}{4M}\tr(\partial_i\rho\,\partial_i\rho)
 +Un-b_As_A+q_{AB}N_{AB}+\varepsilon_c\right],
 \label{eq:hydro-energy}
\end{equation}
in which the third term is the gradient energy of the internal state. The bracket~\eqref{eq:gp-fluid-bracket} gives the continuity equation $\partial_tn+\partial_i(nv_i)=0$, the momentum balance, here in conservation form, and an equation of von Neumann form for the internal state,
\begin{equation}
\begin{gathered}
 \partial_t\pi_i+\partial_j\Pi_{ij}
 =-n\partial_iU+s_A\partial_i b_A-N_{AB}\partial_iq_{AB},\qquad
 i\hbar D_t\rho
 =\frac1n\left[\left.\frac{\delta H}{\delta\rho}\right|_{n,\pi},\rho\right],\\
 \Pi_{ij}=Mn v_iv_j+\frac{\hbar^2\partial_in\,\partial_jn}{4Mn}
 +\frac{\hbar^2n}{2M}\tr(\partial_i\rho\,\partial_j\rho)
 +\delta_{ij}\Bigl(\varepsilon_c-\frac{\hbar^2}{4M}\nabla^2n\Bigr),\\
 \frac1n\left.\frac{\delta H}{\delta\rho}\right|_{n,\pi}
 =-\frac{\hbar^2}{2Mn}\partial_i(n\partial_i\rho)
 -\hbar b_AF_A
 +q_{AB}\frac{\{F_A,F_B\}}2
 +\frac1n\frac{\partial\varepsilon_c}{\partial\rho}.
\end{gathered}
 \label{eq:internal-evolution}
\end{equation}
In $\Pi_{ij}$ the contact pressure $n\,\partial\varepsilon_c/\partial n-\varepsilon_c$ equals $\varepsilon_c$. Wherever $n>0$ and the fields are smooth, the continuity equation, the equation for the internal state and the Josephson relation for the phase are together equivalent to the GP equation~\cite{Kudo2010,Yukawa2012,Foskett2019}: the equation for the internal state is its projection orthogonal to $\psi$, and the other two are the imaginary and the real part of its projection along $\psi$. Because the right-hand side of the equation for $\rho$ is a commutator, it preserves the Hermiticity, the trace and the purity of $\rho$,
\begin{equation}
 D_t(\rho^2-\rho)=\frac1{i\hbar n}
 \left[\left.\frac{\delta H}{\delta\rho}\right|_{n,\pi},\rho^2-\rho\right].
 \label{eq:purity}
\end{equation}
The purity, $\tr\rho^2=1$, is the invariant of a fluid element of the condensate. The spin magnitude per particle, $|\bm s|/n$, is not conserved for $F\geq1$.

By \eqref{eq:internal-evolution} the spin density obeys
\begin{equation}
 J_i^A=\frac{\hbar^2}{M}\operatorname{Im}(\psi^\dagger F_A\partial_i\psi)
 =s^Av_i+\frac{i\hbar^2n}{2M}\tr(F_A[\rho,\partial_i\rho]),\qquad
 \partial_ts^A+\partial_iJ_i^A
 =-\epsilon_{ABC}b_Bs_C+2\epsilon_{ABC}q_{BD}N_{DC};
 \label{eq:spin-current}
\end{equation}
the contact energy is invariant under spin rotations and drops out. The spin current $J_i^A$ has a part that is not convective, produced by the gradient energy of the internal state. On the spin-coherent states, $|\bm s|=\hbar Fn$, the internal state is a function of the spin density, the current is $J_i^A=s^Av_i-(\bm s\times\partial_i\bm s)_A/(2MFn)$, and $|\bm s|/n$ is constant, as in Sec.~\ref{sec:jacobi}; at uniform density and without Zeeman fields the balance~\eqref{eq:spin-current} is then the equation of Landau--Lifshitz type of Refs.~\cite{Lamacraft2008,Kudo2010}.

For $F=1$, in the Cartesian basis $(F_A)_{BC}=-i\epsilon_{ABC}$, the internal state is parametrized by the spin density and the symmetric nematic tensor, with $\tr N=2n$,
\begin{equation}
 \rho_{AB}=\delta_{AB}-\frac{N_{AB}}n-\frac{i}{2\hbar n}\epsilon_{ABC}s_C,\qquad
 J_i^A=s^Av_i-\frac{\epsilon_{ABC}s^B\partial_i s^C}{4Mn}
 -\frac{\hbar^2\epsilon_{ABC}N_{BD}\partial_iN_{CD}}{Mn}.
 \label{eq:spinone-currents}
\end{equation}
The current is \Lcurrent{} of the Letter. Its last term, $-(\hbar^2/Mn)\,\epsilon_{ABC}N_{BD}\partial_iN_{CD}$, is a spin current produced by gradients of the nematic tensor; it does not vanish where $\bm s=0$. The nematic tensor obeys a balance of the same kind, obtained from \eqref{eq:internal-evolution}; with the contact energy of spin-1 atoms the two balances, the momentum balance and \eqref{eq:gp-circulation} are the hydrodynamics of coupled spin and nematic densities of Ref.~\cite{Yukawa2012}.

An example shows that the spin dynamics does not close on $(n,\bm\pi,\bm s)$. Take the polar spin-1 state $\psi(x,0)=\sqrt{n(x)}\,(\cos\alpha(x),\sin\alpha(x),0)^{\mathsf T}$ in the Cartesian basis, with $n$ decaying at infinity. It has $\bm s=0$ and $\bm\pi=0$ but the spin current $J_x^z=-\hbar^2n\,\partial_x\alpha/M$, so that for $b_A=q_{AB}=0$, at $t=0$,
\begin{equation}
 \partial_t s^z=-\partial_xJ_x^z
 =\frac{\hbar^2}{M}\,\partial_x(n\,\partial_x\alpha),\qquad
 \int_{-\infty}^{\infty}\partial_ts^z\,dx=0 .
 \label{eq:polar-counterexample}
\end{equation}
The rate is nonzero wherever $n\,\partial_x\alpha$ varies, although the total spin does not change. The untwisted state with the same $n(x)$ and constant $\alpha$ has the same $(n,\bm\pi,\bm s)$ and the same phase, $\varphi=0$, but $\partial_t s^z=0$. No closed equation for $(n,\bm\pi,\bm s)$ can describe both states. In the twisted state spin appears in fluid elements that had none: the spin magnitude per particle of an element changes reversibly, which \eqref{eq:casimirs} excludes for a fluid whose only internal variable is the spin density. The variable that makes the difference is the nematic tensor.

The equations contain the earlier hydrodynamic descriptions of spinor condensates: on the spin-coherent manifold, for uniform density, incompressible flow and no Zeeman fields, Eqs.~(2a) and~(2b) of Ref.~\cite{Lamacraft2008}, whose gauge field is $\mathcal A_i$; for a fully magnetized spinor with stationary density, the equations of Ref.~\cite{Kudo2010} without the dipolar interaction; for $F=1$ with uniform axial Zeeman fields, Eqs.~(21)--(29) of Ref.~\cite{Yukawa2012}, whose normalized spin vector and nematic tensor are $s_A/(\hbar n)$ and $N_{AB}/n$; and, for the GP energy, the equation for the internal state is Eq.~(118) of Ref.~\cite{Foskett2019}, whose Lie--Poisson bracket, Eqs.~(125) and~(126), becomes~\eqref{eq:gp-fluid-bracket} in the variables $(n,\bm\pi,\rho)$.

\section{Poincar\'e Algebra and Fermi--Walker Transport}
\label{sec:poincare}

We return to the fluid whose only internal variable is the spin density and derive \Lthomas{}--\Lfw{} of the Letter. The bracket fixes how rotations act on the fields; a boost instead relates the descriptions of one state on two equal-time slices, and for matter with spin its generator contains a time--space component of the spin, whose form depends on which point is called the position of a body. Imposing the Poincar\'e algebra fixes this component and requires the Thomas term in the energy; the canonical components then precess at the Thomas rate, while the pure-boost triad to which they refer rotates at the opposite rate, so the spin four-vector is Fermi--Walker (FW) transported. The spin axes are the Cartesian laboratory axes.

\subsection{Charges and the boost generator}
\label{sec:charges}
For an energy--momentum tensor $\Theta^{\mu\nu}$ and a spin current $\Sigma^{\lambda\mu\nu}$ with $\partial_\lambda\Theta^{\lambda\nu}=0$ and $\partial_\lambda\Sigma^{\lambda\mu\nu}=\Theta^{\nu\mu}-\Theta^{\mu\nu}$ the conserved charges are the four-momentum $P^\nu=\int d^3x\,\Theta^{0\nu}$, with the energy $H=P^0$, and the angular momentum $J^{\mu\nu}=\int d^3x\,(x^\mu\Theta^{0\nu}-x^\nu\Theta^{0\mu}+\Sigma^{0\mu\nu})$. With the generators of rotations $J_i=\tfrac12\epsilon_{ijk}J^{jk}$ and of boosts $K_i=J^{0i}$, and with $\partial_tF=\{F,H\}$, they obey
\begin{equation}
\begin{gathered}
 \{J_i,J_j\}=\epsilon_{ijk}J_k,\quad \{J_i,P_j\}=\epsilon_{ijk}P_k,\quad \{J_i,K_j\}=\epsilon_{ijk}K_k,\\
 \{K_i,K_j\}=-\epsilon_{ijk}J_k,\quad \{K_i,P_j\}=-\delta_{ij}H,\quad \{K_i,H\}=-P_i,
\end{gathered}\label{eq:poincare-algebra}
\end{equation}
together with $\{J_i,H\}=\{P_i,H\}=\{P_i,P_j\}=0$. Writing $\mathcal H=\Theta^{00}$ and $\pi_i=\Theta^{0i}$, with the time--space component $\Sigma^{00i}$ of the spin density $\Sigma^{0\mu\nu}$,
\begin{equation}
 K_i=tP_i-\int d^3x\,x_i\mathcal H+\int d^3x\,\Sigma^{00i},\qquad
 \partial_t\Sigma^{00i}+\partial_j\Sigma^{j0i}=\Theta^{i0}-\Theta^{0i} .
 \label{eq:boost-charge}
\end{equation}
The second relation is the local form of $dK_i/dt=0$: the component $\Sigma^{00i}$ is driven by the difference between energy flux and momentum density. The relation $\{K_i,K_j\}=-\epsilon_{ijk}J_k$ states that two boosts in different directions do not commute; their commutator is a rotation. For the fluid, $P_i$ and $J_i$ generate translations and rotations through the bracket, so the relations of \eqref{eq:poincare-algebra} that contain them hold whenever $\mathcal H$ is a rotational scalar and $\Sigma^{00i}$ a vector, neither depending explicitly on $\bm x$. The conditions to be imposed are $\{K_i,H\}=-P_i$, the conservation of the boost charge, and $\{K_i,K_j\}=-\epsilon_{ijk}J_k$.

\subsection{Boosts of a perfect fluid}
\label{sec:fluid-boosts}
Take the fields $(n,s,\pi_i,\bm s)$ with the bracket~\eqref{eq:functional}, $H=\int d^3x\,\mathcal H$, $P_i=\int d^3x\,\pi_i$, $J_i=\int d^3x\,(\epsilon_{ijk}x_j\pi_k+s_i)$ and $K_i$ of \eqref{eq:boost-charge}. We need first the boosts of a fluid without spin, with a gradient-free energy density $\mathcal H_0(n,s,\bm\pi^2)$ and $\Sigma^{00i}=0$. The functional derivatives of $\int d^3x\,x_i\mathcal H_0$ are those of $H_0=\int d^3x\,\mathcal H_0$ multiplied by $x_i$, and \eqref{eq:functional} gives
\begin{equation}
 \Bigl\{\int d^3x\,x_i\mathcal H_0,H_0\Bigr\}=\int d^3x\,(\mathcal H_0+p)\,v_i,\qquad
 \Bigl\{\int d^3x\,x_i\mathcal H_0,\int d^3x\,x_j\mathcal H_0\Bigr\}=\int d^3x\,(x_jv_i-x_iv_j)(\mathcal H_0+p),
\end{equation}
with the pressure $p=Ts+\mu n+v^i\pi_i-\mathcal H_0$. Hence
\begin{equation}
\begin{aligned}
 \{K_i,H\}+P_i&=\int d^3x\,\bigl[\pi_i-(\mathcal H_0+p)v_i\bigr],\\
 \{K_i,K_j\}+\epsilon_{ijk}J_k&=\int d^3x\,\bigl\{x_i\bigl[\pi_j-(\mathcal H_0+p)v_j\bigr]-x_j\bigl[\pi_i-(\mathcal H_0+p)v_i\bigr]\bigr\}+\epsilon_{ijk}\int d^3x\,s_k .
\end{aligned}
 \label{eq:spinless-boosts}
\end{equation}
The orbital parts vanish if and only if $\bm\pi=(\mathcal H_0+p)\bm v$, the Dirac--Schwinger condition~\cite{Schwinger1963}: the momentum density equals the energy flux. It is satisfied by the perfect fluid,
\begin{equation}
 \mathcal H_0=(\varepsilon+p)\gamma^2-p,\qquad \bm\pi=(\varepsilon+p)\gamma^2\bm v,
 \label{eq:enthalpy-densities}
\end{equation}
where $\gamma=(1-\bm v^2)^{-1/2}$ is the Lorentz factor, $\varepsilon$ and $p$ are the energy density and the pressure in the local rest frame, functions of the rest-frame densities $n/\gamma$ and $s/\gamma$, and $\varepsilon+p$ is the rest-frame enthalpy density. The pressure defined above is this rest-frame pressure, and the energy density called $\varepsilon$ in Sec.~\ref{sec:thermo} is $\mathcal H_0$ here. The last term of \eqref{eq:spinless-boosts} remains: the commutator of two boosts rotates the orbital angular momentum but not a spin that is only advected by the flow. The boost generator of matter with spin therefore needs the spin term $\int d^3x\,\Sigma^{00i}$, and the conservation of that term will require a spin-dependent energy.

The bracket of a field $\phi=n,s,\pi_i,s^A$ with $-\int d^3x\,(\bm\beta\cdot\bm x)\mathcal H_0$, the spinless part of the boost of velocity $\bm\beta$ at $t=0$, follows from \eqref{eq:functional}, \eqref{eq:eom-derived} and \eqref{eq:enthalpy-densities}:
\begin{equation}
 \Bigl\{\phi,-\int d^3x\,(\bm\beta\cdot\bm x)\mathcal H_0\Bigr\}=\delta_{\bm\beta}\phi-(\bm\beta\cdot\bm x)\,\partial_t\phi,\qquad
 \delta_{\bm\beta}(n,s,\bm s)=(\bm\beta\cdot\bm v)\,(n,s,\bm s),\qquad
 \delta_{\bm\beta}\bm\pi=(\varepsilon+p)\gamma^2\,[\bm\beta+\bm v\,(\bm\beta\cdot\bm v)],
 \label{eq:spinless-boost-action}
\end{equation}
where $\partial_t$ is the evolution generated by $H_0$. The second term evaluates the field on the tilted slice. The first, $\delta_{\bm\beta}$, is the Lorentz transformation of the fluid at a fixed point: $n$, $s$ and each spin component $s^A$ change as time components of currents, and $\pi_i$ as the components $\Theta^{0i}$ of the energy--momentum tensor of a perfect fluid. Relation~\eqref{eq:spinless-boost-action} holds for any density built from the fields and their spatial derivatives, with $\delta_{\bm\beta}$ acting on functions of the fields by the chain rule and on derivatives by $\delta_{\bm\beta}\partial_j\phi=\partial_j\delta_{\bm\beta}\phi-\beta_j\partial_t\phi$, which follows by differentiating \eqref{eq:spinless-boost-action}. In particular
\begin{equation}
 \delta_{\bm\beta}\frac n\gamma=\delta_{\bm\beta}\frac s\gamma=\delta_{\bm\beta}(\varepsilon+p)=\delta_{\bm\beta}p=0,\qquad
 \delta_{\bm\beta}\gamma=\gamma\,\bm\beta\cdot\bm v,\qquad
 \delta_{\bm\beta}\bm v=\bm\beta-\bm v\,(\bm\beta\cdot\bm v),\qquad
 \delta_{\bm\beta}\nabla p=-\bm\beta\,\partial_tp :
 \label{eq:point-boost}
\end{equation}
rest-frame quantities are invariant, the velocity changes by the relativistic addition of $\bm\beta$, and the boost turns the time derivative of the pressure into a gradient.

\subsection{First order in the spin}
\label{sec:first-order}
We now admit terms linear in the spin, $\mathcal H=\mathcal H_0+\mathcal H_1$ and $\Sigma^{00i}$, and impose the two conditions at first order in $\bm s$. Parity ($\bm s$ axial, $\bm\pi$ polar) and time reversal ($\bm s$ and $\bm\pi$ odd) restrict a gradient-free time--space component to
\begin{equation}
 \Sigma^{00i}=f\,(\bm s\times\bm v)^i ,
 \label{eq:sigma-ansatz}
\end{equation}
with $f$ a function of the rest-frame densities $n/\gamma$ and $s/\gamma$, which $\delta_{\bm\beta}$ leaves unchanged, and of $\gamma$. They forbid a gradient-free $\mathcal H_1$, since no axial vector can be built from $\bm\pi$ alone. We take $\mathcal H_1$ without derivatives of $\bm s$; such derivatives can be removed by an integration by parts, which changes $\Sigma^{00i}$ by a term of the form~\eqref{eq:sigma-ansatz}. The time derivative $\partial_t\Sigma^{00i}=\{\Sigma^{00i},H\}$ includes the precession $(\partial\mathcal H_1/\partial\bm s)\times\bm s$ that $\mathcal H_1$ generates. We define the density $\bm{\mathcal C}$ by
\begin{equation}
 \bm\beta\cdot\bm{\mathcal C}=\beta_i\bigl[\partial_t\Sigma^{00i}+\partial_j(v^j\Sigma^{00i})\bigr]-\delta_{\bm\beta}\mathcal H_1
 \label{eq:conservation-density}
\end{equation}
for all $\bm\beta$. At first order in $\bm s$, the terms $(\bm\beta\cdot\bm x)\,\partial_t\mathcal H_1$ produced in $\{\bm\beta\cdot\bm K,H\}$ by the bracket of $\int d^3x\,(\bm\beta\cdot\bm x)\mathcal H_1$ with $H_0$ and by that of $\int d^3x\,(\bm\beta\cdot\bm x)\mathcal H_0$ with $H_1$ cancel by \eqref{eq:spinless-boost-action}, and the commutator of two boosts collects the same density with a moment of $\bm x$. With the Dirac--Schwinger condition,
\begin{equation}
\begin{aligned}
 \{\bm\beta\cdot\bm K,H\}+\bm\beta\cdot\bm P&=\int d^3x\,\bm\beta\cdot\bm{\mathcal C},\\
 \{\bm\beta\cdot\bm K,\bm\beta'\cdot\bm K\}+(\bm\beta\times\bm\beta')\cdot\bm J&=\int d^3x\,\bigl[(\bm\beta\cdot\bm x)\,\bm\beta'\cdot\bm{\mathcal C}-(\bm\beta'\cdot\bm x)\,\bm\beta\cdot\bm{\mathcal C}+\ell\bigr],
\end{aligned}
 \label{eq:closure-split}
\end{equation}
with the local density
\begin{equation}
 \ell=\beta_i\,\delta_{\bm\beta'}\Sigma^{00i}-\beta'_i\,\delta_{\bm\beta}\Sigma^{00i}+(\bm\beta\cdot\bm v)\,\beta'_i\Sigma^{00i}-(\bm\beta'\cdot\bm v)\,\beta_i\Sigma^{00i}+\epsilon_{ABC}\,s^C\,\frac{\partial(\beta_i\Sigma^{00i})}{\partial s^A}\frac{\partial(\beta'_j\Sigma^{00j})}{\partial s^B}+(\bm\beta\times\bm\beta')\cdot\bm s .
 \label{eq:ell}
\end{equation}
The terms of $\ell$ come, in this order, from the boost of $\Sigma^{00i}$, from the flux $v^j\Sigma^{00i}$ included in $\bm{\mathcal C}$, which integrates to zero in the first line of \eqref{eq:closure-split} but not against $\bm x$, from the spin bracket of $\Sigma^{00i}$ with itself, and from the spin part of $\bm J$.

Both conditions are local. The density $\bm{\mathcal C}$ is linear in $\bm s$ and contains no derivative of $\bm s$, because the flux in \eqref{eq:conservation-density} removes the advection of $\bm s$ from $\partial_t\Sigma^{00i}$; its integral vanishes for all fields only if $\bm{\mathcal C}=0$ pointwise. The moments in the second line of \eqref{eq:closure-split} then vanish, and $\ell$, which depends on $\Sigma^{00i}$ alone, must vanish pointwise as well. The commutator of two boosts therefore fixes $\Sigma^{00i}$ before $\mathcal H_1$ is known, and the conservation of the boost charge then fixes $\mathcal H_1$.

\emph{Gradient-free fields fail.} Without $\mathcal H_1$ the spin is only advected, and \eqref{eq:conservation-density} reduces to $\bm{\mathcal C}=\bm s\times D_t(f\bm v)$. The momentum equation $D_t\pi_i=-\pi_i\partial_jv^j-\partial_ip$ of Sec.~\ref{sec:thermo} and the conservation of $\mathcal H_0$ give the Euler equation
\begin{equation}
 (\varepsilon+p)\gamma^2D_t\bm v=-\nabla p-\bm v\,\partial_tp ,\qquad D_t\gamma=\gamma^3\,\bm v\cdot D_t\bm v ,
 \label{eq:euler}
\end{equation}
so the acceleration has a component normal to $\bm v$ wherever $\nabla p$ has one. Then $\bm{\mathcal C}=0$ requires $f=0$, and $\ell=(\bm\beta\times\bm\beta')\cdot\bm s\neq0$: no gradient-free energy and time--space spin component close the algebra at first order in the spin.

\emph{The commutator of two boosts fixes $\Sigma^{00i}$.} With \eqref{eq:sigma-ansatz} and \eqref{eq:point-boost} the four terms of \eqref{eq:ell} add up to
\begin{equation}
 \ell=(\bm\beta\times\bm\beta')\cdot\Bigl\{\bigl[1-2f+(f-\gamma\partial_\gamma f)\,\bm v^2\bigr]\bm s+\bigl[\gamma\partial_\gamma f-f+f^2\bigr](\bm s\cdot\bm v)\,\bm v\Bigr\},
 \label{eq:ell-sigma}
\end{equation}
with $\partial_\gamma$ taken at fixed rest-frame densities. Both brackets must vanish. Eliminating the derivative with the second, the first becomes $1-2f+f^2\bm v^2=0$, whose roots $f=\gamma/(\gamma\pm1)$ both satisfy the second. Only $\gamma/(\gamma+1)$ is finite at rest:
\begin{equation}
 \Sigma^{00i}=\frac{\gamma}{\gamma+1}\,(\bm s\times\bm v)^i .
 \label{eq:sigma-fixed}
\end{equation}

\emph{The conservation of the boost charge fixes the Thomas term.} With \eqref{eq:sigma-fixed}, the density~\eqref{eq:conservation-density} has three parts, the change of $\Sigma^{00i}$ as the fluid accelerates, its change by the precession that $\mathcal H_1$ generates, and the boost of $\mathcal H_1$:
\begin{equation}
 \bm\beta\cdot\bm{\mathcal C}=\bm\beta\cdot\Bigl[\bm s\times D_t\Bigl(\frac{\gamma\,\bm v}{\gamma+1}\Bigr)+\frac{\gamma}{\gamma+1}\Bigl(\frac{\partial\mathcal H_1}{\partial\bm s}\times\bm s\Bigr)\times\bm v\Bigr]-\delta_{\bm\beta}\mathcal H_1 .
 \label{eq:conservation-parts}
\end{equation}
By \eqref{eq:euler} the first part is
\begin{equation}
 \bm s\times D_t\Bigl(\frac{\gamma\,\bm v}{\gamma+1}\Bigr)=-\frac{\bm s\times\nabla p}{(\varepsilon+p)\gamma(\gamma+1)}-\frac{\gamma\,(\bm v\cdot\nabla p)\,\bm s\times\bm v}{(\varepsilon+p)(\gamma+1)^2}-\frac{\partial_tp\;\bm s\times\bm v}{(\varepsilon+p)(\gamma+1)} .
 \label{eq:sigma-rate}
\end{equation}
It contains $\partial_tp$, which depends on the divergence of the velocity and can be chosen at a point independently of $\nabla p$, $\bm v$ and $\bm s$. Consider energies linear in the gradients of the thermodynamic variables, which we take to be $p$ and the advected ratio $s/n$. Such an energy can cancel this term only through the boost of the pressure gradient, $\delta_{\bm\beta}\nabla p=-\bm\beta\,\partial_tp$; the gradient of $s/n$ boosts into $\delta_{\bm\beta}\nabla(s/n)=\bm\beta\,\bm v\cdot\nabla(s/n)$, since $D_t(s/n)=0$. The scalars linear in $\bm s$ and $\nabla p$ that are even under parity and time reversal are multiples of $\bm s\cdot(\bm v\times\nabla p)$, so $\mathcal H_1$ must contain this term times a function of the rest-frame densities and $\gamma$. Its boost contributes the same function times $\partial_tp\;\bm s\times\bm v$ to $\bm{\mathcal C}$, and $\partial_tp$ cancels only if the function is $1/[(\varepsilon+p)(\gamma+1)]$:
\begin{equation}
 \mathcal H_1=\frac{\bm s\cdot(\bm v\times\nabla p)}{(\varepsilon+p)(\gamma+1)}=\bm\Delta\cdot\nabla p,\qquad
 \bm\Delta=\frac{\bm s\times\bm v}{(\varepsilon+p)(\gamma+1)} .
 \label{eq:thomas-term}
\end{equation}
For this energy the boost and the precession give
\begin{equation}
 -\delta_{\bm\beta}(\bm\Delta\cdot\nabla p)=\bm\beta\cdot\Bigl[\frac{\bm s\times\nabla p+(\bm s\times\bm v)\,\partial_tp}{(\varepsilon+p)(\gamma+1)}+\frac{\gamma\,\bm v\,(\bm\Delta\cdot\nabla p)}{\gamma+1}\Bigr],\qquad
 \frac{\gamma}{\gamma+1}\Bigl(\frac{\partial(\bm\Delta\cdot\nabla p)}{\partial\bm s}\times\bm s\Bigr)\times\bm v=-\frac{\gamma\,(\bm s\cdot\bm v)\,\bm v\times\nabla p}{(\varepsilon+p)(\gamma+1)^2},
\end{equation}
and the three parts add up to
\begin{equation}
 \bm{\mathcal C}=\frac{\gamma}{(\varepsilon+p)(\gamma+1)^2}\bigl[\bm v^2\,\bm s\times\nabla p-(\bm v\cdot\nabla p)\,\bm s\times\bm v-(\bm s\cdot\bm v)\,\bm v\times\nabla p+\bm v\;\bm s\cdot(\bm v\times\nabla p)\bigr]=0 .
\end{equation}
The bracket vanishes for any three vectors: it is the difference of the two expansions of $\bm v\times[\bm v\times(\bm s\times\nabla p)]$, once from the outside and once from the inside.

The conditions~\eqref{eq:closure-split} are linear in $\mathcal H_1$ at fixed $\Sigma^{00i}$, so any other solution differs from $\bm\Delta\cdot\nabla p$ by a term that satisfies them by itself. Among the energies linear in $\nabla p$ and $\nabla(s/n)$ there is no such term: for $\bm s\cdot[\bm v\times\nabla(s/n)]$ times a function of the rest-frame densities and $\gamma$, the condition $\bm{\mathcal C}=0$ requires the function to vanish (at rest, $\bm{\mathcal C}$ is the function times $\bm s\times\nabla(s/n)$). Terms with velocity gradients, whose boosts contain $\partial_t\bm v$ and through \eqref{eq:euler} also $\partial_tp$, are not restricted in the same way; the algebra leaves Lorentz scalars such as a spin--vorticity energy unconstrained (Sec.~\ref{sec:second-order}).

Collected, the first-order closure is
\begin{equation}
 \Sigma^{00i}=\frac{\gamma}{\gamma+1}\,(\bm s\times\bm v)^i,\qquad
 \mathcal H=\mathcal H_0+\bm\Delta\cdot\nabla p=\mathcal H_0+\frac{\bm s\cdot(\bm v\times\nabla p)}{(\varepsilon+p)(\gamma+1)},\qquad
 \bm\Delta=\frac{\bm s\times\bm v}{(\varepsilon+p)(\gamma+1)} ,
 \label{eq:thomas-closure}
\end{equation}
which is \Lthomas{} of the Letter. Since $\bm{\mathcal C}=0$ and $\ell=0$ pointwise, the parts of $\{K_i,H\}+P_i$ and $\{K_i,K_j\}+\epsilon_{ijk}J_k$ that are linear in $\bm s$ vanish identically, to all orders in gradients.

\subsection{Wigner rotation and covariant centroid}
\label{sec:wigner-centroid}
The two terms of \eqref{eq:thomas-closure} have a kinematic meaning. A boost acts on the spin through $\Sigma^{00i}$: by \eqref{eq:spinless-boost-action} and the spin bracket in \eqref{eq:functional}, at $t=0$ and first order in $\bm s$,
\begin{equation}
 \{\bm s,\bm\beta\cdot\bm K\}=(\bm\beta\cdot\bm v)\,\bm s+\frac{\gamma}{\gamma+1}\,(\bm v\times\bm\beta)\times\bm s-(\bm\beta\cdot\bm x)\,\partial_t\bm s ,
 \label{eq:wigner-fluid}
\end{equation}
so that the rest-frame density $\bm s/\gamma$ is rotated by the angle $\gamma\,\bm v\times\bm\beta/(\gamma+1)$. This is the Wigner rotation~\cite{Wigner1939}: with $u^\mu=\gamma(1,\bm v)$ the four-velocity of the fluid and $L(u)$ the rotation-free boost that takes $(1,\bm 0)$ to $u^\mu$, the boost of velocity $\bm\beta$ applied after $L(u)$ differs by this rotation, to first order in $\bm\beta$, from the rotation-free boost to the new four-velocity. The canonical field is therefore the density of the rest-frame components of the spin in the frame reached from the laboratory by $L(u)$. Per particle, $\delta_{\bm\beta}(\bm\pi/n)=(\varepsilon+p)\gamma^2\bm\beta/n$, and this rotation and $\Sigma^{00i}/n$ are the Wigner rotation and the time--space spin component of a body with momentum $\bm\pi/n$, energy $(\varepsilon+p)\gamma^2/n$, mass $(\varepsilon+p)\gamma/n$, the rest-frame enthalpy per particle, and spin $\bm s/n$, placed at its Newton--Wigner (NW) position~\cite{NewtonWigner1949,Pryce1948}. The covariant centroid of such a body (its center of mass in the rest frame), about which its spin tensor is orthogonal to its four-momentum, is displaced from that position by $\bm\Delta$~\cite{Tulczyjew1959,Dixon1970}, and $\bm\Delta\cdot\nabla p$ is the work of the pressure on this displacement: if the perfect-fluid energy is evaluated with the densities of the displaced elements, $n-\nabla\cdot(n\bm\Delta)$ and likewise for $s$ and $\pi_i$, the Gibbs--Duhem relation $dp=n\,d\mu+s\,dT+\pi_i\,dv^i$ shows that it changes by $\int d^3x\,\bm\Delta\cdot\nabla p+O(\bm s^2)$. The pressure thus acts on the covariant centroids, while the canonical fields are densities of NW positions.

\subsection{Proper densities, the spin four-vector and the Fermi--Walker derivative}
\label{sec:proper}
The covariant form of the spin equation uses proper quantities. In the covariant description $n$, $s$ and $s^A$ denote densities in the local rest frame, $\nabla_\mu$ the covariant derivative ($\partial_\mu$ in Cartesian coordinates), a dot the derivative $u^\mu\nabla_\mu$ with respect to proper time, $a^\mu=\dot u^\mu$ the acceleration, $\theta=\nabla_\mu u^\mu$ the expansion rate, and $\omega_A$ the spin potential per unit proper time. The Levi-Civita tensor has $\varepsilon^{0123}=+1/\sqrt{-g}$ and $\varepsilon_{0123}=-\sqrt{-g}$, with the determinant $g$ of the metric $g_{\mu\nu}$ ($\sqrt{-g}=1$ in Cartesian coordinates), and $\Delta^{\mu\nu}=g^{\mu\nu}-u^\mu u^\nu$ projects on the directions normal to $u^\mu$. The vorticity and the shear tensor of the flow are
\begin{equation}
 \Omega^\mu=\tfrac12\varepsilon^{\mu\nu\rho\sigma}u_\nu\nabla_\rho u_\sigma,\qquad
 \sigma_{\mu\nu}=\Delta_\mu{}^\alpha\Delta_\nu{}^\beta\nabla_{(\beta}u_{\alpha)}
 -\tfrac13\theta\Delta_{\mu\nu}.
 \label{eq:fluid-rates}
\end{equation}
In the local rest frame $\theta=\partial_iv^i$, the $\Omega^i$ are the components of $\bm\Omega=\tfrac12\nabla\times\bm v$, and the $\sigma^i{}_j$ are those of the three-dimensional shear tensor $\partial_{(i}v_{j)}-\tfrac13\delta_{ij}\partial_kv^k$, which is written $\sigma_{ij}$, or $\bm\sigma$ as a matrix, wherever only spatial tensors occur. The equal-time bracket uses densities per laboratory volume and the coordinate time. Where both occur we label the latter by ``lab''; with $\gamma=u^0$, convective currents satisfy
\begin{equation}
 (n,s,s^A)_{\rm lab}=\gamma(n,s,s^A),\qquad
 \omega_{A,\rm lab}=\omega_A/\gamma,\qquad
 \dot{}=\gamma D_t,\qquad
 \theta=D_t\gamma+\gamma\partial_iv^i.
 \label{eq:density-time-conversion}
\end{equation}
The first relation is the Lorentz contraction of the volume of a cell, and the others follow from time dilation along the worldline. With $D_t(\gamma s^A)+\gamma s^A\partial_iv^i=\dot s^A+\theta s^A$, the equal-time balance $\partial_t(\gamma s^A)+\partial_i(\gamma s^Av^i)=2\epsilon_{ABC}\omega_{B,\rm lab}\gamma s^C$ becomes
\begin{equation}
 \dot s^A+\theta s^A=2\epsilon_{ABC}\omega_Bs^C.
 \label{eq:proper-time}
\end{equation}
When $\gamma$ is a function of the fields, the brackets and conjugates of the proper densities follow from the chain rules~\eqref{eq:full-chain} and~\eqref{eq:rescale} of Sec.~\ref{sec:triad-brackets}.

Components become a vector in the rest space of a fluid element only once a triad of that space is chosen. For an oriented orthonormal triad $e_A^\mu$ of that space,
\begin{equation}
 e_0^\mu=u^\mu,\quad u^2=1,\quad u\cdot e_A=0,\quad e_A\cdot e_B=-\delta_{AB},\quad
 e^A{}_\mu=-\delta^{AB}g_{\mu\nu}e_B^\nu,\quad
 e_A^\mu e^A{}_\nu=\Delta^\mu{}_{\nu}=\delta^\mu{}_{\nu}-u^\mu u_\nu,
 \label{eq:tetrad}
\end{equation}
components $s^A$ define the four-vector $s^\mu=e_A^\mu s^A$, with $u\cdot s=0$; the triad to which the canonical components refer is identified in Sec.~\ref{sec:fw-spin}. The four-vector is dual to the antisymmetric spin tensor $S^{\mu\nu}$,
\begin{equation}
 S^{\mu\nu}=-\varepsilon^{\mu\nu\rho\sigma}u_\rho s_\sigma,\qquad
 s^\mu=-\frac12\varepsilon^{\mu\nu\rho\sigma}u_\nu S_{\rho\sigma},\qquad
 u_\mu S^{\mu\nu}=0.
 \label{eq:dual}
\end{equation}
In the rest frame $S^{ij}=\epsilon_{ijk}s^k$ and $\frac{1}{2}S_{\mu\nu}S^{\mu\nu}=\bm s^2$. Differentiating $u\cdot e_A=0$ gives $u\cdot\dot e_A=-a\cdot e_A$, so that a triad cannot be constant along an accelerated worldline. The triad that does not rotate in the rest frame is FW transported~\cite{Jantzen1992},
\begin{equation}
 \dot e_{\mathrm{FW},A}^\mu=-u^\mu a_\nu e_{\mathrm{FW},A}^\nu,\qquad
 (\mathcal D_{\mathrm{FW}}s)^\mu=\Delta^\mu{}_{\nu}\dot s^\nu
 =\dot s^\mu+u^\mu a_\nu s^\nu,
 \label{eq:fw}
\end{equation}
where the second relation defines the FW derivative of a transverse vector. FW transport preserves $u\cdot e_A=0$ and $e_A\cdot e_B=-\delta_{AB}$ along any worldline, so that a nonrotating orthonormal triad exists in a fluid with arbitrary acceleration, shear and expansion.

\subsection{Spin equation and its covariant form}
\label{sec:fw-spin}
With $\mathcal H=\mathcal H_0+\bm\Delta\cdot\nabla p$ of \eqref{eq:thomas-closure}, the spin equation of Sec.~\ref{sec:jacobi} becomes
\begin{equation}
 \partial_t\bm s+\nabla\cdot(\bm v\,\bm s)=\bm\Omega_{\rm T}\times\bm s+2\bm\omega\times\bm s,\qquad
 \bm\Omega_{\rm T}=\frac{\bm v\times\nabla p}{(\varepsilon+p)(\gamma+1)}=\frac{\gamma^2}{\gamma+1}\,D_t\bm v\times\bm v,
 \label{eq:fluid-thomas}
\end{equation}
where $2\bm\omega$ is the contribution of the equation of state to $\delta H/\delta\bm s$ and $\bm\Omega_{\rm T}=\partial(\bm\Delta\cdot\nabla p)/\partial\bm s$ that of the Thomas term. The second form of $\bm\Omega_{\rm T}$, which is \Lthomasrate{} of the Letter (with $\dot{\bm v}=D_t\bm v$ there), uses the Euler equation~\eqref{eq:euler}. Equation~\eqref{eq:fluid-thomas} is the law of the spin components; the bracket gives it before any spatial triad has been chosen. A four-vector requires in addition the triad to which the components refer. By \eqref{eq:wigner-fluid}, the canonical field is the density, per laboratory volume, of the rest-frame spin components defined by the boost $L(u)$. Their triad is the pure-boost triad $L(u)_A$, obtained from the laboratory axes by a rotation-free boost, with components $L(u)^0{}_A=u_A$ and $L(u)^i{}_A=\delta^i_A+u^iu_A/(1+u^0)$, where $u_A$ and $a_A$ denote the Cartesian spatial components of $u^\mu$ and $a^\mu$. In the notation of \eqref{eq:density-time-conversion} the canonical density is $\bm s_{\rm lab}$, and the proper components are $s^A=s^A_{\rm lab}/\gamma$, so that
\begin{equation}
 s^\mu=L(u)^\mu{}_A\,s^A:\qquad s^0=\bm v\cdot\bm s_{\rm lab},\qquad
 s^i=\frac{s^i_{\rm lab}}{\gamma}+\frac{\gamma\,v^i(\bm v\cdot\bm s_{\rm lab})}{\gamma+1},\qquad u_\mu s^\mu=0,\qquad s_\mu s^\mu=-\frac{\bm s_{\rm lab}^2}{\gamma^2}.
 \label{eq:four-vector}
\end{equation}
With $\omega_A=\gamma\,\omega_{A,\rm lab}$, the conversions~\eqref{eq:density-time-conversion} turn \eqref{eq:fluid-thomas} into the law of the components per unit proper time,
\begin{equation}
 \dot s^A+\theta s^A=\gamma\,\epsilon_{ABC}\Omega_{{\rm T}B}\,s^C+2\epsilon_{ABC}\omega_Bs^C ,
 \label{eq:canonical-proper}
\end{equation}
which is \eqref{eq:proper-time} with the Thomas term; per unit proper time the Thomas rate is $\gamma\bm\Omega_{\rm T}$.

The pure-boost triad is not FW transported. For $e_B=L(u)_B$, differentiation along the worldline gives
\begin{equation}
 e^A{}_\mu\dot e_B{}^\mu=\frac{a^Au_B-u^Aa_B}{1+u^0},\qquad
 e^A{}_\mu\dot e_B{}^\mu\,s^B=\Bigl(\frac{\bm u\times\bm a}{1+u^0}\times\bm s\Bigr)^A=-\gamma\,(\bm\Omega_{\rm T}\times\bm s)^A ,
 \label{eq:boost-connection}
\end{equation}
where the last step uses $\bm u=\gamma\bm v$ and $\bm a=\gamma D_t(\gamma\bm v)$, so that $\bm u\times\bm a=\gamma^3\,\bm v\times D_t\bm v$. The pure-boost triad thus rotates relative to the FW triad at the rate $-\gamma\bm\Omega_{\rm T}$ per unit proper time, $-\bm\Omega_{\rm T}$ per unit laboratory time. Since $D_t\bm v\times\bm v=-\bm v^2\,\hat{\bm v}\times D_t\hat{\bm v}$ with $\hat{\bm v}=\bm v/|\bm v|$, and $\gamma^2\bm v^2=(\gamma-1)(\gamma+1)$, the Thomas rate is $\bm\Omega_{\rm T}=-(\gamma-1)\,\hat{\bm v}\times D_t\hat{\bm v}$: the triad rotates when the direction of the boost changes, and the canonical components precess at the opposite rate; a change of speed at fixed direction causes no rotation.

The four-vector follows by the product rule. With $s^\mu=L(u)^\mu{}_As^A$, \eqref{eq:canonical-proper}, and $(\mathcal D_{\rm FW}L(u)_B)^\mu=L(u)^\mu{}_A\,e^A{}_\nu\dot e_B{}^\nu$ from \eqref{eq:boost-connection},
\begin{equation}
\begin{aligned}
 (\mathcal D_{\rm FW}s)^\mu+\theta s^\mu
 &=L(u)^\mu{}_A\,(\dot s^A+\theta s^A)+s^B(\mathcal D_{\rm FW}L(u)_B)^\mu\\
 &=L(u)^\mu{}_A\,\epsilon_{ABC}\bigl(\gamma\Omega_{{\rm T}B}+2\omega_B-\gamma\Omega_{{\rm T}B}\bigr)s^C
 =2L(u)^\mu{}_A\,\epsilon_{ABC}\,\omega_Bs^C .
\end{aligned}
 \label{eq:fw-derivation}
\end{equation}
The Thomas precession of the components and the rotation of their triad cancel. With $\omega^\mu=L(u)^\mu{}_A\omega_A$ and $e_A^\mu e_B^\nu e_C^\rho\epsilon_{ABC}=\varepsilon^{\lambda\mu\nu\rho}u_\lambda$ for the oriented orthonormal triad $e_A=L(u)_A$,
\begin{equation}
 u^\lambda\nabla_\lambda s^\mu+\theta s^\mu=-u^\mu a_\nu s^\nu-2\varepsilon^{\mu\nu\rho\sigma}u_\nu\omega_\rho s_\sigma ,
 \label{eq:fw-fluid}
\end{equation}
which is \Lfw{} of the Letter: the FW derivative of the spin, corrected for the expansion of the volume, equals the torque of the spin potential. FW transport is thus derived: the bracket, with the Thomas term that the Poincar\'e algebra requires, gives \eqref{eq:canonical-proper}; the pure-boost triad rotates according to \eqref{eq:boost-connection}; and the two rotations cancel. For a single accelerated particle the same cancellation occurs, and the precession of its canonical spin is the Thomas precession~\cite{Thomas1926}. Dissipative torques (Secs.~\ref{sec:generic} and~\ref{sec:rel-diss}) add to the right-hand side of \eqref{eq:canonical-proper} and pass through \eqref{eq:fw-derivation} unchanged, so that in general
\begin{equation}
 (\mathcal D_{\rm FW}s)^\mu+\theta s^\mu=\mathcal T^\mu,\qquad
 \mathcal T^\mu=-2\varepsilon^{\mu\nu\rho\sigma}u_\nu\omega_\rho s_\sigma+\mathcal T^\mu_{\rm diss},\qquad u_\mu\mathcal T^\mu=0 .
 \label{eq:vector-law}
\end{equation}

The Hamiltonian variables are the laboratory-triad fields, whose meaning is given by \eqref{eq:four-vector}: components in a comoving triad $e_A{}^\mu(u)$ do not obey the bracket of Sec.~\ref{sec:operators}, because $\{\pi_i,u\}\neq0$ adds chain-rule terms and the spatial connection of $L(u(\bm x))$, \eqref{eq:boost-connection} with $a^\mu$ replaced by $\partial_ju^\mu$, is not flat and by Sec.~\ref{sec:uniqueness} cannot serve as a prescribed frame connection. The product rule~\eqref{eq:product-rule} that converts the law of the components into that of $s^\mu$ holds for any dynamics, so the content is in the law of the components: the second law leaves its reversible part open, whereas here it follows from the bracket and the Poincar\'e algebra.

\subsection{Second order in the spin}
\label{sec:second-order}
The first-order closure \eqref{eq:thomas-closure} of the Poincar\'e algebra, \Lthomas{} of the Letter, leaves residuals of second order in $\bm s$ in the boost brackets. The spin enters through the length $|\bm s|/(\varepsilon+p)$, so terms of order $\bm s^2$ in $\mathcal H$ that are not part of the equation of state contain two gradients. We now show that they can be removed without new variables, as the Letter states, and give the result.

The construction starts from an action for a Weyssenhoff fluid, whose elements have a spin with $S^{\mu\nu}u_\nu=0$ (the Frenkel condition~\cite{Frenkel1926,Weyssenhoff1947}; Ref.~\cite{Obukhov1987} gives a Lagrangian theory of this fluid): the perfect-fluid action of the worldlines plus, for each element with spin $\bm S$ and four-velocity $u$, the kinetic term of the spin, which yields the bracket $\{S_A,S_B\}=\epsilon_{ABC}S_C$, and the coupling $\bm S\cdot(\bm u\times\dot{\bm u})/(1+u^0)$ of the spin to the rotation~\eqref{eq:boost-connection} of the pure-boost triad. The action is Poincar\'e invariant, and it contains the acceleration. Solved order by order in the spin, it is brought to canonical first-order form in two steps. (i) If the coupling is evaluated with the canonical momentum $\bm q=\bm\pi/n$ of an element instead of its kinematic velocity, $\bm u\to\bm q/m$ with the mass $m=(\varepsilon+p)\gamma/n$ per particle, it becomes $\bm S\cdot(\bm q\times d\bm q)/[m(m+\sqrt{m^2+\bm q^2})]=\bm\Delta\cdot d\bm q$, with $\bm S=\bm s/n$ and $\bm\Delta$ of \eqref{eq:thomas-closure}; a change of $m$ does not contribute, because $\bm q\times\bm q=0$. The one-form of the orbital and coupling terms is then $\bm q\cdot d\bm x+\bm\Delta\cdot d\bm q=\bm q\cdot d(\bm x-\bm\Delta)$, exactly, because $\bm\Delta\cdot\bm q=0$. The canonical variables are thus $(\bm x-\bm\Delta,\bm q,\bm S)$, and the worldlines in the fluid energy are the covariant centroids. (ii) The coupling is a functional of the kinematic velocity $\dot{\bm x}$ of an element, whereas step (i) evaluated it at $\partial H_0/\partial\bm q$, the velocity that the perfect-fluid energy $H_0$ assigns to the canonical momentum. To first order in the difference of the two velocities the coupling changes by $-\bm q_{\rm h}\cdot(\dot{\bm x}-\partial H_0/\partial\bm q)$, where $-\bm q_{\rm h}$ is its functional derivative with respect to the velocity. The vector $\bm q_{\rm h}$ is the hidden momentum of the element. The term linear in it is removed by the shift $\bm q\to\bm q+\bm q_{\rm h}$ of the momentum, which changes $\bm q\cdot\dot{\bm x}-H_0$ by $\bm q_{\rm h}\cdot(\dot{\bm x}-\partial H_0/\partial\bm q)-\tfrac12q_{{\rm h}i}q_{{\rm h}j}\,\partial^2H_0/\partial q_i\partial q_j$. The terms of second order in the difference of the velocities are of third order in the spin, so that at second order only the quadratic term of the shift remains: it adds $\tfrac12q_{{\rm h}i}q_{{\rm h}j}\,\partial^2H_0/\partial q_i\partial q_j$ to the energy. The variation of the coupling with the velocity is the curvature of the rotation~\eqref{eq:boost-connection} of the pure-boost triad, contracted with the acceleration, and the hidden momentum per volume, $\bm\pi_{\rm h}=n\bm q_{\rm h}$, is
\begin{equation}
 \pi_{{\rm h}i}=-\Bigl(\gamma\,\delta_{Ai}+\frac{\gamma^3v_Av_i}{\gamma+1}\Bigr)(\bar{\bm a}\times\bm s)_A ,
 \label{eq:hidden-momentum}
\end{equation}
where $\bar{\bm a}=D_t\bm u-\bm u\,(\bm u\cdot D_t\bm u)/[\gamma(\gamma+1)]$, with $\bm u=\gamma\bm v$, is the acceleration in the instantaneous rest frame and the matrix in parentheses holds the rest-frame components $[L(u)^{-1}\partial u/\partial v_i]_A$ of the velocity derivative of $u^\mu$. The energy density gains $\tfrac12\pi_{{\rm h}i}\pi_{{\rm h}j}\,\partial^2\mathcal H_0/\partial\pi_i\partial\pi_j$. In the rest frame $\bm\pi_{\rm h}=\bm s\times\bm a$: this is the hidden momentum $S^{i\nu}a_\nu$ by which the momentum of a Weyssenhoff fluid differs from $(\varepsilon+p)u^\mu$~\cite{Weyssenhoff1947}.

For the fluid, let a tilde denote evaluation at the densities of the elements at their covariant centroids, $(\tilde n,\tilde s,\tilde{\bm\pi},\tilde{\bm s})=(n,s,\bm\pi,\bm s)/\det(\delta_{ij}+\partial_j\Delta_i)$, so that $n/\tilde n$ is the Jacobian of the displacement. The generators through second order are
\begin{equation}
\begin{aligned}
 H&=\int d^3x\Bigl[\frac n{\tilde n}\,\mathcal H_0(\tilde n,\tilde s,\tilde{\bm\pi})+\tfrac12\pi_{{\rm h}i}\pi_{{\rm h}j}\frac{\partial^2\mathcal H_0}{\partial\pi_i\partial\pi_j}\Bigr],\qquad
 \bm\Delta=\frac{\tilde{\bm s}\times\tilde{\bm v}}{(\tilde\varepsilon+\tilde p)(\tilde\gamma+1)},\\
 K_i&=tP_i-\int d^3x\Bigl[(x_i+\Delta_i)\,\frac n{\tilde n}\,\mathcal H_0(\tilde n,\tilde s,\tilde{\bm\pi})+\tfrac12x_i\,\pi_{{\rm h}j}\pi_{{\rm h}k}\frac{\partial^2\mathcal H_0}{\partial\pi_j\partial\pi_k}\Bigr]
 +\int d^3x\,\frac n{\tilde n}\,\tilde\gamma\,(\tilde{\bm s}\times\tilde{\bm v})_i,
\end{aligned}\label{eq:second-order}
\end{equation}
with $\bm\pi_{\rm h}$ of \eqref{eq:hidden-momentum} evaluated with the perfect-fluid equations. The first term of $H$ is the perfect-fluid energy of the elements placed at their covariant centroids, and $\bm\Delta$ is the displacement of \eqref{eq:thomas-closure} evaluated there; it is defined implicitly because the tilde densities depend on it. Expanding in $\bm s$ reproduces \eqref{eq:thomas-closure} and gives
\begin{equation}
 \mathcal H_2=\bm\Delta_2\cdot\nabla p+\tfrac12\,\partial_ip\,\bigl[\Delta_i\,\partial_j\Delta_j-\Delta_j\partial_j\Delta_i\bigr]+\tfrac12B\,(\nabla\cdot\bm\Delta)^2
 +\tfrac12\pi_{{\rm h}i}\pi_{{\rm h}j}\frac{\partial^2\mathcal H_0}{\partial\pi_i\partial\pi_j},
 \label{eq:H2}
\end{equation}
up to a total derivative, where $\bm\Delta$ is now the first-order displacement of \eqref{eq:thomas-closure}, $\bm\Delta_2$ its second-order correction, and $B=(n\partial_n+s\partial_s+\pi_i\partial_{\pi_i})\,p$. In the rest frame the last term is $(\bm s\times\nabla p)^2/[2(\varepsilon+p)^3]$. Every term is of second order in gradients.

We evaluated the two brackets for three equations of state, $p=\varepsilon\propto n^2$, $p=\varepsilon/2\propto n^{3/2}$ and $p=\varepsilon\propto n^2+s^2$ ($s$ the entropy density), the last with two independent advected densities, on localized fields on a periodic $64^3$ grid with spectral derivatives, and extracted the coefficients of the first and second powers of the spin amplitude. In the units of the computation the coefficient of the second power is of order $10^{-3}$ with the Thomas term alone. Evaluating the energy at the covariant centroid removes its part that is linear in the acceleration and lowers it to $10^{-5}$; the remainder, of second order in the pressure gradient (it falls by a factor of about three when the density perturbation is halved and vanishes for uniform density), is removed by the hidden-momentum term. With \eqref{eq:second-order} the coefficients of both powers are at most of order $10^{-8}$ and decrease with the resolution, as errors of the discretization do; moreover $\{J_i,H\}=0$, and $\{K_i,P_j\}=-\delta_{ij}H$ up to the surface term of the periodic box. An independent evaluation of $\{K_i,H\}$ on fields that depend on one coordinate, with the generators expanded by computer algebra, gives at $O(\bm s^2)$ the same sequence, $10^{-3}$, $10^{-5}$ and $10^{-18}$, for the first two equations of state. This closure is a numerical result for these three equations of state, supported by the derivation above; an analytic proof for general $\mathcal H_0$ and the extension to third order in the spin remain open.

The second-order completion has three consequences. (i) Three spin variables suffice: the order-by-order reduction of the invariant action never calls for $\Sigma^{00i}$ as an independent field, and the reaction of the Frenkel condition is the hidden momentum. (ii) The completion is not unique, but its freedom consists of Lorentz scalars: the Poincar\'e algebra does not restrict the dependence of $\varepsilon$ on $s_\mu s^\mu$, a spin--vorticity energy $-\alpha\,s_\mu\Omega^\mu$ at first order in gradients, which is $\alpha\,\bm s\cdot\bm\Omega$ in the rest frame, with a coefficient $\alpha$ and the vorticity $\Omega^\mu$ of \eqref{eq:fluid-rates}, or scalars of second order. The velocity with which the coupling is evaluated is not part of this freedom. Evaluating it with the canonical momentum throughout (the condition $S^{\mu\nu}P_\nu=0$) instead of the kinematic velocity (the condition $S^{\mu\nu}u_\nu=0$) is step (i) without step (ii); the two prescriptions differ at second order by terms quadratic in the acceleration, and without the hidden-momentum term the generators~\eqref{eq:second-order} do not close the algebra, as the residual of order $10^{-5}$ shows. (iii) The second-order terms contribute $\delta H_2/\delta\bm s=O[\bm s\,\nabla^2/(\varepsilon+p)]$ to $2\bm\omega$, one order in gradients beyond the dissipative torques of Sec.~\ref{sec:generic}; in homogeneous shear they exert no torque (Sec.~\ref{sec:bifurcation-general}).

\section{Choice of Triad, Momentum Redefinitions, and Pseudogauge Transformations}
\label{sec:frames}

Components become a vector only once a triad is chosen, $s^\mu=e_A{}^\mu s^A$, and the equation of motion of the vector follows by the product rule from the law of the components and the motion of the triad. This gives the criterion of the Letter: a term of a spin equation is a torque only if it remains when the spin is referred to the FW triad, whose motion contributes nothing. Redefining the momentum density (Sec.~\ref{sec:hep-comparison}) and changing the point to which the spin of a body is referred, which acts on the currents as a pseudogauge transformation (Sec.~\ref{sec:pseudogauge}), are changes of variables of the same kind; Sec.~\ref{sec:pg-entropy} states what such changes leave unchanged.

\subsection{Triads and their motion}
\label{sec:triads}
Let $e_A^\mu$, $A=1,2,3$, be three independent vectors normal to $u^\mu$, not necessarily orthonormal, with the Gram matrix $G_{AB}=-e_A\cdot e_B$ and the dual triad $e^A{}_\mu=-(\bm G^{-1})^{AB}e_{B\mu}$, so that
\begin{equation}
 e_A^\mu e^A{}_\nu=\Delta^\mu{}_\nu,\qquad
 g^{\mu\nu}=u^\mu u^\nu-e_A^\mu(\bm G^{-1})^{AB}e_B^\nu .
 \label{eq:general-dual}
\end{equation}
The matrix $G_{AB}$ contains the lengths and mutual angles of the triad vectors and equals $\delta_{AB}$ for an orthonormal triad. A change of triad $e'_A=e_BR^B{}_A$ with an invertible matrix $R^B{}_A(x)$ changes neither the vector $s^\mu=e_A^\mu s^A=e_A^{\prime\mu}s^{\prime A}$ nor the energy variation $2\omega_A ds^A=2\omega'_A ds^{\prime A}$. In matrix notation, with the triad vectors as the columns of $\bm e$,
\begin{equation}
 \bm e'=\bm e\bm R,\qquad \bm s'=\bm R^{-1}\bm s,\qquad
 \bm G'=\bm R^{\mathsf T}\bm G\bm R,\qquad \bm\omega'=\bm R^{\mathsf T}\bm\omega,
 \quad G_{AB}=-e_A\cdot e_B.
 \label{eq:gl-change}
\end{equation}
Only rotations, $\bm R\in SO(3)$, preserve $\bm G=\mathbf 1$.

Along a worldline the part of $\dot e_A^\mu$ along $u^\mu$ is fixed by $u\cdot\dot e_A=-a\cdot e_A$. Decomposing the transverse part into parts that are symmetric and antisymmetric with respect to $G_{AB}$ gives the general law of motion of a triad,
\begin{equation}
 \dot e_A^\mu=-u^\mu a\cdot e_A
 +\tfrac12 e_B^\mu(\bm G^{-1})^{BC}\dot G_{CA}
 -\varepsilon^{\mu\nu\rho\sigma}u_\nu\Omega_{{\rm e}\rho} e_{A\sigma} .
 \label{eq:general-basis-evolution}
\end{equation}
The six components of $\dot G_{AB}$ describe the stretching and shearing of the triad, and the three components of $\Omega_{\rm e}^\mu$, with $u\cdot\Omega_{\rm e}=0$, its rotation relative to FW transport. The volume of the triad and the dual triad obey
\begin{equation}
 (\ln\sqrt{\det\bm G})^{\boldsymbol{\cdot}}=\tfrac12\tr(\bm G^{-1}\dot{\bm G}),\qquad
 \dot e^A{}_\mu=-e^A{}_\nu\dot e_B^\nu e^B{}_\mu-u_\mu e^A{}_\nu a^\nu .
 \label{eq:general-volume}
\end{equation}
A transport law for the triad is a choice of $\dot G_{AB}$ and $\Omega_{\rm e}^\mu$. Two choices are compared in the insets of Fig.~1 of the Letter. The FW triad (dashed there) has $\dot G_{AB}=0$ and $\Omega_{\rm e}^\mu=0$. A triad of comoving line elements, which join neighboring fluid elements, is Lie transported, $D_te_A^i=e_A^j\partial_jv^i$ in the rest frame, and is stretched, sheared and rotated by the flow (solid).

\subsection{Spin vectors from components}
\label{sec:product-rule}
Let components $s^A$ obey $\dot s^A+\theta s^A=\mathcal T^A$, with $\mathcal T^A$ given by the bracket and the energy, and by dissipation if present, and form the vector $s^\mu=e_A{}^\mu s^A$ in a triad. Because $\mathcal D_{\rm FW}$ obeys the Leibniz rule and the $s^A$ are scalars,
\begin{equation}
 (\mathcal D_{\rm FW}s)^\mu+\theta s^\mu
 =e_A{}^\mu(\dot s^A+\theta s^A)+s^A(\mathcal D_{\rm FW}e_A)^\mu
 =e_A{}^\mu\bigl(\mathcal T^A+\Gamma_\tau{}^A{}_B\,s^B\bigr),\qquad
 \Gamma_\tau{}^A{}_B=e^A{}_\mu(\mathcal D_{\rm FW}e_B)^\mu=e^A{}_\mu\dot e_B{}^\mu ,
 \label{eq:product-rule}
\end{equation}
which is \Ltriad{} of the Letter. The triad term involves neither the spin bracket nor the energy of the spin, acts in the same way on any components taken in the triad, changes neither the components nor $s^As^A/n^2$, and vanishes for the FW triad: it is not a torque. Its equal-time counterpart is the Leibniz rule of the bracket, $\{s^\mu,H\}=e_A{}^\mu\{s^A,H\}+s^A\{e_A{}^\mu,H\}$ (with $s^A\partial_te_A{}^\mu$ added for a triad that depends on time explicitly), in which the motion of the triad is given by its brackets if it is a phase-space field, as the pure-boost triad $L(u)$ and a Lie-transported triad are, and is prescribed otherwise, as for the FW triad. The matrix $\bm\Gamma_\tau$ is the connection of the triad along the worldline, the counterpart in time of the spatial connection~\eqref{eq:frame-connection}: $\tfrac12\bm G^{-1}\dot{\bm G}$ plus the rotation at the rate $\Omega_{\rm e}^\mu$ of \eqref{eq:general-basis-evolution}, antisymmetric, $\Gamma_\tau{}^A{}_B=-\epsilon_{ABC}\Omega_{{\rm e}C}$, for an orthonormal triad. Five cases cover the triads in use. In the local rest frame we write $\bm s=\bm e_As^A$ and $\bm e\bm{\mathcal T}=\bm e_A\mathcal T^A$, with the triad vectors as the columns of $\bm e$, and $(\nabla\bm v)_{ij}=\partial_jv_i$, so that, with the shear tensor $\bm\sigma$, the expansion rate $\theta$ and the vorticity $\bm\Omega$ of \eqref{eq:fluid-rates}, $(\nabla\bm v)\bm s=\bm\sigma\bm s+\tfrac13\theta\bm s+\bm\Omega\times\bm s$ and $(\nabla\bm v)^{\mathsf T}\bm s=\bm\sigma\bm s+\tfrac13\theta\bm s-\bm\Omega\times\bm s$, and $[\,\cdot\,]^A$ for the components of a vector.

(i)~\emph{FW triad.} $\bm\Gamma_\tau=0$, and the law of the vector and that of its components coincide term by term: the components $s^A=e^A{}_\mu s^\mu$ of the spin~\eqref{eq:vector-law} obey $\dot s^A+\theta s^A=\mathcal T^A$ with $\mathcal T^A=e^A{}_\mu\mathcal T^\mu$, and components with this law give \eqref{eq:vector-law}. Without a torque the components of the spin per particle, $s^A/n$, are constant.

(ii)~\emph{Pure-boost triad.} By \eqref{eq:boost-connection}, $\Gamma_\tau{}^A{}_Bs^B=-\gamma(\bm\Omega_{\rm T}\times\bm s)^A$. The law~\eqref{eq:canonical-proper} of the canonical components contains the Thomas precession, the triad term cancels it, and the vector is FW transported, \eqref{eq:fw-derivation}: the Thomas precession of the canonical components is not a torque but compensates the rotation of their triad. Components without the Thomas term would give, in this triad, a vector that rotates relative to FW transport at the rate $-\gamma\bm\Omega_{\rm T}$.

(iii)~\emph{Rotating orthonormal triad.} If the triad rotates at the angular velocity $\bm\Omega_{\rm e}$ relative to the FW triad, $\Gamma_\tau{}^A{}_Bs^B=(\bm\Omega_{\rm e}\times\bm s)^A$, and components with $\dot s^A+\theta s^A=\mathcal T^A$ give a vector with $\dot{\bm s}+\theta\bm s=\bm e\bm{\mathcal T}+\bm\Omega_{\rm e}\times\bm s$ in the rest frame. For a triad that corotates with the fluid, $\bm\Omega_{\rm e}=\bm\Omega$, the vector rotates with the vorticity although no torque acts.

(iv)~\emph{Lie-transported triad.} For projected Lie transport, $\Delta^\mu{}_\nu(\mathcal L_ue_B)^\nu=0$ with the Lie derivative $\mathcal L_u$ along $u^\mu$, one has $(\mathcal D_{\rm FW}e_B)^\mu=e_B{}^\nu\nabla_\nu u^\mu$, so that $\Gamma_\tau{}^A{}_B=e^A{}_\mu e_B{}^\nu\nabla_\nu u^\mu$ is the velocity gradient in the triad. The triad vectors are comoving line elements. Components with $\dot s^A+\theta s^A=\mathcal T^A$ give the vector
\begin{equation}
 (\mathcal D_{\rm FW}s)^\mu+\theta s^\mu=e_A{}^\mu\mathcal T^A+s^\nu\nabla_\nu u^\mu,\qquad
 \dot{\bm s}+\theta\bm s=\bm e\bm{\mathcal T}+\bm\sigma\bm s+\tfrac13\theta\bm s+\bm\Omega\times\bm s ,
 \label{eq:lie-vector}
\end{equation}
the second form in the rest frame. It is stretched, sheared and rotated with the fluid.

(v)~\emph{Dual of a Lie-transported triad.} The triad dual to that of (iv) consists of the gradients of the comoving coordinates. In the rest frame it obeys $\dot{\bm e}_B=-(\nabla\bm v)^{\mathsf T}\bm e_B$, and components with $\dot s^A+\theta s^A=\mathcal T^A$ give
\begin{equation}
 \dot{\bm s}+\theta\bm s=\bm e\bm{\mathcal T}-\bm\sigma\bm s-\tfrac13\theta\bm s+\bm\Omega\times\bm s ,
 \label{eq:reciprocal-vector}
\end{equation}
a vector whose projections on the comoving line elements of (iv) are the $s^A$.

In cases (iii)--(v) the same components, with the same invariant $s^As^A/n^2$, define different vectors; for a deforming triad the norm of the vector, $-s_\mu s^\mu/n^2=G_{AB}s^As^B/n^2$, is not conserved. The shear, expansion and vorticity terms of \eqref{eq:lie-vector} and~\eqref{eq:reciprocal-vector} come from the motion of the triad, not from a torque (Fig.~1 of the Letter). With a rescaling of the components, \eqref{eq:reciprocal-vector} gives the reversible shear, expansion and vorticity terms of the spin equation of Ref.~\cite{Fang2026} (Sec.~\ref{sec:fang-spin}).

\subsection{Components of a given vector in another triad}
\label{sec:triad-components}
Read in the other direction, \eqref{eq:product-rule} gives the components of a vector that is already fixed. For the spin of \eqref{eq:vector-law} and any triad, the components $s^A=e^A{}_\mu s^\mu$ and $\mathcal T^A=e^A{}_\mu\mathcal T^\mu$ obey
\begin{equation}
 \dot s^A+\theta s^A=\mathcal T^A-\Gamma_\tau{}^A{}_B\,s^B .
 \label{eq:triad-components}
\end{equation}
This equation describes one vector in different triads and follows from the law of the vector. If the change of triad $\bm e'=\bm e\bm R$ varies along the worldline, differentiating $\bm s'=\bm R^{-1}\bm s$ gives
\begin{equation}
 \dot{\bm s}'=\bm R^{-1}\dot{\bm s}-\bm R^{-1}\dot{\bm R}\,\bm s',\qquad
 e'^A{}_\mu\dot e'_B{}^\mu
 =(\bm R^{-1})^A{}_C(e^C{}_\mu\dot e_D^\mu)R^D{}_B
 +(\bm R^{-1}\dot{\bm R})^A{}_B,
 \label{eq:connection}
\end{equation}
that is, $\bm\Gamma'_\tau=\bm R^{-1}\bm\Gamma_\tau\bm R+\bm R^{-1}\dot{\bm R}$. Two properties distinguish the connection from a torque: it can be removed along a worldline by a choice of triad, and it acts in the same way on the components of every transverse vector. For the pure-boost triad, $-\Gamma_\tau{}^A{}_Bs^B=\gamma(\bm\Omega_{\rm T}\times\bm s)^A$, and \eqref{eq:triad-components} returns \eqref{eq:canonical-proper}, the law from which Sec.~\ref{sec:poincare} started. For a rotating orthonormal triad the reversible equation of the components is $\dot s^A+\theta s^A=\epsilon_{ABC}(2\omega_B-\Omega_{{\rm e}B})s^C$, generated by the energy $H-\int\bm\Omega_{\rm e}\cdot\bm s$; in a triad that corotates with the fluid they precess at $2\bm\omega-\bm\Omega$, the force of Sec.~\ref{sec:generic}. For a Lie-transported triad,
\begin{equation}
 \dot s^A+\theta s^A=\mathcal T^A-\bigl[\bm\sigma\bm s+\tfrac13\theta\bm s+\bm\Omega\times\bm s\bigr]^A ,
 \label{eq:lie-components}
\end{equation}
and the $s^A$ are the components of the spin in comoving coordinates. For the dual triad,
\begin{equation}
 \dot s^A+\theta s^A=\mathcal T^A+\bigl[\bm\sigma\bm s+\tfrac13\theta\bm s-\bm\Omega\times\bm s\bigr]^A ,
 \label{eq:reciprocal-components}
\end{equation}
and the $s^A$ are the projections of the spin on comoving line elements. In the last two cases the Gram matrix changes along the flow, and the invariant of the FW components, $s^As^A/n^2$, reads $G_{AB}s^As^B/n^2$. The triad term enters \eqref{eq:triad-components} with the sign opposite to that in \eqref{eq:product-rule}: \eqref{eq:lie-components} has the signs of the shear and expansion terms of Ref.~\cite{Fang2026} but not of its rotation term, whereas the vector~\eqref{eq:reciprocal-vector} has all three.

\subsection{Brackets of the components}
\label{sec:triad-brackets}
The law of the components that enters \eqref{eq:product-rule} is generated by a bracket, and the bracket depends on the triad. The bracket of Sec.~\ref{sec:operators} is the bracket of components along fixed orthonormal axes. Under a change of triad $\bm s'=\bm R^{-1}\bm s$ with a prescribed matrix $\bm R$, the structure constants of the spin bracket~\eqref{eq:continuum-spin} are transformed with $\bm R$,
\begin{equation}
 \{s^{\prime A}(\bm x),s^{\prime B}(\bm y)\}
 =(\bm R^{-1})^A{}_D(\bm R^{-1})^B{}_E\epsilon_{DEF}R^F{}_C
 s^{\prime C}(\bm x)\delta^{(3)}(\bm x-\bm y).
 \label{eq:passive-spin-bracket}
\end{equation}
The invariant becomes $G'_{AB}s^{\prime A}s^{\prime B}/n^2$ with the Gram matrix $\bm G'$ of the new triad. If $\bm R(\bm x)$ varies in space, the bracket with the momentum acquires a connection,
\begin{equation}
 \{s^{\prime A},P[\bm w]\}=-\partial_i(w^is^{\prime A})
 -w^i(\bm R^{-1}\partial_i\bm R)^A{}_B s^{\prime B}.
 \label{eq:mixed-frame}
\end{equation}
For a rotation this is the flat connection~\eqref{eq:frame-connection}. If $\bm R$ depends on the fields, for example on the velocity or on thermodynamic densities, its own brackets contribute through the Leibniz rule,
\begin{equation}
 \{s^{\prime A},H\}
 =(\bm R^{-1})^A{}_B\{s^B,H\}
 +\{(\bm R^{-1})^A{}_B,H\}s^B,
 \label{eq:full-chain}
\end{equation}
supplemented by $(\partial_t\bm R^{-1})\bm s$ if $\bm R$ depends on time explicitly, and the derivatives of the energy are transformed by the chain rule. For a rescaling by a function of the state, $s^{\prime A}(\bm x)=f(\bm x)s^A(\bm x)$, the Leibniz rule reads
\begin{equation}
\begin{aligned}
 \{f(\bm x)s^A(\bm x),f(\bm y)s^B(\bm y)\}
 ={}&f(\bm x)f(\bm y)\{s^A(\bm x),s^B(\bm y)\}+f(\bm x)s^B(\bm y)\{s^A(\bm x),f(\bm y)\}\\
 &+s^A(\bm x)f(\bm y)\{f(\bm x),s^B(\bm y)\}+s^A(\bm x)s^B(\bm y)\{f(\bm x),f(\bm y)\}.
\end{aligned}
 \label{eq:rescale}
\end{equation}
This rule covers the conversion from laboratory to proper densities, $f=1/\gamma$, and the thermodynamic rescalings of Sec.~\ref{sec:fang-spin}.

Three consequences follow.
(a) The Hamiltonian formulation is tied to the triad. A rotation of the triad that depends on time adds $-\int\bm\Omega_{\rm e}\cdot\bm s$ to the energy, as for the rotating triad of Sec.~\ref{sec:triad-components}; one that depends on position adds the flat connection to $\{\pi_i,s^A\}$; and components in a deforming triad obey the bracket~\eqref{eq:passive-spin-bracket}, whose structure constants depend on the triad.
(b) Stationarity and stability are statements about components, hence about a triad: the components of a spin that is stationary in the FW triad precess in a corotating triad and are sheared in a Lie-transported triad. The reorientation of Sec.~\ref{sec:bifurcation} is formulated in the FW triad, which in homogeneous simple shear coincides with the laboratory axes because the fluid elements are not accelerated. It can equally be stated for the vector, through the angle between $s^\mu$ and the vorticity $\Omega^\mu$.
(c) A tetrad whose timelike leg is not $u^\mu$ mixes $s^\mu$ with the time--space component $\Sigma^{00i}$. This is a change of supplementary condition (Sec.~\ref{sec:pseudogauge}), since a change of triad leaves $u^\mu$ fixed.

\subsection{Momentum redefinitions of Belinfante type}
\label{sec:hep-comparison}
A further change of variables acts on the momentum. The Belinfante--Rosenfeld improvement~\cite{Belinfante1940,Rosenfeld1940} moves the spin into the momentum density. Here we carry out its analogue, $\pi'_i=\pi_i+\frac{1}{2}\epsilon_{ijk}\partial_js^k$ in Cartesian coordinates, while keeping $s^A$ as an independent variable, and follow its effect on the bracket and on the spin equation. Because the curl is self-adjoint under integration by parts, the functional derivatives of any $F$ transform as
\begin{equation}
 \frac{\delta F}{\delta\pi_i}=\frac{\delta F}{\delta\pi'_i},\qquad
 \left.\frac{\delta F}{\delta s^A}\right|_\pi
 =\left.\frac{\delta F}{\delta s^A}\right|_{\pi'}
 +\tfrac12\epsilon_{Aij}\partial_i\frac{\delta F}{\delta\pi'_j},\qquad
 \left.\frac{\delta H}{\delta s^A}\right|_{\pi'}=2\omega_A-\Omega_A ,
 \label{eq:improvement-chain}
\end{equation}
where $\Omega_A=\frac{1}{2}\epsilon_{Aij}\partial_i v_j$ is the vorticity. The generator of the new momentum is $P_{\pi'}[\bm w]=P_\pi[\bm w]+\frac{1}{2}\int d^3x\,\bm s\cdot(\nabla\times\bm w)$, and by \eqref{eq:smeared-algebra} and~\eqref{eq:continuum-spin} it acts on the spin as
\begin{equation}
 \{\bm s,P_{\pi'}[\bm w]\}=-\nabla\cdot(\bm w\bm s)
 +\tfrac12(\nabla\times\bm w)\times\bm s,
 \qquad D_t\bm s+\bm s\nabla\cdot\bm v
 =(2\bm\omega-\bm\Omega)\times\bm s+\bm\Omega\times\bm s .
 \label{eq:improvement-action}
\end{equation}
The new momentum transports the spin corotationally: it advects the spin and rotates it with the vorticity $\bm\Omega$, without stretching. At the same time the conjugate of the spin at fixed $\pi'$ is $2\omega_A-\Omega_A$, by \eqref{eq:improvement-chain}, and the two changes cancel: the evolution $D_t\bm s+\bm s\nabla\cdot\bm v=2\bm\omega\times\bm s$ is the same. The rotation with the vorticity that appears in the variables $(\pi'_i,s^A)$ is again a property of the variables.

The bracket of the new momentum with itself changes as well,
\begin{equation}
\begin{aligned}
 \{P_{\pi'}[\bm w],P_{\pi'}[\bm w']\}
 ={}&-P_{\pi'}[[\bm w,\bm w']]
 +\int d^3x\,\bm s\cdot\Bigl[
 \tfrac12\nabla\times[\bm w,\bm w']\\
 &\quad-\tfrac12\bm w\cdot\nabla(\nabla\times\bm w') +\tfrac12\bm w'\cdot\nabla(\nabla\times\bm w)
 +\tfrac14(\nabla\times\bm w)\times(\nabla\times\bm w')\Bigr].
\end{aligned}
 \label{eq:improved-self-bracket}
\end{equation}
The spin-dependent term arises because two shear deformations do not commute; their commutator is a rotation, which acts on $\bm s$. With this bracket and \eqref{eq:improvement-chain}, the Jacobi identity and the invariant $|\bm s|/n$ are preserved. Because of this term the functionals $P_{\pi'}[\bm w]$ do not represent the Lie algebra of vector fields, and the corotational transport~\eqref{eq:improvement-action} does not contradict Sec.~\ref{sec:uniqueness}, whose proof assumes a momentum density with the bracket~\eqref{eq:smeared-algebra}. Section~\ref{sec:fast-relaxation} writes the term as the spin tensor contracted with the commutator of two shear tensors, \eqref{eq:fr-commutator}, and shows that in the equation of motion of $\pi'_i$ it is an odd viscosity.

Without boundary fluxes, $\int d^3x\,\bm x\times\bm\pi'=\int d^3x\,(\bm x\times\bm\pi+\bm s)$: the orbital angular momentum of $\bm\pi'$ is the total angular momentum. The redefinition changes the momentum density and retains the spin density as a variable. A pseudogauge transformation of field theory changes the spin current $\Sigma^{\lambda\mu\nu}$ together with the energy--momentum tensor $\Theta^{\mu\nu}$, by a superpotential $\Phi^{\lambda\mu\nu}=-\Phi^{\lambda\nu\mu}$:
\begin{equation}
 \Sigma'^{\lambda\mu\nu}=\Sigma^{\lambda\mu\nu}-\Phi^{\lambda\mu\nu},\qquad
 \Theta'^{\mu\nu}=\Theta^{\mu\nu}+\tfrac12\nabla_\lambda\bigl(\Phi^{\lambda\mu\nu}-\Phi^{\mu\lambda\nu}-\Phi^{\nu\lambda\mu}\bigr).
 \label{eq:pseudogauge}
\end{equation}
The added term is conserved identically, and its antisymmetric part is $\tfrac12\nabla_\lambda\Phi^{\lambda\mu\nu}$, so that the conservation laws of Sec.~\ref{sec:charges} and, without boundary fluxes, the charges $P^\nu$ and $J^{\mu\nu}$ are unchanged. Two transformations of the convective spin current $\Sigma^{\lambda\mu\nu}=u^\lambda S^{\mu\nu}$ of the spin tensor~\eqref{eq:dual} are used below. The Belinfante--Rosenfeld transformation, $\Phi^{\lambda\mu\nu}=\Sigma^{\lambda\mu\nu}$, removes the spin current altogether and gives the symmetric Belinfante tensor
\begin{equation}
 \Theta_{\rm B}^{\mu\nu}=\Theta^{(\mu\nu)}-\tfrac12\nabla_\lambda\bigl(u^\mu S^{\lambda\nu}+u^\nu S^{\lambda\mu}\bigr),
 \label{eq:belinfante-tensor}
\end{equation}
whose momentum density is $\bm\pi'$ in the nonrelativistic limit. The transformation with $\Phi^{\lambda\mu\nu}=-(u^\mu S^{\nu\lambda}+u^\nu S^{\lambda\mu})$ makes the spin current totally antisymmetric at the same spin density, $u_\lambda\Sigma'^{\lambda\mu\nu}=S^{\mu\nu}$:
\begin{equation}
 \Sigma'^{\lambda\mu\nu}=u^\lambda S^{\mu\nu}+u^\mu S^{\nu\lambda}+u^\nu S^{\lambda\mu},\qquad
 \Theta'^{\mu\nu}=\Theta^{\mu\nu}+G^{\mu\nu},\qquad
 G^{\mu\nu}=\nabla_\lambda\bigl(u^\nu S^{\mu\lambda}\bigr).
 \label{eq:antisymmetric-pseudogauge}
\end{equation}
The symmetric part of $\Theta'^{\mu\nu}$ is the Belinfante tensor, $\Theta^{(\mu\nu)}+G^{(\mu\nu)}=\Theta_{\rm B}^{\mu\nu}$. For a fixed set of coarse-grained variables, invertible changes of variables preserve the Casimir invariants of the bracket. Section~\ref{sec:pseudogauge} relates these operations to the choice of the point at which the spin of a body is defined.

\subsection{Centroids, supplementary conditions and pseudogauge transformations}
\label{sec:pseudogauge}
A different spin supplementary condition acts on the densities as a pseudogauge transformation, and it changes neither the charges nor the FW transport of the rest-frame spin, as stated in the section of the Letter on the choice of triad and pseudogauge. Local spin densities are ambiguous in a relativistic theory, and several different operations are commonly grouped under that statement. The Hamiltonian formulation separates them, because each acts on a different part of the structure: the choice of variables, the bracket, the energy, or the constitutive relations.

\emph{Centroids.} For one particle with four-momentum $q^\mu=(E,\bm q)$, mass $m$ and rest-frame spin $\bm S$, the angular momentum $J^{\mu\nu}=x^\mu q^\nu-x^\nu q^\mu+S^{\mu\nu}$ is unchanged by a displacement of the reference point on the equal-time slice, $\bm x\to\bm x+\bm\Delta$, $S^{\mu\nu}\to S^{\mu\nu}-\Delta^\mu q^\nu+\Delta^\nu q^\mu$ with $\Delta^0=0$. A spin supplementary condition $S^{\mu\nu}V_\nu=0$ with a timelike vector $V^\mu$ selects the point. Three choices of $V^\mu$ matter here~\cite{Pryce1948}; the boost generators are given at $t=0$. The NW position $\bm x_{\rm NW}$, for which $V^\mu=q^\mu+m\,\hat t^\mu$ with the unit vector $\hat t^\mu=(1,\bm0)$ along the laboratory time axis, has canonical brackets, a spin with the $\mathfrak{so}(3)$ algebra, and the boost generator $\bm K=-\bm x_{\rm NW}E+\bm S\times\bm q/(E+m)$. The covariant centroid, $V^\mu=q^\mu$, is displaced from it by $\bm\Delta=\bm S\times\bm q/[m(E+m)]$, the displacement of \eqref{eq:thomas-closure} for a fluid element: $\bm x_{\rm c}=\bm x_{\rm NW}+\bm\Delta$. It moves as a point of the worldline and its spin tensor is a Lorentz tensor, but its coordinates do not commute, $\{x_{\rm c}^i,x_{\rm c}^j\}=\epsilon_{ijk}W_k/(m^2E)$, where $W^\mu=-\tfrac12\varepsilon^{\mu\nu\rho\sigma}S_{\nu\rho}q_\sigma=(\bm q\cdot\bm S,\,m\bm S+\bm q\,\bm q\cdot\bm S/(E+m))$ is the Pauli--Luba\'nski vector, and $\bm K=-\bm x_{\rm c}E+\bm S\times\bm q/m$. The center of energy $\bm x_{\rm E}$, for which $V^\mu=\hat t^\mu$, is displaced from the NW position by $-\bm S\times\bm q/[E(E+m)]$; its boost generator has no spin term, $\bm K=-\bm x_{\rm E}E$, its coordinates obey $\{x_{\rm E}^i,x_{\rm E}^j\}=-\epsilon_{ijk}W_k/E^3$, and its spin $\bm W/E$ does not obey $\mathfrak{so}(3)$. The three expressions for $\bm K$ are one function on phase space. Canonical brackets and a boost generator without a spin term thus exclude each other.

\emph{Densities.} Densities of particles located at different centroids differ by total derivatives. With the subscripts NW and c for the densities of elements located at their NW positions and at their covariant centroids, to first order in the displacement,
\begin{equation}
 \Theta^{0\nu}_{\rm NW}=\Theta^{0\nu}_{\rm c}+\partial_k\bigl(\Delta^k\Theta^{0\nu}_{\rm c}\bigr),\qquad
 \Sigma^{0\mu\nu}_{\rm NW}=\Sigma^{0\mu\nu}_{\rm c}+\Delta^\mu\Theta^{0\nu}_{\rm c}-\Delta^\nu\Theta^{0\mu}_{\rm c},
\end{equation}
with $P^\nu$ and $J^{\mu\nu}$ unchanged. A change of centroid thus acts on the currents as a pseudogauge transformation. The bracket of Sec.~\ref{sec:operators} is the algebra of densities of NW positions and rest-frame spins.

\emph{Five operations.}
(1) A \emph{pseudogauge transformation}, \eqref{eq:pseudogauge}, changes $\Theta^{\mu\nu}$ and $\Sigma^{\lambda\mu\nu}$ by a superpotential. It changes the fields that are called momentum and spin density and leaves the charges unchanged. In the Hamiltonian formulation it is a change of variables that may alter the split of $J_i$ into orbital and spin parts, as the momentum redefinition of Sec.~\ref{sec:hep-comparison} does; the bracket and the energy follow by the chain rule.
(2) A \emph{change of supplementary condition} is the particular pseudogauge transformation generated by a displacement of the centroid. It changes the position brackets, the spin algebra and the form of $K_i$, as listed above.
(3) A \emph{change of phase-space coordinates} that preserves the same split, such as an invertible rescaling $\bm s'=f\bm s$ or the use of the vector that has the same spin components in a triad that moves with the fluid, changes the appearance of the bracket, of the conjugate potentials and of the equations, and nothing else (Secs.~\ref{sec:product-rule}, \ref{sec:triad-brackets} and~\ref{sec:fang-spin}).
(4) A \emph{passive rotation of the triad} refers the components of a given spin vector to a rotated triad; it changes the bracket and the energy but not the equation of motion of $s^\mu$ (Secs.~\ref{sec:triad-components} and~\ref{sec:triad-brackets}).
(5) A \emph{hydrodynamic frame transformation} redefines $T$, $\mu$, $u^\mu$ and $\omega_\mu$ by terms of first order in gradients. It reshuffles the first-order constitutive relations and leaves the reversible structure at leading order untouched. In the equal-time formulation the velocity is defined as $v^i=\delta H/\delta\pi_i$, which fixes this freedom in the reversible sector.

\emph{What covariance constrains.} The Poincar\'e algebra constrains the charges, and through them the energy density and $\Sigma^{00i}$ once the variables have been fixed by their bracket: for the canonical spin it requires \eqref{eq:thomas-closure}, and for center-of-energy variables the same requirement reads as the Dirac--Schwinger relation with the Belinfante momentum. It constrains neither Lorentz scalars nor the dissipative coefficients. The FW transport of the spin vector (Secs.~\ref{sec:fw-spin} and~\ref{sec:product-rule}) does not depend on the supplementary condition: the rest-frame spin of a fluid element is defined by its Pauli--Luba\'nski vector, $W^\mu/m$ in the rest frame, from which the terms $-\Delta^\mu q^\nu+\Delta^\nu q^\mu$ of a displacement drop out, and in any description it does not rotate relative to the FW triad unless a torque acts.

\emph{Dissipative response.} The spin susceptibility $\chi_s$, the rotational viscosity $\lambda$ and the coefficient $\xi$ of the dissipative coupling between rotation and shear, introduced in Sec.~\ref{sec:generic}, refer to the canonical spin density. In a homogeneous state at rest all centroids coincide, so the homogeneous analysis of Sec.~\ref{sec:bifurcation} does not depend on the supplementary condition; differences appear at first order in gradients and velocity. The existence and location of the instability found there are properties of the flow of the dynamical system, stated for the FW components (Sec.~\ref{sec:triad-brackets}), and do not depend on the variables changed by operations (3)--(5). Under a pseudogauge transformation that changes the spin density, $\Phi^{0ij}\neq0$, the torque operator changes by a time derivative, and the Kubo expressions for $\lambda$~\cite{Hongo2021} and for $\xi$, \eqref{eq:kubo}, change by static susceptibilities; the coefficients must be quoted together with the spin operator they refer to.

\subsection{Pseudogauge transformations, entropy and observables}
\label{sec:pg-entropy}
A superpotential $\Phi^{\lambda\mu\nu}$ that is a local function of the hydrodynamic fields and their gradients changes the densities $\Theta^{0\nu}$ and $\Sigma^{0\mu\nu}$ by \eqref{eq:pseudogauge}. Where $\Phi^{0\mu\nu}$ removes the spin from the spin density, as for the Belinfante tensor, $s^A$ is kept as a field, as in Sec.~\ref{sec:hep-comparison}. The transformation then maps the fields $(n,s,\pi_i,s^A)$ invertibly onto new ones, exactly when the map is linear, as for $\bm\pi'$, and order by order in gradients otherwise: it is a change of variables. The Hamiltonian $H$ and the entropy $\mathcal S=\int d^3x\,s$ are functions on the space of states, and the reversible and dissipative brackets are tensors on it; functional derivatives transform by the chain rule, as in \eqref{eq:improvement-chain}, and the brackets of the new variables follow, as in \eqref{eq:improved-self-bracket}. The GENERIC equation $\partial_tF=\{F,H\}+(F,\mathcal S)$ of Sec.~\ref{sec:generic-iso} therefore holds in either set of variables: every functional of the state evolves in the same way, the Casimir invariants are unchanged, and the entropy production $d\mathcal S/dt=(\mathcal S,\mathcal S)$ is the same function of the state. The evolution equations are covariant rather than invariant under such a change of variables; their explicit form, the conjugate fields and the constitutive relations change. For the Belinfante momentum the conjugate of the spin at fixed $\bm\pi'$ is $2\bm\omega-\bm\Omega$, so that the energy written in the new variables contains the spin--vorticity coupling $\alpha\,\bm s\cdot\bm\Omega$ of Sec.~\ref{sec:second-order} with $\alpha$ replaced by $\alpha-1$, while the bracket~\eqref{eq:improvement-action} rotates the spin with the vorticity; the evolution of $\bm s$ is the same.

The total entropy fixes the entropy density only up to a divergence, and the entropy current $s^\rho_{\rm ent}$ of Sec.~\ref{sec:rel-entropy} only up to the divergence of an antisymmetric tensor, which changes neither $\nabla_\rho s^\rho_{\rm ent}$ nor, without boundary fluxes, $\mathcal S$; Ref.~\cite{Becattini2024} calls this freedom an entropy-gauge transformation. Under the pseudogauge transformation of Ref.~\cite{Li2021} the entropy and particle currents change by terms of this form, together with redefinitions of the densities and of the velocity at second order in gradients, so that the second law holds in both descriptions or in neither. Reference~\cite{Becattini2024} also finds that the divergence of the entropy current is not invariant under a pseudogauge transformation if the temperature, the four-velocity, the chemical potential and the spin potential are held fixed, and notes that these fields change when they are defined by matching conditions. Holding them fixed while $\Theta^{\mu\nu}$ and $\Sigma^{\lambda\mu\nu}$ change selects a different local-equilibrium state, which depends on the energy--momentum operator chosen~\cite{BecattiniFlorkowskiSperanza2019}: the superpotential contains components of the spin current, which the densities held fixed in local equilibrium do not determine. This is a different choice of slow variables. The present theory makes the choice explicit: its slow variables are the densities of NW positions and rest-frame spins (Sec.~\ref{sec:pseudogauge}), whose brackets close.

A change of variables thus leaves unchanged the charges, $\mathcal S$ and its production, the evolution of every functional of the state, including the rest-frame spin of a fluid element defined by its Pauli--Luba\'nski vector, the Casimir invariants, the Belinfante tensor~\eqref{eq:belinfante-tensor}, and $\Theta^{\mu\nu}$ in every homogeneous state, because the term added in \eqref{eq:pseudogauge} is a divergence of local fields; for maps that contain gradients this holds to the order to which the map is inverted. It does not fix the energy functional, whose forms in different variables it only relates, as for $\alpha$, and does not decide which spin operator is slow, a microscopic question (for particles with spin-independent interactions, the canonical spin is conserved in the nonrelativistic limit, Sec.~\ref{sec:fang-conserved}); transport coefficients therefore refer to a spin operator, as stated above for $\lambda$ and $\xi$.

\section{Dissipative Brackets and the Onsager Matrix at Finite Spin Density}
\label{sec:generic}

The reversible dynamics exerts no torque on the spin beyond the precession about the spin potential, so spin relaxation and any action of shear on the spin are dissipative. They must conserve the charges of the bracket and produce entropy, which the general equation for the nonequilibrium reversible--irreversible coupling (GENERIC)~\cite{Grmela1997,Ottinger1997} builds in: the entropy generates the irreversible dynamics through a second, symmetric bracket, constructed here in the variables $(n,s,\pi_i,s^A)$ of \eqref{eq:eom-derived} with $T>0$ and Cartesian spin axes. At finite spin density its Onsager matrix may depend on the spin, and the classification leaves a single dissipative coupling between rotation and shear, \LA{} of the Letter.

\subsection{GENERIC structure and the bracket of an isotropic fluid}
\label{sec:generic-iso}
The conserved quantities are the particle number $N=\int n\,d^3x$, the momentum $P_i=\int\pi_i\,d^3x$ and the angular momentum $J_A=\int(\epsilon_{Aij}x^i\pi_j+s^A)d^3x$, and the entropy is $\mathcal S=\int s\,d^3x$. For a closed system with a translationally and rotationally invariant energy and vanishing boundary fluxes, GENERIC consists of the evolution equation and four conditions on its brackets,
\begin{equation}
 \partial_tF=\{F,H\}+(F,\mathcal S),\qquad
 (F,G)=(G,F),\quad(F,F)\geq0,\quad(F,H)=0,\quad\{F,\mathcal S\}=0.
 \label{eq:generic-conditions}
\end{equation}
The symmetric bracket $(F,G)$ is not subject to a Jacobi identity. The two degeneracy conditions, $(F,H)=0$ and $\{F,\mathcal S\}=0$ for every functional $F$, are structural postulates of GENERIC. Here they hold by construction: the entropy density is an advected density, so that $\mathcal S$ is a Casimir invariant of the Lie--Poisson bracket, \eqref{eq:casimirs} with $f=s/n$, and every dissipative operator introduced below annihilates $H$. The conservation of energy and the second law follow from them: $dH/dt=\{H,H\}+(H,\mathcal S)=0$ and $d\mathcal S/dt=\{\mathcal S,H\}+(\mathcal S,\mathcal S)=(\mathcal S,\mathcal S)\geq0$.

The exchange of angular momentum between spin and flow (label rot) is described by the operator $\mathscr{D}_A^{\mathrm{rot}}[F]$, with the vorticity $\Omega_A=\frac{1}{2}\epsilon_{Aij}\partial_iv_j$ and the rotational viscosity $\lambda\geq0$:
\begin{equation}
\mathscr{D}_A^{\mathrm{rot}}[F]
 =\frac{\delta F}{\delta s^A}-\tfrac12\epsilon_{Aij}\partial_i\frac{\delta F}{\delta\pi_j}-\frac{2\omega_A-\Omega_A}{T}\frac{\delta F}{\delta s},\qquad (F,G)_{\rm rot}=\int d^3x\,T\lambda\,
 \mathscr{D}_A^{\mathrm{rot}}[F]\mathscr{D}_A^{\mathrm{rot}}[G].
 \label{eq:generic-rot}
\end{equation}
The first two terms of $\mathscr{D}_A^{\mathrm{rot}}$ annihilate $J_B$, since $\frac{1}{2}\epsilon_{Aij}\epsilon_{Bij}=\delta_{AB}$, and the third makes the energy degenerate, $\mathscr{D}_A^{\mathrm{rot}}[H]=2\omega_A-\Omega_A-(2\omega_A-\Omega_A)=0$. With $\mathscr{D}_A^{\mathrm{rot}}[N]=\mathscr{D}_A^{\mathrm{rot}}[P_i]=0$ and $\mathscr{D}_A^{\mathrm{rot}}[\mathcal S]=-(2\omega_A-\Omega_A)/T$, the bracket conserves $H$, $N$, $P_i$ and $J_A$ and produces the entropy $(\mathcal S,\mathcal S)_{\rm rot}=\int d^3x\,\lambda(2\omega_A-\Omega_A)^2/T\geq0$ for any $\lambda\geq0$, which may depend on the state. For the densities, $(\partial_t F)_{\rm rot}=(F,\mathcal S)_{\rm rot}$ gives
\begin{equation}
 (\partial_ts^A)_{\rm rot}=-\lambda (2\omega_A-\Omega_A),\quad
 (\partial_ts)_{\rm rot}=\lambda (2\omega_A-\Omega_A)^2/T,\quad
 (\partial_t\pi_j)_{\rm rot}=-\partial_i\Pi^{\rm rot}_{ij},\quad
 \Pi^{\rm rot}_{ij}=\tfrac12\epsilon_{Aij}\lambda (2\omega_A-\Omega_A),
 \label{eq:rot-output}
\end{equation}
where the first index of $\Pi^{\rm rot}_{ij}$ is the direction of the flux. The spin potential $2\omega_A$ relaxes toward the vorticity $\Omega_A$, and the antisymmetric stress exerts the compensating torque on the orbital motion~\cite{Stark2005,Hattori2019}, $\partial_t\int d^3x\,\epsilon_{Aij}x^i\pi_j=\int d^3x\,\lambda(2\omega_A-\Omega_A)=-\partial_t\int d^3x\,s^A$. This is the relaxation $\partial_ts^A=-\lambda{(2\omega_A-\Omega_A)}$ quoted in the Letter. With $\Pi^{\rm rot}_{ij}\partial_iv_j=\lambda(2\omega_A-\Omega_A)\Omega_A$, the local rates obey the energy balance
\begin{equation}
 T(\partial_ts)_{\rm rot}+2\omega_A(\partial_ts^A)_{\rm rot}
 +v_j(\partial_t\pi_j)_{\rm rot}
 =-\partial_i(v_j\Pi^{\rm rot}_{ij}).
 \label{eq:rot-energy}
\end{equation}
For the spin energy $\varepsilon_{\rm spin}=\bm s^2/2\chi_s$ with susceptibility $\chi_s>0$ one has $2\bm\omega=\bm s/\chi_s$, and the spin density relaxes toward $\chi_s\bm\Omega$ at the rate $\Gamma_s=\lambda/\chi_s$~\cite{Hongo2021}.

The other dissipative processes of an isotropic fluid, shear and bulk viscosity, heat conduction, particle diffusion and spin diffusion, have the operators
\begin{equation}
\begin{gathered}
 \mathscr{D}_{ij}^{\mathrm{shear}}[F]
 =\tfrac12\left(\partial_i\frac{\delta F}{\delta\pi_j}+\partial_j\frac{\delta F}{\delta\pi_i}\right)-\tfrac13\delta_{ij}\partial_k\frac{\delta F}{\delta\pi_k}-\frac{\sigma_{ij}}{T}\frac{\delta F}{\delta s},\\
 \mathscr{D}^{\mathrm{bulk}}[F]
 =\partial_i\frac{\delta F}{\delta\pi_i}-\frac{\partial_iv^i}{T}\frac{\delta F}{\delta s},\qquad \mathscr{D}_i^{\mathrm{heat}}[F]=\partial_i\!\left(\frac1T\frac{\delta F}{\delta s}\right),\\
 \mathscr{D}_i^{\mathrm{diff}}[F]=\partial_i\!\left(\frac{\delta F}{\delta n}-\frac\mu T\frac{\delta F}{\delta s}\right),\qquad \mathscr{D}_{iA}^{\mathrm{spin\,diff}}[F]=\partial_i\!\left(\frac{\delta F}{\delta s^A}-\frac{2\omega_A}T\frac{\delta F}{\delta s}\right),
\end{gathered}
 \label{eq:generic-operators}
\end{equation}
where $\sigma_{ij}=\partial_{(i}v_{j)}-\frac{1}{3}\delta_{ij}\partial_kv^k$; each is degenerate on the energy. The sum of the quadratic forms is the dissipative bracket
\begin{equation}
\begin{aligned}
 (F,G)={}&(F,G)_{\rm rot}+\int d^3x\,\Bigl[
 2T\eta\,\mathscr{D}_{ij}^{\mathrm{shear}}[F]\mathscr{D}_{ij}^{\mathrm{shear}}[G]
 +T\zeta\,\mathscr{D}^{\mathrm{bulk}}[F]\mathscr{D}^{\mathrm{bulk}}[G]\\
 &\quad+\kappa T^2\,\mathscr{D}_i^{\mathrm{heat}}[F]\mathscr{D}_i^{\mathrm{heat}}[G]
 +L_n\,\mathscr{D}_i^{\mathrm{diff}}[F]\mathscr{D}_i^{\mathrm{diff}}[G]
 +L_s\,\mathscr{D}_{iA}^{\mathrm{spin\,diff}}[F]\mathscr{D}_{iA}^{\mathrm{spin\,diff}}[G]\Bigr],
\end{aligned}
 \label{eq:generic-full}
\end{equation}
with the shear viscosity $\eta$, the bulk viscosity $\zeta$, the thermal conductivity $\kappa$ and the coefficients $L_n$ and $L_s$ of particle and spin diffusion, all non-negative. Every operator in \eqref{eq:generic-operators} annihilates $H$, $N$, $P_i$ and $J_A$: symmetric velocity gradients vanish for the rigid rotation $\delta J_A/\delta\pi_j=\epsilon_{Aij}x^i$, and gradients vanish for uniform conjugates. An integration by parts in $\partial_t F=\{F,H\}+(F,\mathcal S)$ gives the dissipative stress $\Pi_{ij}$ and the diffusive fluxes $j_i^\varepsilon$, $j_i^n$ and $j_{iA}^{\rm s}$,
\begin{equation}
 \Pi_{ij}=\tfrac12\epsilon_{Aij}\lambda (2\omega_A-\Omega_A)-2\eta\sigma_{ij}-\zeta\delta_{ij}\partial_kv^k,
 \quad j_i^\varepsilon=-\kappa\partial_iT,\quad
 j_i^n=-L_n\partial_i(\mu/T),\quad
 j_{iA}^{\rm s}=-L_s\partial_i(2\omega_A/T).
 \label{eq:generic-fluxes}
\end{equation}
With the reversible part~\eqref{eq:eom-derived} and a local equation of state, the hydrodynamic equations are
\begin{equation}
\begin{gathered}
 \partial_tn+\partial_i(nv^i+j_i^n)=0,\qquad \partial_t\pi_j+\partial_i(v^i\pi_j+p\delta_{ij}+\Pi_{ij})=0,\\
 \partial_t\varepsilon+\partial_i[(\varepsilon+p)v^i+v_j\Pi_{ij}+j_i^\varepsilon]=0,\qquad \partial_ts^A+\partial_i(v^is^A+j_{iA}^{\rm s})
 =2\epsilon_{ABC}\omega_Bs^C-\lambda (2\omega_A-\Omega_A) ,
\end{gathered}
 \label{eq:generic-combined}
\end{equation}
and the entropy obeys
\begin{equation}
\begin{split}
 \partial_ts+\partial_i\left[sv^i+\frac{j_i^\varepsilon-\mu j_i^n-2\omega_Aj_{iA}^{\rm s}}{T}\right]
 ={}&\frac{\lambda (2\omega_A-\Omega_A)^2+2\eta\sigma_{ij}\sigma_{ij}+\zeta(\partial_iv^i)^2}{T}
 +\kappa\frac{|\nabla T|^2}{T^2}\\
 &+L_n|\nabla(\mu/T)|^2+L_s\sum_A|\nabla(2\omega_A/T)|^2 \geq 0.
\end{split}\label{eq:generic-entropy}
\end{equation}
For a one-component Galilean fluid, whose momentum density is the mass current, one has $L_n=0$.

In the variables of Sec.~\ref{sec:hep-comparison}, $\pi'_i=\pi_i+\frac{1}{2}\epsilon_{ijk}\partial_js^k$, the rotational operator is $\mathscr{D}_A^{\mathrm{rot}}[F]=\delta F/\delta s^A|_{\pi'}-(2\omega_A-\Omega_A)(\delta F/\delta s)/T$. Since $\pi'_i$ contains the spin, the relaxation takes place at fixed $\pi'_i$, with the same dynamics and the same entropy production.

With these isotropic brackets shear does not act on the spin.

\subsection{Classification of the Onsager matrix at finite spin density}
\label{sec:onsager}
A fluid with a finite spin density has a preferred axis, and its dissipative coefficients may depend on $\bm s$. We classify the first-order couplings among $2\omega_A-\Omega_A$, the shear tensor and the expansion rate; exactly one of them couples rotation to shear dissipatively, \LA{} of the Letter.

The thermodynamic forces are ${2\omega_A-\Omega_A}$, $\sigma_{ij}$ and $\partial_kv_k$, conjugate to the torque density, the traceless symmetric stress and the isotropic stress; the Onsager matrix $M_{ab}$ expresses the fluxes as linear functions of the forces, with $a$ and $b$ labeling the components of all three. Under time reversal $s^A\to-s^A$ and $\pi_i\to-\pi_i$, so $\omega_A$, $v_i$, $\Omega_A$ and $\sigma_{ij}$, and with them the three forces, are all odd. For forces of equal signature the Onsager--Casimir relations~\cite{Onsager1931,Casimir1945} read $M_{ab}(\bm s)=M_{ba}(-\bm s)$: the symmetric part of the matrix, which produces entropy, is even in $\bm s$, and the antisymmetric part, which produces none, is odd. Under parity the three forces and the axial vector $\bm s$ are even, so parity imposes nothing further within this sector and decouples it from the polar forces $\nabla T$ and $\nabla\mu$.

The matrix commutes with rotations about $\hat{\bm s}=\bm s/|\bm s|$, so it is block diagonal in the helicity $m$. The components are ${(2\omega-\Omega)_\parallel}={(2\bm\omega-\bm\Omega)}\cdot\hat{\bm s}$, $\sigma_\parallel=\hat{\bm s}\cdot\bm\sigma\hat{\bm s}$ and $\partial_kv_k$ for $m=0$; ${(2\bm\omega-\bm\Omega)_\perp}$ and $(\bm\sigma\hat{\bm s})_\perp$ for $|m|=1$; and the part of $\bm\sigma$ transverse to $\hat{\bm s}$ for $|m|=2$. Counting the couplings that are even and odd in $\bm s$ gives Table~\ref{tab:onsager}: eight dissipative and six reversible coefficients, each a function of $\bm s^2$ and of the thermodynamic state. The viscous part is the set of five shear viscosities, a bulk viscosity and a cross viscosity familiar from magnetized plasmas~\cite{Braginskii1965}, with $\bm s$ in the role of the magnetic field.

\begin{table}[t]
 \centering
 \begingroup\renewcommand{\arraystretch}{1.25}
 \begin{tabular}{@{}l>{\raggedright\arraybackslash}p{0.27\textwidth}>{\raggedright\arraybackslash}p{0.60\textwidth}@{}}\toprule
 $m$ & Dissipative & Reversible\\\midrule
 $0$ & $\lambda_\parallel$, $\eta_0$, $\zeta$, $\zeta_\times$ ($\sigma_\parallel$ with $\partial_kv_k$) & $(2\omega-\Omega)_\parallel$ with $\sigma_\parallel$; $(2\omega-\Omega)_\parallel$ with $\partial_kv_k$\\
 $\pm1$ & $\lambda_\perp$, $\eta_1$, $\xi$ & $(2\bm\omega-\bm\Omega)_\perp$ with $\hat{\bm s}\times(2\bm\omega-\bm\Omega)_\perp$; $(2\bm\omega-\bm\Omega)_\perp$ with $(\bm\sigma\hat{\bm s})_\perp$; odd viscosity\\
 $\pm2$ & $\eta_2$ & odd viscosity\\\bottomrule
 \end{tabular}
 \endgroup
 \caption{First-order coefficients of a fluid with spin density $\bm s$ in the sector of the forces $({2\omega_A-\Omega_A},\sigma_{ij},\partial_kv_k)$, by helicity $m$ about $\hat{\bm s}$. Dissipative coefficients are even in $\bm s$ and enter the symmetric part of the Onsager matrix; reversible ones are odd and enter the antisymmetric part.}
 \label{tab:onsager}
\end{table}

For the block that couples rotation to shear the classification can be made explicit. A tensor $N_{Aij}(\bm s)$ that is symmetric and traceless in $(ij)$ and covariant under rotations is a combination, with coefficients depending on $\bm s^2$, of
\begin{equation}
\begin{aligned}
 N^{(1)}_{Aij}&=s_i\delta_{Aj}+s_j\delta_{Ai}-\tfrac23\delta_{ij}s_A,\\
 N^{(2)}_{Aij}&=\epsilon_{iAk}s_ks_j+\epsilon_{jAk}s_ks_i,\\
 N^{(3)}_{Aij}&=s_A\bigl(s_is_j-\tfrac13\delta_{ij}\bm s^2\bigr),
\end{aligned}\label{eq:N-basis}
\end{equation}
and there are no others: the space of such tensors that are homogeneous polynomials of degree $1,2,3,4$ in $\bm s$ has dimension $1,1,2,1$. Contracted with the shear tensor, $N^{(1)}_{Aij}\sigma_{ij}=2(\bm\sigma\bm s)_A$ stretches the spin, $N^{(3)}_{Aij}\sigma_{ij}=s_A\,\bm s\cdot\bm\sigma\bm s$ changes its length, and $N^{(2)}_{Aij}\sigma_{ij}=2[\bm s\times(\bm\sigma\bm s)]_A$ rotates it. $N^{(1)}$ and $N^{(3)}$ are odd in $\bm s$ and can appear only in the reversible part; $N^{(2)}$ is even. The dissipative coupling between rotation and shear is therefore unique,
\begin{equation}
 \Xi_{Aij}=\xi(\bm s^2)\,N^{(2)}_{Aij},\qquad \Xi_{Aij}\sigma_{ij}=2\xi\,[\bm s\times(\bm\sigma\bm s)]_A,\qquad \Xi_{Aij}s_A=0 .
 \label{eq:A-supp}
\end{equation}

The reversible coefficients are not transport coefficients in a Hamiltonian theory. They are produced by the Poisson bracket and the energy, and Sec.~\ref{sec:uniqueness} shows that the bracket of the canonical spin generates neither $N^{(1)}$ nor $N^{(3)}$; with an isotropic equation of state and without gradient energies all six vanish. They reappear when the spin or the momentum is described by other variables: the vector with the FW components in the dual of a Lie-transported triad, \eqref{eq:reciprocal-vector}, acquires the stretching term $N^{(1)}$ and a rotation with the vorticity, and a thermodynamic rescaling adds a coupling to the expansion rate; the reversible terms of the spin equation of Ref.~\cite{Fang2026} have this form but act on the spin density itself (Sec.~\ref{sec:fang}). The flux of the momentum $\pi'_i$ of Sec.~\ref{sec:hep-comparison}, which contains the spin, has both odd viscosities, fixed by the spin density (Sec.~\ref{sec:fast-relaxation}). A spin--vorticity energy $\alpha\,\bm s\cdot\bm\Omega$ would contribute a precession about the vorticity.

\subsection{Cross bracket, Onsager-reciprocal stress, and entropy production}
\label{sec:cross}
With the operators of \eqref{eq:generic-rot} and~\eqref{eq:generic-operators}, define
\begin{equation}
 (F,G)_{\rm cross}=\int d^3x\,T\,\Xi_{Aij}\Bigl(\mathscr{D}^{\rm rot}_A[F]\,\mathscr{D}^{\rm shear}_{ij}[G]+\mathscr{D}^{\rm shear}_{ij}[F]\,\mathscr{D}^{\rm rot}_A[G]\Bigr).
 \label{eq:cross-bracket}
\end{equation}
It is symmetric, and since both operators annihilate $H$, $N$, $P_i$ and $J_A$ it conserves them. Using $\mathscr{D}^{\rm rot}_A[\mathcal S]=-{(2\omega_A-\Omega_A)}/T$ and $\mathscr{D}^{\rm shear}_{ij}[\mathcal S]=-\sigma_{ij}/T$, the contribution to $\partial_tF$ is $-\int \Xi_{Aij}(\sigma_{ij}\mathscr{D}^{\rm rot}_A[F]+{(2\omega_A-\Omega_A)}\mathscr{D}^{\rm shear}_{ij}[F])$, which gives
\begin{equation}
\begin{gathered}
 (\partial_ts^A)_{\rm cross}=-\Xi_{Aij}\sigma_{ij}=-2\xi\,[\bm s\times(\bm\sigma\bm s)]_A,\qquad
 T(\partial_ts)_{\rm cross}=2(2\omega_A-\Omega_A)\Xi_{Aij}\sigma_{ij},\\
 (\partial_t\pi_j)_{\rm cross}=-\partial_i\Pi^{\rm cross}_{ij},\qquad
 \Pi^{\rm cross}_{ij}=-\Xi_{Aij}(2\omega_A-\Omega_A)+\tfrac12\epsilon_{Aij}\Xi_{Akl}\sigma_{kl},\\
 \Xi_{Aij}(2\omega_A-\Omega_A)=\xi\bigl\{[(2\bm\omega-\bm\Omega)\times\bm s]_is_j+[(2\bm\omega-\bm\Omega)\times\bm s]_js_i\bigr\}.
\end{gathered}\label{eq:cross-output}
\end{equation}
The symmetric part of $\Pi^{\rm cross}$ is the stress reciprocal to the torque. The antisymmetric part arises from the momentum derivative inside $\mathscr{D}^{\rm rot}$; it satisfies $\epsilon_{Bij}\Pi^{\rm cross}_{ij}=\Xi_{Bkl}\sigma_{kl}=-(\partial_ts^B)_{\rm cross}$, so that spin and orbital angular momentum are exchanged locally. The energy balance of \eqref{eq:rot-energy} holds with $\Pi^{\rm rot}\to\Pi^{\rm cross}$. The torque is perpendicular to $\bm s$: at fixed $\bm\sigma$ it conserves $\bm s^2$ and $\bm s\cdot\bm\sigma\bm s$. It is of second order in $\bm s$, does not create spin in a fluid without spin density, and is distinct from the polarization induced by shear in the momentum distribution~\cite{Fu2021,Becattini2021}, which is of first order in $\bm\sigma$ at $\bm s=0$ and has no net spin. For an isotropic equation of state and $\bm\Omega=0$ the cross term in the entropy production vanishes ($\Xi_{Aij}s_A=0$) although the torque does not.

Together with the rotational and shear parts of the dissipative bracket the entropy production is
\begin{equation}
\begin{aligned}
 T(\partial_ts)_{\rm rot+shear+cross}&=\lambda(2\bm\omega-\bm\Omega)^2+2\eta\,\sigma_{ij}\sigma_{ij}+2\Xi_{Aij}(2\omega_A-\Omega_A)\sigma_{ij}\\
 &=\lambda\Bigl|2\bm\omega-\bm\Omega+\frac{\bm\Xi{:}\bm\sigma}{\lambda}\Bigr|^2+\sigma_{ij}\Bigl(2\eta\,\delta_{ik}\delta_{jl}-\frac{\Xi_{Aij}\Xi_{Akl}}{\lambda}\Bigr)\sigma_{kl},
\end{aligned}
 \label{eq:cross-entropy}
\end{equation}
with $(\bm\Xi{:}\bm\sigma)_A=\Xi_{Aij}\sigma_{ij}$ for the tensor $\bm\Xi$ with components $\Xi_{Aij}$. Since $\Xi_{Aij}\Xi_{Bij}=2\xi^2\bm s^2(\bm s^2\delta_{AB}-s_As_B)$ has eigenvalues $0$ and $2\xi^2(\bm s^2)^2$ (twice), the quadratic form is non-negative if and only if
\begin{equation}
 \lambda\ge0,\qquad\eta\ge0,\qquad \bigl[\xi(\bm s^2)\,\bm s^2\bigr]^2\le\lambda\eta ;
 \label{eq:psd-bound}
\end{equation}
with anisotropic coefficients the last condition reads $\xi^2(\bm s^2)^2\le\lambda_\perp\eta_1$. For constant $\xi$ the bound restricts the states to $(\bm s^2)^2\le\lambda\eta/\xi^2$; a function with $\sup \xi(\bm s^2)\bm s^2\le\sqrt{\lambda\eta}$ satisfies the bound everywhere. The second form of \eqref{eq:cross-entropy} shows in what sense $\xi$ is independent: it can be exchanged for an anisotropic, spin-dependent shear viscosity together with a shifted relaxation target, ${2\bm\omega-\bm\Omega}\to-\bm\Xi{:}\bm\sigma/\lambda$, but no redefinition of the fields removes it, because redefinitions act on the Onsager matrix by congruence with the identity at leading order.

\subsection{Green--Kubo formula}
\label{sec:kubo}
The Onsager matrix is the time integral of the correlations of the fluxes conjugate to $({2\omega_A-\Omega_A},\sigma_{ij})$, evaluated in a state with spin density $\bm s$ maintained by a conjugate source. With the torque density $\tau_A=\epsilon_{Aij}\widehat\Theta^{[ij]}$, which obeys $\partial_t\widehat s^A+\partial_i\widehat\jmath^{\,\rm s}_{iA}=-\tau_A$, and the traceless symmetric stress $\widehat\pi_{ij}$,
\begin{equation}
 T\,\Xi_{Aij}(\bm s)=\lim_{\omega\to0}\int_0^\infty dt\,e^{i\omega t}\int d^3x\,
 \bigl(\tau_A(t,\bm x);\widehat\pi_{ij}(0,\bm 0)\bigr)_{\bm s},
 \label{eq:kubo}
\end{equation}
where $(\,;\,)_{\bm s}$ is the Kubo canonical correlation in the ensemble with spin density $\bm s$ and the symmetric part under the exchange of the two fluxes is understood. In an expansion of the right-hand side in powers of $\bm s$, rotational covariance removes the term of zeroth order and time reversal the term of first order. The coefficient $\xi(0)$ is fixed by the term of second order, which is a combination of three- and four-point functions of the fluid at zero spin density. Three properties of $\xi$ follow. The sign of $\xi$ is not fixed; only the bound~\eqref{eq:psd-bound} is. If the Onsager matrix is analytic in $\bm s$ on the scale of the saturation density $s_{\rm m}$, then $|\xi|\lesssim\sqrt{\lambda\eta}/s_{\rm m}^2$; since $\tau_A$ is of first order in the interaction that exchanges spin and orbital angular momentum and $\widehat\pi_{ij}$ of zeroth order, $\xi$ is of first order in that interaction, as is $\sqrt{\lambda\eta}$. Finally, $\tau_A$ depends on the spin operator; $\xi$, like $\lambda$~\cite{Hongo2021}, is defined for the canonical spin density (Sec.~\ref{sec:pseudogauge}), and the source that maintains $\bm s$ contributes contact terms that must be separated.

\subsection{Relation to earlier work on cross effects}
\label{sec:cross-effects}
Reference~\cite{Hu2023} classifies cross effects in spin hydrodynamics about the state of zero spin density, where the coupling between shear and spin relaxation vanishes, consistent with $\Xi_{Aij}=O(\bm s^2)$. In spin magnetohydrodynamics~\cite{FangHattoriHu2025} a cross rotational viscosity $\xi_\parallel$ appears with the bound $\gamma_\parallel\eta_\parallel\ge\xi_\parallel^2$, which has the form of~\eqref{eq:psd-bound} with $\xi\bm s^2$ and $\lambda$ in the roles of $\xi_\parallel$ and $\gamma_\parallel$. There the anisotropy is that of the magnetic field at zero background spin, and all linear modes are damped. The tensor $N^{(2)}$ coincides with the cross-viscosity term of polar active fluids, Eq.~(18) of Ref.~\cite{Markovich2019}, with the polarization in place of $\bm s$, including both stresses of \eqref{eq:cross-output}; there the polarization is a vector order parameter distinct from the angular-momentum density, the coupling is linear in the dynamical angular momentum, and eliminating the angular momentum renormalizes the viscosity. In polar liquid crystals a reversible coupling that changes the magnitude of the polarization under shear produces a shear-induced first-order transition between the isotropic and the polar state~\cite{MarkovichPRL2019}; that polarization is even under time reversal, and for the canonical spin density a coupling of this kind is excluded by the bracket (Sec.~\ref{sec:uniqueness}). What is specific to the present case is that the axis of anisotropy is the angular-momentum density on which the torque acts. The coupling is then quadratic in the dynamical variable, and the torque, $\bm s\times(-2\xi\bm\sigma\bm s)$, has the algebraic form of the Euler equations of a rigid body~\cite{LandauMechanics}, with the traceless matrix $-2\xi\bm\sigma$ in place of the positive inverse inertia tensor. This quadratic torque and the bifurcation it produces (Sec.~\ref{sec:bifurcation}) therefore do not arise in the works cited.

\subsection{Power counting}
\label{sec:counting}
In the counting of Ref.~\cite{Fang2026} the torque is of first order in gradients, as the Letter states; here we show this and identify the regime in which the shear-driven spin reorientation of Sec.~\ref{sec:bifurcation} lies within a first-order theory. Gradients and spin amplitude are independent small parameters. Let $\delta$ measure gradients and rates in units of the microscopic relaxation rate, and $\mathsf P=|\bm s|/s_{\rm m}$ the spin density in units of its saturation value; write $\hat\chi_s$, $\hat\lambda$, $\hat \xi$ for $\chi_s$, $\lambda$, $\xi\,s_{\rm m}^2$ in the corresponding units. Table~\ref{tab:counting} lists the terms of the spin equation.

\begin{table}[t]
 \centering
 \begingroup\renewcommand{\arraystretch}{1.25}
 \begin{tabular}{@{}lll@{}}\toprule
 Term & Expression & Order\\\midrule
 advection & $\nabla\cdot(\bm v\bm s)$ & $\mathsf P\,\delta$\\
 Thomas precession & $\bm\Omega_{\rm T}\times\bm s$, $\bm\Omega_{\rm T}=\bm v\times\nabla p/[(\varepsilon+p)(\gamma+1)]$ & $\mathsf P\,\delta$; zero for an unaccelerated element\\
 second-order Poincar\'e terms & $(\delta H_2/\delta\bm s)\times\bm s$ & $\mathsf P^2\delta^2\,[s_{\rm m}/(\varepsilon+p)]$\\
 relaxation & $-\lambda(\bm s/\chi_s-\bm\Omega)$ & $\hat\lambda\,\mathsf P/\hat\chi_s$ and $\hat\lambda\,\delta$\\
 shear torque & $-2\xi\,\bm s\times(\bm\sigma\bm s)$ & $\hat \xi\,\mathsf P^2\delta$\\
 spin diffusion & $D_s\nabla^2\bm s$ & $\mathsf P\,\delta^2$\\\bottomrule
 \end{tabular}
 \endgroup
 \caption{Orders of the terms in $\partial_t\bm s$, in units of $s_{\rm m}$ times the microscopic rate. $\delta$: gradient; $\mathsf P$: spin density in units of $s_{\rm m}$; $D_s=L_s/T\chi_s$: spin diffusivity.}
 \label{tab:counting}
\end{table}

The power counting distinguishes three regimes. (A) \emph{Fast-relaxing spin mode}: $\hat\chi_s,\hat\lambda=O(1)$, so the spin relaxes at the microscopic rate to $\mathsf P=\hat\chi_s\delta$; the torque is $O(\delta^3)$ and plays no role, and the theory reduces to the conventional hydrodynamics of Sec.~\ref{sec:fast-relaxation}. (B) \emph{Finite spin with small spin potential}: $\mathsf P=O(1)$ and $\omega_A=s_A/2\chi_s=O(\delta)$, the counting of Ref.~\cite{Fang2026}. It requires $\hat\chi_s^{-1}=O(\delta)$, a large susceptibility, and it is thermodynamically consistent: the spin energy $\bm s^2/2\chi_s$ is $O(\delta)$, and the equilibrium condition $2\bm\omega=\bm\Omega$ is satisfied by a finite spin density at small vorticity. With $\hat\lambda,\hat \xi=O(1)$ the relaxation rate $\Gamma_s$ is $O(\delta)$, and relaxation, torque and the stresses~\eqref{eq:cross-output} are all of first order. With the shear rate $\sigma=(\tfrac12\sigma_{ij}\sigma_{ij})^{1/2}$ and its critical value $\sigma_c=\Gamma_s/\sqrt{2\xi\lambda}$ for the reorientation, the ratio $\sigma^2/\sigma_c^2=2\xi\chi_s^2\sigma^2/\lambda$ is then $O(1)$, and the bifurcation lies within a first-order theory. Because $\mathsf P$ is not small, no expansion in the spin is available: the functions $\chi_s(\bm s^2)$, $\lambda_{\perp,\parallel}(\bm s^2)$, $\xi(\bm s^2)$ and the spin dependence of the viscosities must be retained. (C) \emph{Natural susceptibility with slow relaxation}: $\hat\chi_s=O(1)$, $\mathsf P=\hat\chi_s\delta$, and $\sigma\sim\sigma_c$ would require $\Gamma_s=O(\delta^2)$; relaxation and torque are then both $O(\delta^3)$ and compete with second-order transport terms in the spin equation that have not been classified. The bifurcation is parametrically controlled in regime (B), or when $\hat \xi$ is anomalously large. At threshold $\Gamma_s/\sigma_c=\sqrt{2\xi\lambda}$, which is of order one in regime (B) and of order $\delta$ in regime (C); in both, $\sigma_c$ lies far below the microscopic rate.

\section{Covariant First-Order Constitutive Relations}
\label{sec:rel-diss}

At first order in gradients the equal-time brackets of Sec.~\ref{sec:generic} correspond to the covariant constitutive relations~\Lcross{} of the Letter, which we check against the Ward identity of angular momentum and the second law. In Minkowski spacetime, consider a rest-frame energy density $\varepsilon(n,s,-s\cdot s)$ that is isotropic in the spin, with $T>0$, $\beta=1/T$, and the local relations, which hold at global equilibrium and are assumed here away from it (see Ref.~\cite{Becattini2024} for their status),
\begin{equation}
 d\varepsilon=Tds+\mu dn-2\omega_{\rho} ds^\rho,
 \qquad \varepsilon+p=Ts+\mu n-2\omega_{\rho} s^\rho .
 \label{eq:rel-first-law}
\end{equation}
Variations $ds^\mu$ at fixed $u^\mu$ are transverse, $u\cdot ds=0$, and isotropy makes $\omega^\mu$ parallel to $s^\mu$. The precession about the spin potential therefore vanishes, and $\mathcal D_{\rm FW}s^\mu+\theta s^\mu$ is determined by the dissipative exchange with the orbital motion and by spin diffusion.

\subsection{Ward identity of angular momentum}
\label{sec:ward}
The spin tensor $S^{\mu\nu}$ of \eqref{eq:dual} is the intrinsic part of the conserved angular momentum. This constrains the spin current $\Sigma^{\lambda\mu\nu}$ and the energy--momentum tensor $\Theta^{\mu\nu}$ beyond the three components of the spin equation~\cite{Hattori2019,Gallegos2021}. With $\Sigma^{\lambda\mu\nu}=u^\lambda S^{\mu\nu}+\Sigma_\perp^{\lambda\mu\nu}$ and $u_\lambda\Sigma_\perp^{\lambda\mu\nu}=0$, the conservation law $\nabla_\lambda J^{\lambda\mu\nu}=0$ requires
\begin{equation}
 \nabla_\lambda\Sigma^{\lambda\mu\nu}=-2\Theta^{[\mu\nu]},\qquad
 -S^{\mu\nu}a_\nu+u_\nu\nabla_\lambda\Sigma_\perp^{\lambda\mu\nu}
 =-2u_\nu \Theta^{[\mu\nu]}.
 \label{eq:ward-longitudinal}
\end{equation}
The second relation is the contraction of the first with $u_\nu$. A convective spin current, $\Sigma_\perp^{\lambda\mu\nu}=0$, together with a symmetric energy--momentum tensor would require $S^{\mu\nu}a_\nu=0$, which an accelerated fluid with spin density does not satisfy. Either a spin current $\Sigma_\perp^{\lambda\mu\nu}$ or an antisymmetric component $u_\nu\Theta^{[\mu\nu]}$, such as an energy flux $q_\varepsilon^\mu$, is therefore needed; both are constructed below.

The spatial dual of the first relation in \eqref{eq:ward-longitudinal} is the spin equation,
\begin{equation}
 \Delta^\mu{}_\nu\dot s^\nu+\theta s^\mu
 =\varepsilon^{\mu\nu\rho\sigma}u_\nu
 (\Theta_{[\rho\sigma]}+\tfrac12\nabla_\lambda\Sigma_\perp^\lambda{}_{\rho\sigma}) ,
 \label{eq:ward-transverse}
\end{equation}
and the second relation constrains the energy flux $q_\varepsilon^\mu=\Delta^\mu{}_\alpha\Theta^{\alpha\beta}u_\beta$.

\subsection{Constitutive relations and the spin equation}
\label{sec:rel-constitutive}
We take the spin current $\Sigma_\perp^{\lambda\mu\nu}$ to be the dual of a diffusive flux $j_{\rm s}^{\lambda\mu}$ that is transverse to $u^\mu$ in both indices, and fix the four-velocity by $u_\mu\Theta^{\mu\nu}=\varepsilon u^\nu$:
\begin{equation}
\Sigma_\perp^{\lambda\mu\nu}=-\varepsilon^{\mu\nu\rho\sigma}u_\rho j_{{\rm s}\sigma}^{\lambda},
 \qquad u_\lambda j_{\rm s}^{\lambda\mu}=u_\mu j_{\rm s}^{\lambda\mu}=0,\quad u_\mu\Theta^{\mu\nu}=\varepsilon u^\nu,\qquad
 q_\varepsilon^\mu=S^{\mu\nu}a_\nu-u_\nu\nabla_\lambda\Sigma_\perp^{\lambda\mu\nu} .
 \label{eq:rel-current-choice}
\end{equation}
The expression for $q_\varepsilon^\mu$ follows from the second relation in \eqref{eq:ward-longitudinal} with $\Theta^{\alpha\beta}u_\beta=\varepsilon u^\alpha+q_\varepsilon^\alpha$ and $u_\beta u_\alpha\Theta^{\alpha\beta}=\varepsilon$; there is no momentum density in the rest frame, $\Delta^\mu{}_\alpha u_\beta\Theta^{\beta\alpha}=0$. Since $u_\nu\Sigma_\perp^{\lambda\mu\nu}=0$, we have $-u_\nu\nabla_\lambda\Sigma_\perp^{\lambda\mu\nu}=\Sigma_\perp^{\lambda\mu\nu}\nabla_\lambda u_\nu$, which is of second order in gradients for $j_{\rm s}^{\lambda\mu}=O(\partial)$. We keep this term in $q_\varepsilon^\mu$, so that the longitudinal balance holds identically. With the particle current $N^\mu$ and the diffusion current $j_n^\mu$, the first-order constitutive relations are
\begin{equation}
\begin{gathered}
 \Theta^{\mu\nu}=\varepsilon u^\mu u^\nu-p\Delta^{\mu\nu}
 +2\eta\sigma^{\mu\nu}+\zeta\theta\Delta^{\mu\nu}+q_\varepsilon^\mu u^\nu
 -\frac{\lambda}{2}\varepsilon^{\mu\nu\rho\sigma}u_\rho(2\omega_{\sigma}-\Omega_\sigma),\\
 N^\mu=nu^\mu+j_n^\mu,\qquad
 j_n^\rho=L_n\Delta^{\rho\kappa}\nabla_\kappa(\beta\mu)+\frac{n}{\varepsilon+p}q_\varepsilon^\rho,\qquad j_{\rm s}^{\lambda\mu}=L_s\Delta^{\lambda\kappa}\Delta^{\mu\nu}
 \nabla_\kappa(2\beta\omega_{\nu})+\frac{s^\mu}{\varepsilon+p}q_\varepsilon^\lambda ,
\end{gathered}
 \label{eq:rel-constitutive}
\end{equation}
with $\eta,\zeta,\lambda,L_n,L_s\geq0$. The rotational relaxation is counted as of first order, either through $\lambda=O(\partial)$ or through $2\omega_\mu-\Omega_\mu=O(\partial)$~\cite{Hongo2021}.

These relations satisfy all projections of the Ward identity:
\begin{equation}
\varepsilon^{\mu\nu\rho\sigma}u_\nu\Theta_{[\rho\sigma]}=-\lambda(2\omega^\mu-\Omega^\mu),\quad \tfrac12\varepsilon^{\mu\nu\rho\sigma}u_\nu\nabla_\lambda\Sigma_\perp^\lambda{}_{\rho\sigma}
 =-\Delta^\mu{}_\nu\nabla_\lambda j_{\rm s}^{\lambda\nu},\quad 2u_\nu\Theta^{[\mu\nu]}=S^{\mu\nu}a_\nu-u_\nu\nabla_\lambda\Sigma_\perp^{\lambda\mu\nu} .
 \label{eq:rel-ward-check}
\end{equation}
In \eqref{eq:ward-transverse} they give the spin equation
\begin{equation}
 \mathcal D_{\rm FW}s^\mu+\theta s^\mu
 =-\lambda(2\omega^\mu-\Omega^\mu)-\Delta^\mu{}_\nu\nabla_\lambda j_{\rm s}^{\lambda\nu}.
 \label{eq:rel-spin-eom}
\end{equation}
This is the covariant form of the spin equation in \eqref{eq:generic-combined} for an isotropic equation of state.

\subsection{Entropy production}
\label{sec:rel-entropy}
For the second law we evaluate the divergence of the entropy current $s_{\rm ent}^\rho=su^\rho+\beta q_\varepsilon^\rho-\beta\mu j_n^\rho+2\beta\omega_{\nu} j_{\rm s}^{\rho\nu}$. With $\nabla_\rho(su^\rho)=\beta[\nabla_\rho(\Theta^{\rho\nu}u_\nu)-\mu\nabla_\rho N^\rho+2\omega_\nu\nabla_\lambda j_{\rm s}^{\lambda\nu}-2\omega_\nu\lambda(2\omega^\nu-\Omega^\nu)-(p+\varepsilon-Ts-\mu n+2\omega_\nu s^\nu)\theta]$, the terms proportional to $q_\varepsilon^\mu/(\varepsilon+p)$ in $j_n^\mu$ and $j_{\rm s}^{\mu\nu}$ combine with $\nabla_\mu(\beta q_\varepsilon^\mu)$, by the Gibbs--Duhem relation, into
\begin{equation}
 q_\varepsilon^\rho\left[\nabla_\rho\beta-\frac{n}{\varepsilon+p}\nabla_\rho(\beta\mu)
 +\frac{s^\nu}{\varepsilon+p}\nabla_\rho(2\beta\omega_{\nu})\right]
 =-\frac{\beta}{\varepsilon+p}q_\varepsilon^\rho\nabla_\rho p=O(\partial^3).
 \label{eq:rel-drift-cancellation}
\end{equation}
The estimate holds because the ideal momentum balance gives $\Delta^{\mu\nu}\nabla_\nu p=(\varepsilon+p)a^\mu+O(\partial^2)$, $q_\varepsilon^\mu=S^{\mu\nu}a_\nu+O(\partial^2)$, and $a_\mu S^{\mu\nu}a_\nu=0$ by antisymmetry. Through second order in gradients the entropy production is therefore
\begin{equation}
\begin{split}
 \nabla_\rho s_{\rm ent}^\rho={}&\beta[2\eta\sigma_{\rho\nu}\sigma^{\rho\nu}
 +\zeta\theta^2-\lambda(2\omega_{\rho}-\Omega_\rho)(2\omega^\rho-\Omega^\rho)]\\
 &-L_n\Delta^{\rho\kappa}\nabla_\rho(\beta\mu)\nabla_\kappa(\beta\mu)
 +L_s\Delta^{\lambda\kappa}\Delta^{\rho\nu}
 \nabla_\lambda(2\beta\omega_{\rho})\nabla_\kappa(2\beta\omega_{\nu})+O(\partial^3)\geq 0,
\end{split}\label{eq:rel-entropy}
\end{equation}
since in the signature $(+,-,-,-)$ a contraction of two transverse vectors with $\Delta^{\mu\nu}$, as in the $\lambda$ and $L_n$ terms, is negative semidefinite, and a contraction over two pairs of transverse indices, as in the $\eta$ and $L_s$ terms, is positive semidefinite.

\subsection{Covariant form of the rotation--shear coupling}
\label{sec:rel-cross}
The coupling of Sec.~\ref{sec:onsager} is added to \eqref{eq:rel-constitutive} and~\eqref{eq:rel-spin-eom} as follows. Define the transverse vectors
\begin{equation}
 Z^\mu=-2\xi\,\varepsilon^{\mu\nu\rho\sigma}u_\nu s_\rho\,\sigma_{\sigma\lambda}s^\lambda,\qquad
 Y^\mu=\varepsilon^{\mu\nu\rho\sigma}u_\nu(2\omega_\rho-\Omega_\rho)s_\sigma ,
\end{equation}
whose components in the local rest frame are $2\xi[\bm s\times(\bm\sigma\bm s)]$ and $-{(2\bm\omega-\bm\Omega)}\times\bm s$ in terms of the three-dimensional quantities of Sec.~\ref{sec:cross} (the components $\sigma^{ij}$ are minus those of the three-dimensional shear tensor, and lower spatial components change sign). Then
\begin{equation}
 \Theta^{\mu\nu}_{\rm cross}=-\tfrac12\varepsilon^{\mu\nu\rho\sigma}u_\rho Z_\sigma+\xi\bigl(Y^\mu s^\nu+s^\mu Y^\nu\bigr),\qquad
 \Delta^\mu{}_\nu\dot s^\nu+\theta s^\mu=-\lambda{(2\omega^\mu-\Omega^\mu)}-Z^\mu-\Delta^\mu{}_\nu\nabla_\lambda j_{\rm s}^{\lambda\nu},
\end{equation}
the Ward identity $\nabla_\lambda\Sigma^{\lambda\mu\nu}=-2\Theta^{[\mu\nu]}$ is preserved, and the entropy production of \eqref{eq:rel-entropy} acquires the term $-2\beta{(2\omega_\mu-\Omega_\mu)}Z^\mu$. The sum of $2\beta\eta\,\sigma_{\mu\nu}\sigma^{\mu\nu}$, $-\beta\lambda{(2\omega_\mu-\Omega_\mu)(2\omega^\mu-\Omega^\mu)}$ and this term is non-negative under the condition~\eqref{eq:psd-bound} with $\bm s^2\to-s_\mu s^\mu$. In the notation of \eqref{eq:vector-law}, the dissipative torque is $\mathcal T^\mu_{\rm diss}=-\lambda{(2\omega^\mu-\Omega^\mu)}-Z^\mu-\Delta^\mu{}_\nu\nabla_\lambda j_{\rm s}^{\lambda\nu}$.

At first order, where $q_\varepsilon^\mu=S^{\mu\nu}a_\nu$, the relations~\eqref{eq:rel-constitutive} with $\Theta^{\mu\nu}_{\rm cross}$ added are \Lcross{} of the Letter.

\subsection{Covariant circulation law}
\label{sec:kelvin-cov}
The law \eqref{eq:kelvin-law} is written in the canonical variables of one frame. It holds as it stands for the relativistic energy of Sec.~\ref{sec:poincare}, but $\hat{\bm s}$ is then the direction of the canonical spin, which a change of frame subjects to a Wigner rotation that depends on position, and the Thomas term of the energy is part of the spin potential. In covariant variables the corresponding phase appears explicitly. We use the constitutive relations \eqref{eq:rel-constitutive} without dissipation, with an equation of state that is isotropic in the spin, $d\varepsilon=T\,ds+\mu\,dn+|2\omega|\,d|\bm s|$, where $|\bm s|=(-s_\rho s^\rho)^{1/2}$ and $|2\omega|=\partial\varepsilon/\partial|\bm s|$ is the magnitude of the spin potential, which is parallel to the spin:
\begin{equation}
 \Theta^{\mu\nu}=\varepsilon u^\mu u^\nu-p\Delta^{\mu\nu}+S^{\mu\rho}a_\rho u^\nu,\qquad
 N^\mu=n\Bigl(u^\mu+\frac{S^{\mu\nu}a_\nu}{\varepsilon+p}\Bigr),\qquad
 \Sigma^{\lambda\mu\nu}=\frac{N^\lambda}{n}\,S^{\mu\nu} .
 \label{eq:kelvin-ideal}
\end{equation}
The particle current is not parallel to $u^\mu$: with the velocity fixed by \eqref{eq:rel-current-choice}, an accelerated element with spin has the energy flux $S^{\mu\nu}a_\nu$, its hidden momentum, and the particles drift relative to $u^\mu$ at $S^{\mu\nu}a_\nu/(\varepsilon+p)$. Without dissipation the entropy current of Sec.~\ref{sec:rel-entropy} is $sN^\mu/n$, and the spin tensor is advected by the same current: particles, entropy and spin move together. Consequently $s/n$ and $|\bm s|/n$ are constant along $N^\mu$ and, by \eqref{eq:rel-spin-eom}, the direction of the spin is FW transported, all to the order of the constitutive relations. With the unit vector $\hat s^\mu=s^\mu/|\bm s|$ define
\begin{equation}
 \mathcal B_{\mu\nu}=-\frac{|\bm s|}{n}\,\varepsilon_{\alpha\beta\rho\sigma}\,u^\alpha\hat s^\beta
 \bigl(\partial_\mu\hat s^\rho\,\partial_\nu\hat s^\sigma-\partial_\mu u^\rho\,\partial_\nu u^\sigma\bigr),\qquad
 W_{\mu\nu}=\partial_\mu\frac{(\varepsilon+p)u_\nu}{n}-\partial_\nu\frac{(\varepsilon+p)u_\mu}{n}+\mathcal B_{\mu\nu} .
 \label{eq:kelvin-cov-berry}
\end{equation}
In the rest frame of a uniform flow, $\mathcal B_{ij}=\epsilon_{ijk}\mathcal B_k$ with the curvature \eqref{eq:kelvin-berry}. To the order of the constitutive relations,
\begin{equation}
 N^\mu W_{\mu\nu}=-n\Bigl[T\,\partial_\nu\frac sn+|2\omega|\,\partial_\nu\frac{|\bm s|}{n}\Bigr] .
 \label{eq:kelvin-cov}
\end{equation}
For uniform $s/n$ and $|\bm s|/n$ the two-form $W_{\mu\nu}$ is closed and annihilated by the particle current. For every loop $C$ that moves with the particles and every surface $S$ that spans it,
\begin{equation}
 \oint_C\frac{\varepsilon+p}{n}\,u_\mu\,dx^\mu+\frac12\int_S\mathcal B_{\mu\nu}\,dx^\mu\wedge dx^\nu
 \label{eq:kelvin-cov-loop}
\end{equation}
is then constant. Without spin, \eqref{eq:kelvin-cov} is the relativistic form of Kelvin's theorem, $u^\mu\{\partial_\mu[{(\varepsilon+p)}u_\nu/n]-\partial_\nu[{(\varepsilon+p)}u_\mu/n]\}=-T\partial_\nu(s/n)$. In the nonrelativistic limit $[(\varepsilon+p)/n]\,u_i\to-mv_i$ for a particle of mass $m$, and \eqref{eq:kelvin-cov-loop} is minus the conserved quantity of \eqref{eq:kelvin-law}.

The derivation has two steps. (i) For arbitrary fields and any equation of state of the above form,
\begin{equation}
\begin{aligned}
 N^\mu\Bigl[\partial_\mu\frac{(\varepsilon+p)u_\nu}{n}-\partial_\nu\frac{(\varepsilon+p)u_\mu}{n}\Bigr]
 ={}&E_\nu-nT\,\partial_\nu\frac sn-n|2\omega|\,\partial_\nu\frac{|\bm s|}{n}-S^{\mu\rho}a_\rho\,\partial_\nu u_\mu\\
 &+u_\nu\Bigl[T\,N^\mu\partial_\mu\frac sn+|2\omega|\,N^\mu\partial_\mu\frac{|\bm s|}{n}+\frac{S^{\mu\rho}a_\rho\,\partial_\mu p}{\varepsilon+p}\Bigr],
\end{aligned}
 \label{eq:kelvin-cov-identity}
\end{equation}
where $E_\nu=(\varepsilon+p)a_\nu-\Delta_\nu{}^\lambda\partial_\lambda p+S^{\mu\rho}a_\rho\,\partial_\mu u_\nu$ vanishes by the momentum balance $\nabla_\mu\Theta^{\mu\nu}=0$ of \eqref{eq:kelvin-ideal}. The bracket in the second line vanishes to the order considered: $s/n$ and $|\bm s|/n$ are constant along $N^\mu$, and $S^{\mu\rho}a_\rho\partial_\mu p=(\varepsilon+p)\,S^{\mu\rho}a_\rho a_\mu+O(\partial^3)=O(\partial^3)$ by antisymmetry, as in \eqref{eq:rel-drift-cancellation}. (ii) For a spin direction that is FW transported, $u^\lambda\partial_\lambda\hat s^\rho$ is parallel to $u^\rho$, so that only the velocity term of $\mathcal B_{\mu\nu}$ survives the contraction with $u^\mu$: $n\,u^\mu\mathcal B_{\mu\nu}=S^{\mu\rho}a_\rho\,\partial_\nu u_\mu$. The remainder $(N^\mu-nu^\mu)\mathcal B_{\mu\nu}$ and the correction from the drift of the spin relative to $u^\mu$ are of third order in gradients. The last term of the first line of \eqref{eq:kelvin-cov-identity} thus cancels against $N^\mu\mathcal B_{\mu\nu}$, which gives \eqref{eq:kelvin-cov}.

The second term of $\mathcal B_{\mu\nu}$ contains no derivative of the spin. It is the curvature of the hyperboloid of four-velocities, projected on the spin: the geometric origin of the Thomas precession and of the Wigner rotation of Sec.~\ref{sec:poincare}. Contracted with $u^\mu$, as in step (ii), it is the force that the hidden momentum exerts; its spatial part is a phase that the spins acquire from the variation of the velocity alone. For a fluid in rigid rotation with angular velocity $\Omega_0$ and spins along the axis, the flux of $\mathcal B_{\mu\nu}$ through a disk of radius $r$ normal to the axis is $-2\pi(|\bm s|/n)[\gamma(r)-1]$ with $\gamma(r)=(1-\Omega_0^2r^2)^{-1/2}$, that is, $|\bm s|/n$ times the angle of the Thomas precession per revolution. In the canonical variables of one frame this term is moved from $\mathcal B_{\mu\nu}$ into the momentum per particle, which is why \eqref{eq:kelvin-law} keeps its form there but is not covariant.

Equation~\eqref{eq:kelvin-cov} was obtained to the order of the constitutive relations of Sec.~\ref{sec:rel-constitutive} and for an isotropic equation of state. An anisotropic one adds a precession about the spin potential and the corresponding force, which combine as in \eqref{eq:kelvin-law}; the covariant form of that case is not derived here.

Two earlier results concern relativistic fluids. For a relativistic fluid whose spin has a constant magnitude along each flow line, Ref.~\cite{Maugin1976} derived from an action principle a Helmholtz theorem: the curl of a four-vector that contains, besides the enthalpy current per particle, the spin tensor contracted with the acceleration is transported by the flow where the entropy per particle is uniform. For a relativistic superfluid with spin at zero temperature, Ref.~\cite{Son2026} constructed the closed two-form $\mathcal B_{\mu\nu}$, including its velocity term, and the relativistic Mermin--Ho relation, which corresponds to $W_{\mu\nu}=0$. Equation~\eqref{eq:kelvin-cov} is the statement for a normal fluid, in which $|\bm s|/n$ is not quantized and $W_{\mu\nu}$ is conserved along the particle current instead of vanishing.

\section{Fast Spin Relaxation and Odd Viscosity}
\label{sec:fast-relaxation}

When the relaxation rate exceeds the rates of the flow, regime (A) of Sec.~\ref{sec:counting}, the spin is no longer an independent variable, and the limit is conventional hydrodynamics for the momentum density $\pi'_i$ of Sec.~\ref{sec:hep-comparison}~\cite{Li2021,Hongo2021}, in which the spin term of the bracket of $\pi'_i$ with itself appears as the odd viscosity~\Lodd{} of the Letter. The limit is a constraint enforced by friction, which no reversible dynamics could enforce; it therefore differs from a fluid with the same spin density inserted into an action, by exactly the odd viscosity.

\subsection{The limit and the torque of constraint}
\label{sec:fr-limit}
Consider a one-component Galilean fluid of particles of mass $m$, $\bm\pi=mn\bm v$, with the energy $H=\int d^3x\,[\bm\pi^2/2mn+\varepsilon_0(n,s)+\bm s^2/2\chi_s]$, so that $2\bm\omega=\bm s/\chi_s$; the susceptibility $\chi_s>0$ may depend on $n$ and $s$. We keep the rotational part $(F,G)_{\rm rot}$ of \eqref{eq:generic-full}; viscosity, heat conduction and diffusion add their usual terms and do not affect the argument. By \eqref{eq:generic-combined},
\begin{equation}
 \partial_t\bm s+\nabla\cdot(\bm v\,\bm s)=\bm{\mathcal T},\qquad
 mn\,D_t\bm v=-\nabla p-\tfrac12\nabla\times\bm{\mathcal T},\qquad
 \bm{\mathcal T}=-\lambda(2\bm\omega-\bm\Omega)=-\Gamma_s(\bm s-\chi_s\bm\Omega) ,
 \label{eq:fr-full}
\end{equation}
with $[\nabla\cdot(\bm v\,\bm s)]^A=\partial_i(v^is^A)$. The torque density $\bm{\mathcal T}$ on the spin is taken from the orbital motion through the antisymmetric stress $\Pi^{\rm rot}_{ij}$ of \eqref{eq:rot-output}, and the dissipated power is $\lambda{(2\bm\omega-\bm\Omega)^2}=\bm{\mathcal T}^2/\lambda$. The pressure $p$ is that of Sec.~\ref{sec:thermo} and contains the spin energy.

The relaxation drives the spin toward $2\bm\omega=\bm\Omega$, that is, $\bm s=\chi_s\bm\Omega$: the equilibrium of the spin with the vorticity. No reversible dynamics can keep it there. By \eqref{eq:casimirs} the bracket conserves $|\bm s|/n$ on every fluid element, for every energy, whereas $\chi_s|\bm\Omega|/n$ changes whenever the flow stretches the vorticity. A reversible coupling of the spin to the vorticity, however strong, makes the spin precess and leaves its magnitude unchanged. The condition $\bm s=\chi_s\bm\Omega$ is maintained by the rotational viscosity $\lambda$ alone.

When $\Gamma_s$ is large compared with the rates of the flow, ${2\bm\omega-\bm\Omega}=O(\Gamma_s^{-1})$ while the torque $\bm{\mathcal T}=-\lambda{(2\bm\omega-\bm\Omega)}$ remains finite and is fixed by the spin equation:
\begin{equation}
 mn\,D_t\bm v=-\nabla p-\tfrac12\nabla\times\bm{\mathcal T}_{\rm c},\qquad
 \bm{\mathcal T}_{\rm c}=\partial_t(\chi_s\bm\Omega)+\nabla\cdot(\bm v\,\chi_s\bm\Omega) .
 \label{eq:fr-limit}
\end{equation}
The torque is no longer given by a constitutive relation. It is a torque of constraint, the torque that the spin $\chi_s\bm\Omega$ requires in order to follow the vorticity, and with the equations of $n$ and $s$ \eqref{eq:fr-limit} is a closed equation for the velocity. The constraint does no work: the torque delivers the power $2\bm\omega\cdot\bm{\mathcal T}_{\rm c}$ to the spin and takes $\bm\Omega\cdot\bm{\mathcal T}_{\rm c}$ from the flow, and their difference ${(2\bm\omega-\bm\Omega)}\cdot\bm{\mathcal T}_{\rm c}$ vanishes. The energy
\begin{equation}
 H=\int d^3x\,\bigl[\tfrac12mn\bm v^2+\varepsilon_0(n,s)+\tfrac12\chi_s\bm\Omega^2\bigr]
 \label{eq:fr-energy}
\end{equation}
is conserved. At large finite $\Gamma_s$ the deviation $\delta\bm s=\bm s-\chi_s\bm\Omega$ obeys, by \eqref{eq:fr-full} and the definition of $\bm{\mathcal T}_{\rm c}$, $\partial_t\delta\bm s+\nabla\cdot(\bm v\,\delta\bm s)=-\Gamma_s\,\delta\bm s-\bm{\mathcal T}_{\rm c}$. After a transient of duration $\Gamma_s^{-1}$ the spin therefore lags behind $\chi_s\bm\Omega$ by $\delta\bm s=-\bm{\mathcal T}_{\rm c}/\Gamma_s+O(\Gamma_s^{-2})$, and the torque is $\bm{\mathcal T}=-\Gamma_s\,\delta\bm s=\bm{\mathcal T}_{\rm c}+O(\Gamma_s^{-1})$. The solutions of \eqref{eq:fr-full} thus deviate from those of \eqref{eq:fr-limit} at order $\Gamma_s^{-1}$, and the energy decays at the rate $\int d^3x\,\bm{\mathcal T}^2/\lambda=\Gamma_s^{-1}\int d^3x\,\bm{\mathcal T}_{\rm c}^2/\chi_s+O(\Gamma_s^{-2})$.

The limit is controlled in the variables $(\pi'_i,s^A)$ of Sec.~\ref{sec:hep-comparison}. There the relaxation acts on the spin at fixed $\pi'_i$ (end of Sec.~\ref{sec:generic-iso}), and by \eqref{eq:improvement-chain} ${2\bm\omega-\bm\Omega}=\delta H/\delta\bm s$ at fixed $\pi'_i$. The states with $2\bm\omega=\bm\Omega$ are therefore the minima of the energy with respect to the spin at fixed $n$, $s$ and $\pi'_i$, and the relaxation is a descent toward them. Linearized about a uniform state, a perturbation of the spin with wave vector $\bm k$ decays at the rate $\Gamma_s(1+\chi_sk^2/4mn)$ if it is transverse to $\bm k$ and at $\Gamma_s$ if it is longitudinal. The manifold of these states attracts at a rate that stays finite at long wavelength, and the slow motion along it is \eqref{eq:fr-limit}.

\subsection{Belinfante momentum and odd viscosity}

The limit is simplest for the momentum density $\pi'_i=\pi_i+\tfrac12\epsilon_{ijk}\partial_js^k$. It contains the spin, the relaxation leaves it unchanged, and the torque drops out of its equation of motion. What takes the place of the torque is the spin term of the bracket \eqref{eq:improved-self-bracket}. Pointwise, and for compressible fields $\bm w$ and $\bm w'$, the integrand of that term is the spin tensor contracted with the commutator of the shear tensors of the two fields,
\begin{equation}
 \{P_{\pi'}[\bm w],P_{\pi'}[\bm w']\}=-P_{\pi'}[[\bm w,\bm w']]
 +\frac12\int d^3x\,S_{ij}\,[\bm\sigma_{\bm w},\bm\sigma_{\bm w'}]_{ij},\qquad S_{ij}=\epsilon_{ijk}s^k ,
 \label{eq:fr-commutator}
\end{equation}
where $(\bm\sigma_{\bm w})_{ij}=\partial_{(i}w_{j)}-\tfrac13\delta_{ij}\partial_k w^k$, so that $\bm\sigma_{\bm v}=\bm\sigma$, and $[\bm\sigma_{\bm w},\bm\sigma_{\bm w'}]$ is the commutator of the two matrices. The commutator of two symmetric matrices is antisymmetric: two shear deformations applied in opposite orders differ by a rotation, and a rotation changes the direction of the spin. The isotropic parts of the velocity gradients commute with everything and do not contribute. In $\partial_tP_{\pi'}[\bm w]=\{P_{\pi'}[\bm w],H\}$ the second field is the velocity, $\delta H/\delta\pi'_i=v^i$, and since $\tfrac12S_{ij}[\bm\sigma_{\bm w},\bm\sigma]_{ij}=(\bm\sigma_{\bm w})_{ij}\,\Pi^{\rm odd}_{ij}$, an integration by parts shows that the spin term adds to the flux of $\pi'_j$ the stress
\begin{equation}
 \Pi^{\rm odd}_{ij}=\tfrac12\bigl(S_{ik}\sigma_{kj}-\sigma_{ik}S_{kj}\bigr)
 =\tfrac12\bigl(\epsilon_{ikl}\sigma_{kj}+\epsilon_{jkl}\sigma_{ki}\bigr)s^l,\qquad
 \Pi^{\rm odd}_{ij}\,\sigma_{ij}=0 .
 \label{eq:fr-odd}
\end{equation}
It is symmetric, traceless and linear in the shear tensor, and it does no work: an odd viscosity~\cite{Banerjee2017,Markovich2021} of magnitude $|\bm s|/2$. For $\bm s$ along $\hat{\bm z}$, $\Pi^{\rm odd}_{xx}=-\Pi^{\rm odd}_{yy}=|\bm s|\,\sigma_{xy}$ and $\Pi^{\rm odd}_{xy}=\tfrac12|\bm s|(\sigma_{yy}-\sigma_{xx})$ in the plane normal to the spin, and $\Pi^{\rm odd}_{xz}=\tfrac12|\bm s|\,\sigma_{yz}$, $\Pi^{\rm odd}_{yz}=-\tfrac12|\bm s|\,\sigma_{xz}$, $\Pi^{\rm odd}_{zz}=0$: both odd viscosities of Table~\ref{tab:onsager} are present, in a fixed ratio. That the momentum $\pi'_i$ of a fluid with an intrinsic angular-momentum density has this stress, with these brackets as its origin, is the result of Ref.~\cite{Markovich2021}; \eqref{eq:fr-commutator} writes it in a form that displays the mechanism.

For every energy $H$, and for every torque density $\bm{\mathcal T}$ that is exchanged with the orbital motion through the antisymmetric stress, the equations \eqref{eq:eom-derived} with this torque give
\begin{equation}
 \partial_t\pi'_j+\partial_i(v^i\pi'_j)+\pi'_i\,\partial_jv^i
 =-n\,\partial_j\mu-s\,\partial_jT-s^A\partial_j(2\omega_A-\Omega_A)-\tfrac12\bigl\{\nabla\times\bigl[\bm s\times(2\bm\omega-\bm\Omega)\bigr]\bigr\}_j-\partial_i\Pi^{\rm odd}_{ij},
 \label{eq:fr-belinfante}
\end{equation}
in which the torque does not appear. The spin acts on $\pi'_i$ through three forces: the force of the field ${2\bm\omega-\bm\Omega}=\delta H/\delta\bm s|_{\pi'}$ on the spin, the antisymmetric stress of the reactive torque $\bm s\times{(2\bm\omega-\bm\Omega)}$, and the odd viscosity. In the limit of fast relaxation $2\bm\omega=\bm\Omega$, and only the last one remains:
\begin{equation}
 \partial_t\pi'_j+\partial_i(v^i\pi'_j)+\pi'_i\,\partial_jv^i
 =-n\,\partial_j\mu-s\,\partial_jT-\partial_i\Pi^{\rm odd}_{ij}\Big|_{\bm s=\chi_s\bm\Omega},\qquad
 \bm\pi'=mn\bm v+\tfrac12\nabla\times(\chi_s\bm\Omega) .
 \label{eq:fr-pi}
\end{equation}
This is \eqref{eq:fr-limit} in the variables $(n,s,\pi'_i)$, with $\mu=\delta H/\delta n$ and $T=\delta H/\delta s$ taken at fixed $\pi'_i$ and $s^A$ and evaluated at $\bm s=\chi_s\bm\Omega$. Equation~\Lodd{} of the Letter is $\Pi^{\rm odd}_{ij}$ at this value of the spin.

\subsection{A spin held by relaxation and a spin inserted into an action}
\label{sec:fr-action}

Equation~\eqref{eq:fr-pi} has the form of an Euler--Poincar\'e equation~\cite{Holm1998} with one additional force. Consider a fluid without a spin variable whose Lagrangian contains the rotational energy of its elements,
\begin{equation}
 \mathscr L[\bm v,n,s]=\int d^3x\,\bigl[\tfrac12mn\bm v^2+\tfrac12\chi_s\bm\Omega^2-\varepsilon_0(n,s)\bigr] .
 \label{eq:fr-lagrangian}
\end{equation}
This is what the insertion of the spin density $\chi_s\bm\Omega$ into an action gives, as in Ref.~\cite{Montenegro2017}, where a polarization proportional to the vorticity is substituted into the Lagrangian before the variation. Its Euler--Poincar\'e equation is
\begin{equation}
 \partial_t\pi'_j+\partial_i(v^i\pi'_j)+\pi'_i\,\partial_jv^i
 =n\,\partial_j\frac{\delta\mathscr L}{\delta n}+s\,\partial_j\frac{\delta\mathscr L}{\delta s},\qquad
 \pi'_j=\frac{\delta\mathscr L}{\delta v^j}=mnv_j+\tfrac12\bigl[\nabla\times(\chi_s\bm\Omega)\bigr]_j .
 \label{eq:fr-ep}
\end{equation}
The momentum is the same $\bm\pi'$ as in \eqref{eq:fr-pi}, and the Legendre transform of $\mathscr L$ is the energy \eqref{eq:fr-energy}. Moreover $\delta\mathscr L/\delta n=-\mu$ and $\delta\mathscr L/\delta s=-T$ with the potentials of \eqref{eq:fr-pi}: a Legendre transformation reverses the sign of the derivatives with respect to $n$ and $s$, and the energy is stationary with respect to the spin at $2\bm\omega=\bm\Omega$, so that the elimination of the spin does not change these derivatives. The two fluids thus have the same variables, the same energy and the same momentum, and for every equation of state and every $\chi_s(n,s)$ their equations of motion differ by one term,
\begin{equation}
 \text{spin held at $\chi_s\bm\Omega$ by relaxation}\;=\;\text{Euler--Poincar\'e equation of $\mathscr L$}\;-\;\partial_i\Pi^{\rm odd}_{ij}\Big|_{\bm s=\chi_s\bm\Omega} .
 \label{eq:fr-difference}
\end{equation}
For incompressible flow with uniform $mn$ and $\chi_s$, where $p$ in each case is the pressure that keeps $\nabla\cdot\bm v=0$, \eqref{eq:fr-pi} and \eqref{eq:fr-ep} read
\begin{equation}
\begin{aligned}
 \bigl(mn-\tfrac14\chi_s\nabla^2\bigr)\partial_t\bm v&=-mn\,\bm v\cdot\nabla\bm v-\tfrac12\chi_s\nabla\times(\bm v\cdot\nabla\bm\Omega)-\nabla p
 &&\text{(relaxation)},\\
 \partial_t\bm\pi'+\bm v\cdot\nabla\bm\pi'+\pi'_i\nabla v^i&=-\nabla p,\qquad\bm\pi'=\bigl(mn-\tfrac14\chi_s\nabla^2\bigr)\bm v
 &&\text{(action)}.
\end{aligned}
 \label{eq:fr-incompressible}
\end{equation}

A direct integration of the incompressible form of \eqref{eq:fr-full} confirms the expansion in $\Gamma_s^{-1}$ and shows that the limit is \eqref{eq:fr-limit}, not \eqref{eq:fr-ep}. On a periodic $24^3$ grid, from a state with $\bm s=\chi_s\bm\Omega$ ($\chi_s=0.8$, $mn=1$, root-mean-square velocity $1$) and over one unit of time, $\Gamma_s$ times the root-mean-square distance of its velocity from the solution of \eqref{eq:fr-limit}, in units of the root-mean-square velocity, stays between $0.11$ and $0.15$ for $\Gamma_s$ from $5$ to $320$. In addition, $\Gamma_s$ times the energy lost approaches the predicted $\int dt\int d^3x\,\bm{\mathcal T}_{\rm c}^2/\chi_s=0.202$, in units of the initial energy and evaluated on the solution of \eqref{eq:fr-limit} (it is $0.206$ at $\Gamma_s=320$). By contrast, the distance from the solution of \eqref{eq:fr-ep} remains at $0.21$, the distance between the two limits. In a second test, for small perturbations of a rigidly rotating fluid whose spin has its equilibrium value, the frequencies of \eqref{eq:fr-full} approach those of \eqref{eq:fr-limit} with an error proportional to $\Gamma_s^{-2}$, the waves decay at the rate that corresponds to the dissipated power $\bm{\mathcal T}_{\rm c}^2/\lambda$, and the fast modes have the rates $\Gamma_s(1+\chi_sk^2/4mn)$ and $\Gamma_s$ of Sec.~\ref{sec:fr-limit}.

The difference \eqref{eq:fr-difference} has a counterpart in mechanics. A condition on the velocities that cannot be integrated, such as rolling without slipping, can be imposed on the virtual displacements, as strong friction does (the Lagrange--d'Alembert principle), or on the paths compared in the action; the two differ by a gyroscopic force, the momentum of the constrained coordinates times the curvature of the condition~\cite{Bloch2005}. Here the constrained coordinates are the internal orientations of the fluid elements, their momentum is the spin, and the condition is $2\bm\omega=\bm\Omega$. It is not integrable: four successive deformations without vorticity, $e^{\bm\sigma_1}$, $e^{\bm\sigma_2}$, $e^{-\bm\sigma_1}$ and $e^{-\bm\sigma_2}$, rotate a fluid element by $[\bm\sigma_2,\bm\sigma_1]$ to leading order, while an orientation that follows the vorticity does not rotate. The curvature is thus the commutator of two shear tensors, the gyroscopic force is the odd-viscous force, and \eqref{eq:fr-difference} is the Lagrange--d'Alembert equation of the fluid. In two dimensions Ref.~\cite{Monteiro2023} showed that the bracket of a fluid with odd viscosity satisfies the Jacobi identity only if the intrinsic angular momentum per particle is constant; the present case is the three-dimensional counterpart.

\subsection{Second-order transport}

In a relativistic fluid the limit is the theory of Refs.~\cite{Li2021,Hongo2021}: beyond the relaxation time the spin is in equilibrium with the vorticity, and a pseudogauge transformation that absorbs the spin current leaves conventional hydrodynamics with nondissipative terms of second order in gradients. With the convective spin current $\Sigma^{\lambda\mu\nu}=u^\lambda S^{\mu\nu}$ the Belinfante tensor~\eqref{eq:belinfante-tensor} differs from $\Theta^{(\mu\nu)}$ by $-\tfrac12\nabla_\lambda(u^\mu S^{\lambda\nu}+u^\nu S^{\lambda\mu})$, the symmetric part of the tensor $G^{\mu\nu}$ of \eqref{eq:antisymmetric-pseudogauge}. In the local rest frame the spatial components of this term contain the velocity gradient contracted with the spin tensor on the index of the derivative,
\begin{equation}
 \Theta_{\rm B}^{ij}-\Theta^{(ij)}=-\tfrac12\bigl[(\partial_kv_i)S_{kj}+(\partial_kv_j)S_{ki}\bigr]
 =\Pi^{\rm odd}_{ij}+\tfrac12(s_i\Omega_j+s_j\Omega_i)-\delta_{ij}\,\bm s\cdot\bm\Omega ,
 \label{eq:fr-dragged}
\end{equation}
where the shear part of the velocity gradient gives the odd viscosity \eqref{eq:fr-odd} and its antisymmetric part the remaining terms; the expansion drops out. These are the symmetric transverse components of $G^{\mu\nu}$, the term that the pseudogauge transformation to the totally antisymmetric spin current of Ref.~\cite{Fang2026} adds to the energy--momentum tensor (Sec.~\ref{sec:fang-counterpart}): for such a current the symmetric part of the energy--momentum tensor is the Belinfante tensor. This is the stress that Ref.~\cite{Son2026} obtains from the same transformation for a spin of constant magnitude per particle [its Eq.~(31)]. At $\bm s=\chi_s\bm\Omega$,
\begin{equation}
 \Theta_{\rm B}^{ij}-\Theta^{(ij)}=\tfrac12\chi_s\bigl(\epsilon_{ikl}\sigma_{kj}+\epsilon_{jkl}\sigma_{ki}\bigr)\Omega_l
 +\chi_s\bigl(\Omega_i\Omega_j-\delta_{ij}\bm\Omega^2\bigr) ,
 \label{eq:fr-second-order}
\end{equation}
both terms being of second order in gradients. In the conventions of Ref.~\cite{Baier2008}, in which the stress contains $\lambda_2\,\sigma^{\langle\mu}{}_\lambda\Omega^{\nu\rangle\lambda}+\lambda_3\,\Omega^{\langle\mu}{}_\lambda\Omega^{\nu\rangle\lambda}$ with a shear tensor twice the one used here and the metric signature $(-,+,+,+)$, the traceless parts of the two terms are $\delta\lambda_2=\chi_s/2$ and $\delta\lambda_3=-\chi_s$; the isotropic part depends on how the pressure is defined. The two terms have different origins. The vorticity-squared term comes from the rotational energy $\tfrac12\chi_s\bm\Omega^2$ and is present in the fluid with Lagrangian~\eqref{eq:fr-lagrangian} as well. The shear--vorticity term is the odd viscosity, and by \eqref{eq:fr-difference} it is absent there: that fluid has $\delta\lambda_2=0$ and the same $\delta\lambda_3$. This is consistent with the classification of Ref.~\cite{Haehl2015}, in which a neutral conformal fluid has one nondissipative second-order combination that vanishes in equilibrium and that no fluid derived from a Lagrangian possesses; in the conventions of Ref.~\cite{Baier2008}, with the relaxation time $\tau_\Pi$ of the shear stress and the coefficient $\kappa$ of the terms that contain the curvature of spacetime, it is $\lambda_2-2\eta\tau_\Pi+2\kappa$, and the spin held by relaxation shifts it by $\chi_s/2$.

Reference~\cite{Li2021} carries out the transformation for $S^{\mu\nu}=\chi\omega^{\mu\nu}$, its Eq.~(20), and matches the vorticity-squared terms to the general form that the second law allows. The shear--vorticity term is contained in its transformed tensor, Eq.~(25) of that work, $\tfrac12\chi[(\theta^\mu{}_\alpha+\omega^\mu{}_\alpha)\omega^{\alpha\nu}+(\mu\leftrightarrow\nu)]$, where $\theta_{\mu\nu}$ and $\omega_{\mu\nu}$ are the symmetric and antisymmetric parts of the transverse velocity gradient, but it is not discussed there. With the definitions of that work, $u_\mu u^\mu=-1$ and $\omega_{\mu\nu}=\tfrac12(\partial^\perp_\mu u_\nu-\partial^\perp_\nu u_\mu)$, the derivative index of $\partial_\lambda u^\mu$ is the one contracted with the spin tensor, as in \eqref{eq:fr-dragged}, which gives $(-\theta^\mu{}_\alpha+\omega^\mu{}_\alpha)$ in place of $(\theta^\mu{}_\alpha+\omega^\mu{}_\alpha)$. The vorticity-squared terms agree with ours; the shear--vorticity term has the same magnitude and the opposite sign in the arXiv version of that work, whereas our sign agrees with Eq.~(31) of Ref.~\cite{Son2026} and with the nonrelativistic stress \eqref{eq:fr-odd}.

\subsection{Circulation}

Which circulation is conserved depends on how the spin enters. For an independent spin without dissipation it is $\oint_C(\pi_i/n)\,dx^i-(|\bm s|/n)\,\Phi_C$, by \eqref{eq:kelvin-law}; relaxation changes both of its terms, whereas the circulation of $\pi'_i/n$, driven by ${2\bm\omega-\bm\Omega}$ and by the odd viscosity, \eqref{eq:fr-belinfante}, does not contain the torque. In the limit of fast relaxation, \eqref{eq:fr-pi} gives
\begin{equation}
 \frac{d}{dt}\oint_C\frac{\pi'_j}{n}\,dx^j=-\oint_C\frac sn\,dT-\oint_C\frac1n\,\partial_i\Pi^{\rm odd}_{ij}\,dx^j .
 \label{eq:fr-circulation}
\end{equation}
The fluid then produces no entropy and conserves its energy, but no circulation: the odd-viscous force is a source, as Ref.~\cite{Markovich2021} noted for odd viscosity in three dimensions, except in planar incompressible flow with uniform $mn$ and $\chi_s$, where the torque of constraint vanishes and \eqref{eq:fr-limit} is the Euler equation, which conserves the circulation of the velocity. For the fluid with Lagrangian~\eqref{eq:fr-lagrangian}, \eqref{eq:fr-ep} conserves the circulation of $\pi'_i/n$ around a loop that moves with $\bm v$ for uniform $s/n$, its Kelvin--Noether theorem~\cite{Holm1998}. Reference~\cite{Montenegro2017} observed that the polarization is a source for the circulation of the velocity; in this Galilean counterpart the conserved circulation is that of $\pi'_i/n$. Only the relaxation removes the conservation law.

\subsection{Scope}

Three qualifications apply. (i) The statements of this section hold for the isotropic relaxation and, apart from \eqref{eq:fr-dragged} and~\eqref{eq:fr-second-order}, for the Galilean fluid; at fixed $\xi$ the coupling of Sec.~\ref{sec:onsager} does not change the limit, in which its torque is of third order in gradients (Sec.~\ref{sec:counting}). (ii) The bracket that the condition $\bm s=\chi_s\bm\Omega$ induces on functionals of $(n,s,\pi'_i)$ is antisymmetric, and it generates \eqref{eq:fr-pi} with the energy \eqref{eq:fr-energy}, which is therefore conserved. It does not satisfy the Jacobi identity, as expected for a condition that is not integrable (Sec.~\ref{sec:fr-action}) and as we checked numerically; this is the three-dimensional counterpart of the result of Ref.~\cite{Monteiro2023}. Whether \eqref{eq:fr-limit} has a Hamiltonian formulation with another bracket remains open. (iii) The coefficient $\chi_s/2$ is the contribution of the bracket; other nondissipative contributions to $\lambda_2$ exist~\cite{Baier2008}.

\section{Comparison with the Entropy-Current Construction of Ref.~\texorpdfstring{\cite{Fang2026}}{[1]}}
\label{sec:fang}

Reference~\cite{Fang2026} constructs first-order spin hydrodynamics by an entropy-current analysis for the totally antisymmetric spin current of the canonical tensors of the Dirac field. It uses the same spin tensor $S^{\mu\nu}$, the same local relations~\eqref{eq:rel-first-law} and the same power counting, $S^{\mu\nu}=O(1)$ and $\omega^\mu=O(\partial)$, and at global equilibrium its energy--momentum tensor is that of the present theory after the pseudogauge transformation~\eqref{eq:antisymmetric-pseudogauge} (Secs.~\ref{sec:fang-counterpart} and~\ref{sec:fang-equilibrium}). Away from equilibrium the constitutive relations differ further in two respects: by reversible couplings odd in the spin, which no change of variables generates and which remain when the spin is conserved (Secs.~\ref{sec:fang-nonpg} and~\ref{sec:fang-conserved}), and by the absence of the dissipative couplings that depend on the direction of the spin (Sec.~\ref{sec:fang-dissipative}). The reversible shear, expansion and vorticity terms of its spin equation are the transport law of rescaled FW components in the dual of a Lie-transported triad, a variable that differs from the spin density of its currents (Sec.~\ref{sec:fang-spin}).

\subsection{Constitutive relations of Ref.~\texorpdfstring{\cite{Fang2026}}{[1]}}
\label{sec:fang-relations}
Reference~\cite{Fang2026} takes the spin current $\Sigma_{\rm F}^{\lambda\mu\nu}=u^\lambda S^{\mu\nu}+u^\mu S^{\nu\lambda}+u^\nu S^{\lambda\mu}+O(\partial)$, which is $\Sigma'^{\lambda\mu\nu}$ of \eqref{eq:antisymmetric-pseudogauge} up to the spin diffusion, of second order in this counting; here and below a subscript F marks quantities of Ref.~\cite{Fang2026}. It writes $\Theta_{\rm F}^{\mu\nu}=(\varepsilon_{\rm F}+p_{\rm F})u^\mu u^\nu-p_{\rm F}g^{\mu\nu}+2h^{(\mu}u^{\nu)}+\pi^{\mu\nu}+2q^{[\mu}u^{\nu]}+\phi^{\mu\nu}$, with $h^\mu$, $q^\mu$, the symmetric $\pi^{\mu\nu}$ and the antisymmetric $\phi^{\mu\nu}$ transverse to $u^\mu$, and closes the entropy production with four isotropic coefficients: the heat conductivity, the bulk and shear viscosities, and the rotational viscosity $\gamma_\phi$. The spin-dependent corrections enter $\Theta_{\rm F}^{\mu\nu}$ as $-\pi_s^{\mu\nu}-\phi_s^{\mu\nu}$ and the heat flux as $h^\mu-(\varepsilon+p)j_n^\mu/n+h_s^\mu$, where $j_n^\mu$ is the diffusion current. In vector form,
\begin{equation}
\begin{gathered}
 -\pi_s^{\mu\nu}-\phi_s^{\mu\nu}=2\omega^\mu s^\nu-2(1-2v_n^2)\Delta^{\mu\nu}s_\lambda\omega^\lambda+v_n^2\theta S^{\mu\nu},\\
 h_s^\mu=-\tfrac12\Delta^\mu{}_\nu\nabla_\lambda S^{\nu\lambda}+\tfrac12S^{\mu\nu}a_\nu+\xi_{\rm F} S^{\mu\nu}(\beta^{-1}\nabla_\nu\beta+a_\nu),
\end{gathered}
 \label{eq:fang-corrections}
\end{equation}
with $v_n^2=(\partial p/\partial\varepsilon)_{n,s^A}$ and a parameter $\xi_{\rm F}$ that the entropy analysis leaves free. The vector form fixes the order of the indices in the products $S^{\mu\lambda}\omega_\lambda{}^\nu$ of Ref.~\cite{Fang2026}, where $\omega_{\mu\nu}=-\varepsilon_{\mu\nu\rho\sigma}u^\rho\omega^\sigma$ is the spin-potential tensor dual to $\omega^\mu$, such that the corrections, inserted into the Ward identity~\eqref{eq:ward-transverse}, give the spin equation of Ref.~\cite{Fang2026}, its Eq.~(13), which in our conventions for the metric and the Levi-Civita tensor reads
\begin{equation}
\dot s^\mu={}-u^\mu a_\nu s^\nu
 +\left(\varepsilon^{\mu\nu\rho\sigma}s_\nu u_\rho
       -2\beta\gamma_\phi g^{\mu\sigma}\right)(2\omega_\sigma-\Omega_\sigma)
 -s_\nu\partial^{\langle\mu}u^{\nu\rangle}
 -\bigl(\tfrac13+2v_n^2\bigr)\theta s^\mu .
 \label{eq:fang-covariant}
\end{equation}
In the rest frame, where $s_j=-s^j$ and $\partial^{\langle i}u^{j\rangle}=-\sigma_{ij}$,
\begin{equation}
 \dot{\bm s}=(\bm\Omega-2\bm\omega)\times\bm s-\bm\sigma\bm s
 -\bigl(\tfrac13+2v_n^2\bigr)\theta\bm s
 -2\beta\gamma_\phi(2\bm\omega-\bm\Omega).
 \label{eq:fang-rest}
\end{equation}
The last term is the relaxation of Sec.~\ref{sec:generic-iso}, with $\gamma_\phi=\lambda/2\beta$. Compared with the reversible transport law~\eqref{eq:fw-fluid}, the equation contains four reversible terms that are absent there or differ: the shear term $-\bm\sigma\bm s$, an expansion term whose coefficient $-(\tfrac13+2v_n^2)$ depends on the equation of state, the rotation $\bm\Omega\times\bm s$ with the vorticity, and a precession about the spin potential with the opposite sign, $-2\bm\omega\times\bm s$ instead of $+2\bm\omega\times\bm s$.

\subsection{Pseudogauge transformation to the totally antisymmetric spin current}
\label{sec:fang-counterpart}
The convective spin current~\eqref{eq:rel-current-choice} and $\Sigma_{\rm F}^{\lambda\mu\nu}$ contain the same spin density, $u_\lambda\Sigma^{\lambda\mu\nu}=u_\lambda\Sigma_{\rm F}^{\lambda\mu\nu}=S^{\mu\nu}$, and the transformation~\eqref{eq:antisymmetric-pseudogauge} maps the first onto the second up to the spin diffusion. It adds to the energy--momentum tensor the identically conserved term $G^{\mu\nu}$; since $\nabla_\lambda\Phi^{\lambda\mu\nu}=2G^{[\mu\nu]}$, the charges and the Ward identity~\eqref{eq:ward-longitudinal} are unchanged. This term is fixed: at fixed spin density the superpotential $\Phi^{\lambda\mu\nu}=\Sigma^{\lambda\mu\nu}-\Sigma_{\rm F}^{\lambda\mu\nu}$ is determined by the two spin currents, and \eqref{eq:pseudogauge} then determines $\Theta'^{\mu\nu}-\Theta^{\mu\nu}$ up to identically conserved symmetric terms of second order in gradients. Reference~\cite{Daher2023} relates the canonical formulation with a totally antisymmetric spin current to the formulation with a convective one by this transformation, and Ref.~\cite{YangYan2024} obtains the same term as a nondissipative solution of an entropy-current analysis. The term decomposes as $u_\mu G^{\mu\nu}=2(s_\lambda\Omega^\lambda)u^\nu$ and
\begin{equation}
 \Delta^\mu{}_\alpha G^{\alpha\nu}u_\nu=\Delta^\mu{}_\nu\nabla_\lambda S^{\nu\lambda},\qquad
 \Delta^\mu{}_\alpha\Delta^\nu{}_\beta G^{\alpha\beta}=\Omega^\mu s^\nu-\Delta^{\mu\nu}s_\lambda \Omega^\lambda+\tfrac13\theta S^{\mu\nu}+S^{\mu\lambda}\sigma_\lambda{}^\nu .
 \label{eq:fang-G}
\end{equation}

The antisymmetric part of $G^{\mu\nu}$ preserves the FW transport of the spin. Through the Ward identity~\eqref{eq:ward-transverse}, the components $u^\mu S^{\nu\lambda}+u^\nu S^{\lambda\mu}$ of $\Sigma'^{\lambda\mu\nu}$ exert on the spin the torque $-(\partial_iv_j)s_j+\theta s_i$ in the rest frame. Alone, this torque would Lie-transport $s_\mu$ as a one-form, $\dot s_i=-(\partial_iv_j)s_j=(\bm\Omega\times\bm s-\bm\sigma\bm s-\tfrac13\theta\bm s)_i$; the antisymmetric part of $G^{\mu\nu}$ exerts the opposite torque, $(\partial_iv_j)s_j-\theta s_i$. In Ref.~\cite{Fang2026} the antisymmetric stress $\phi_s^{\mu\nu}$ takes the place of $G^{[\mu\nu]}$. By \eqref{eq:fang-corrections} its torque is $-2\bm\omega\times\bm s-2v_n^2\theta\bm s$, which agrees with that of $G^{[\mu\nu]}$ only at global equilibrium, where both are $-\bm\Omega\times\bm s$; away from it the Lie transport of the one-form remains in \eqref{eq:fang-rest}.

For free Dirac particles the canonical tensor can be computed. For a Dirac field $\Psi$ of mass $m$, with $\overleftrightarrow{\partial}=\overrightarrow{\partial}-\overleftarrow{\partial}$ and $\hbar=1$, the Gordon identity gives on shell
\begin{equation}
 \tfrac i2\bar\Psi\gamma^\mu\overleftrightarrow{\partial}{}^{\nu}\Psi
 =-\frac{1}{4m}\bar\Psi\overleftrightarrow{\partial}{}^{\mu}\overleftrightarrow{\partial}{}^{\nu}\Psi
 +\partial_\lambda\Bigl(\bar\Psi\,\tfrac i4[\gamma^\mu,\gamma^\lambda]\,\frac{i\overleftrightarrow{\partial}{}^{\nu}}{2m}\,\Psi\Bigr).
 \label{eq:gordon}
\end{equation}
The first term is symmetric. In a fluid of nonrelativistic particles the expectation value of the term in parentheses is $S^{\mu\lambda}u^\nu$, up to the spin diffusion current, so that the canonical energy--momentum tensor is a symmetric tensor plus $G^{\mu\nu}$: its antisymmetric part is built from the velocity gradient, whereas the antisymmetric stress of Ref.~\cite{Fang2026} is built from the spin potential.

\subsection{Heat flux and equilibrium}
\label{sec:fang-equilibrium}
The longitudinal projection of $\Theta'^{\mu\nu}=\Theta^{\mu\nu}+G^{\mu\nu}$ satisfies $\Theta'^{[\mu\nu]}u_\nu=q^\mu$ with $q^\mu=\tfrac12(S^{\mu\nu}a_\nu+\Delta^\mu{}_\nu\nabla_\lambda S^{\nu\lambda})$, the relation that Ref.~\cite{Fang2026} uses to fix $q^\mu$; in the frame of~\eqref{eq:rel-current-choice}, $h^\mu=q^\mu$. The frame-invariant combination $h^\mu-(\varepsilon+p)j_n^\mu/n$ is $\tfrac12\Delta^\mu{}_\nu\nabla_\lambda S^{\nu\lambda}-\tfrac12S^{\mu\nu}a_\nu$ plus heat conduction in both constructions, provided $\xi_{\rm F}=0$. The term with $\xi_{\rm F}$ is a reversible Hall current of heat; it is absent here, because the bracket advects the particle number and the entropy with the momentum density alone.

Since $u_\mu u_\nu\Theta'^{\mu\nu}=\varepsilon+2s_\lambda\Omega^\lambda$, the energy density of Ref.~\cite{Fang2026} is $\varepsilon_{\rm F}=\varepsilon+2s_\lambda\Omega^\lambda$, and its pressure is $p_{\rm F}=p+2v_n^2s_\lambda\Omega^\lambda$. With this identification, $\Theta'^{\mu\nu}$ at global equilibrium ($\sigma^{\mu\nu}=0$, $\theta=0$, $2\omega^\mu=\Omega^\mu$) reproduces the equilibrium energy--momentum tensor of Ref.~\cite{Fang2026}, with the spin-dependent corrections $\pi^{\mu\nu}_{\rm eq}+\phi^{\mu\nu}_{\rm eq}=\Omega^{\langle\mu}s^{\nu\rangle}+2(v_n^2-\tfrac13)\Delta^{\mu\nu}s_\lambda\Omega^\lambda+\Omega^{[\mu}s^{\nu]}$, which are the terms~\eqref{eq:fang-G} of $G^{\mu\nu}$ with the pressure written as $p_{\rm F}$.

\subsection{Reversible couplings odd in the spin}
\label{sec:fang-nonpg}
With the same identification,
\begin{equation}
 \Theta_{\rm F}^{\mu\nu}-\Theta'^{\mu\nu}=D^{\mu\nu}\equiv(2\omega^\mu-\Omega^\mu)s^\nu-(1-2v_n^2)\Delta^{\mu\nu}s_\lambda(2\omega^\lambda-\Omega^\lambda)+\bigl(v_n^2-\tfrac13\bigr)\theta S^{\mu\nu}-S^{\mu\lambda}\sigma_\lambda{}^\nu ,
 \label{eq:fang-D}
\end{equation}
apart from the isotropic dissipative terms, which coincide when the shear viscosity of Ref.~\cite{Fang2026} is identified with $2\eta$ and $\gamma_\phi$ with $\lambda/2\beta$. The tensor $D^{\mu\nu}$ vanishes at global equilibrium, is linear in the thermodynamic forces ${2\omega^\mu-\Omega^\mu}$, $\sigma^{\mu\nu}$ and $\theta$ with coefficients odd in the spin, and produces no entropy: in the rest frame $-D^{ij}\partial_iv_j-2\bm\omega\cdot\dot{\bm s}_D=0$ for every $\bm\omega$, where $(\dot{\bm s}_D)_k=-\epsilon_{kij}D^{ij}$ is its torque. It thus belongs to the antisymmetric part of the Onsager matrix (Table~\ref{tab:onsager}), whose entries the bracket of the canonical spin sets to zero (Sec.~\ref{sec:onsager}). Its torque,
\begin{equation}
 \dot{\bm s}_D=(\bm\Omega-2\bm\omega)\times\bm s-\bm\sigma\bm s+\bigl(\tfrac23-2v_n^2\bigr)\theta\bm s ,
 \label{eq:fang-D-torque}
\end{equation}
is the difference between~\eqref{eq:fang-rest} and the spin equation~\eqref{eq:rel-spin-eom} of the present theory. Its symmetric part contains the stress ${(2\omega-\Omega)}^{\langle\mu}s^{\nu\rangle}$ reciprocal to the shear term, the pressure $(\tfrac23-2v_n^2)s_\lambda{(2\omega^\lambda-\Omega^\lambda)}$ reciprocal to the expansion term, and an odd-viscous stress, $-S^{(\mu|\lambda|}\sigma_\lambda{}^{\nu)}$, which in the rest frame is minus the odd viscosity~\eqref{eq:fr-odd}: with a totally antisymmetric spin current the symmetric part of the energy--momentum tensor is the Belinfante tensor, which contains this odd viscosity through the term $S^{\mu\lambda}\sigma_\lambda{}^\nu$ of $G^{\mu\nu}$ (Sec.~\ref{sec:fast-relaxation}), and the symmetric stress of Ref.~\cite{Fang2026} does not contain it.

\emph{The second law does not select $D^{\mu\nu}$.} Because $D^{\mu\nu}$ produces no entropy and vanishes at global equilibrium, the second law admits it with any coefficient, including zero. In Ref.~\cite{Fang2026} it arises from two terms of indefinite sign in the entropy production: $2\beta\omega_{\mu\nu}S^{\lambda\mu}\nabla_\lambda u^\nu$, generated by the components $u^\mu S^{\nu\lambda}+u^\nu S^{\lambda\mu}$ of the spin current, and $-\dot\beta\,S^{\mu\nu}\nabla_\mu u_\nu$. Each is a product of two quantities of first order, and either factor can be taken as the force. Reference~\cite{Fang2026} takes the velocity gradient as the force in both. Taking instead the difference between $2\beta$ times the spin-potential tensor and the thermal vorticity as the force of the first, and the expansion rate as the force of the second, gives fluxes built from the velocity gradient, the same entropy production and the same equilibrium, and FW transport, $\dot{\bm s}=-\theta\bm s$ without relaxation; the stress then differs from $\Theta'^{\mu\nu}$ only by the odd viscosity, which produces no entropy either.

\emph{No change of variables generates $D^{\mu\nu}$.} In a homogeneous state at rest with $\bm s\neq0$, $\bm\Omega=0$, $\sigma_{ij}=0=\theta$ and $2\bm\omega=\bm s/\chi_s$, every term that a pseudogauge transformation built from the fields adds vanishes (Sec.~\ref{sec:pg-entropy}), whereas the constitutive relations of Ref.~\cite{Fang2026} give
\begin{equation}
 \Theta^{ij}_{\rm F}=p\,\delta_{ij}+\frac{1}{\chi_s}\bigl[s_is_j-(1-2v_n^2)\,\bm s^2\delta_{ij}\bigr],
 \label{eq:fang-static}
\end{equation}
that is, $p\,\delta_{ij}+D^{ij}$: the pressure along $\bm s$ exceeds that across it by $\bm s^2/\chi_s$. The present theory and all its pseudogauge-equivalent forms give $p\,\delta_{ij}$. Without spin--orbit interaction this state is an equilibrium, whose Gibbs density operator, with the spin potential conjugate to the total spin, is invariant under rotations of space alone; the stress operator does not act on the spins, and the stress is isotropic. For an ideal gas of spin-$\tfrac12$ particles, $\chi_s=n/4T$ in units of $\hbar$, and the anisotropy~\eqref{eq:fang-static} is $\mathsf P^2$ times the pressure, with $\mathsf P=|\bm s|/s_{\rm m}$ of Sec.~\ref{sec:counting}.

\subsection{Limit of conserved spin}
\label{sec:fang-conserved}
Without spin--orbit interaction, the interaction that exchanges spin and orbital angular momentum, the spin of every fluid element is conserved: rotations of the spins alone are a symmetry, separate from rotations of space, and $\nabla_\lambda(u^\lambda S^{\mu\nu})=0$ up to spin diffusion. For particles with spin-independent interactions this is the nonrelativistic limit~\cite{Hongo2021}. In the present theory the reversible dynamics exerts no torque for any interaction, and the dissipative torques carry $\lambda$ and $\xi$, which are of second and first order in the spin--orbit interaction (Sec.~\ref{sec:kubo}); the spin is conserved in this limit. The torque~\eqref{eq:fang-D-torque} has coefficients that are numbers and $v_n^2$: it does not depend on the spin--orbit interaction, contains no power of $1/c$, and remains. For $\gamma_\phi=0$ and an equation of state isotropic in the spin, \eqref{eq:fang-rest} integrated over a region that moves with a flow of uniform velocity gradient gives
\begin{equation}
 \frac{d}{dt}\int d^3x\,s_i=-(\partial_iv_j)\int d^3x\,s_j+(1-2v_n^2)\,\theta\int d^3x\,s_i ,
 \label{eq:fang-total-spin}
\end{equation}
which conservation of the spin requires to vanish. In simple shear, $\bm v=2\sigma(y,0,0)$, it gives $\frac{d}{dt}\int d^3x\,s_y=-2\sigma\int d^3x\,s_x$; in rigid rotation the spin rotates with the fluid, $\dot{\bm s}=\bm\Omega\times\bm s$; and in isotropic expansion $d\ln(|\bm s|/n)/dt=(\tfrac23-2v_n^2)\theta$. For a nonrelativistic gas with spin-independent collisions the spin density obeys instead $\partial_t\bm s+\nabla\cdot(\bm v\bm s)=0$ up to spin diffusion, as in the present theory, and its canonical tensors are those of Sec.~\ref{sec:fang-counterpart}.

\emph{Rotation.} In a rotating fluid one energy fixes both the precession of the spin and its equilibrium value. A spin--vorticity energy $\alpha\,\bm s\cdot\bm\Omega$ in the canonical variables (Sec.~\ref{sec:second-order}) makes the spin precess about the vorticity at the rate $\alpha\bm\Omega$ in the laboratory and gives the rotating equilibrium $\bm s=(1-\alpha)\chi_s\bm\Omega$; the change of variables of Sec.~\ref{sec:pg-entropy} shifts $\alpha$ and the bracket together and leaves both results unchanged. Equation~\eqref{eq:fang-rest} combines the precession of $\alpha=1$, rotation with the fluid, with the equilibrium $\bm s=\chi_s\bm\Omega$ of $\alpha=0$ [Eq.~(17) of Ref.~\cite{Fang2026}]. Nuclear magnetic resonance on rotating solids, in which the position of the line measures the precession of the nuclear spins, distinguishes the two: a sample that rotates together with the detection coil shows a shift of the line by the rotation frequency, with the sign of the nuclear magnetic moment~\cite{Chudo2014}, and a sample that rotates in a fixed coil shows none~\cite{Chudo2015}, as in routine magic-angle spinning. Both observations are those of spins on which the rotation exerts no torque, $\alpha=0$; spins that rotate with the sample would show the opposite.

\emph{Spin per particle.} The bracket conserves $|\bm s|/n$ in reversible motion; \eqref{eq:fang-rest} does not. In irrotational shear, $\bm v=2\sigma(y,x,0)$, which has neither vorticity nor expansion, the component of the spin along the axis of compression, $(1,-1,0)/\sqrt2$, grows as $e^{(2\sigma-\Gamma_s)t}$. For $2\sigma>\Gamma_s$, which the counting of Ref.~\cite{Fang2026} allows, the spin density grows past its saturation value $s_{\rm m}$.

\subsection{Dissipative couplings that depend on the direction of the spin}
\label{sec:fang-dissipative}
The coupling $\xi$ and the anisotropic viscosities of Table~\ref{tab:onsager} are absent in Ref.~\cite{Fang2026}, although its counting places them at the order of its isotropic coefficients: with $S^{\mu\nu}=O(1)$ the tensor $\Xi_{Aij}$ is an $O(1)$ coefficient multiplying the $O(\partial)$ forces. Any one of three assumptions excludes them: an Onsager matrix built from $u^\mu$ and $\Delta^{\mu\nu}$ alone, a small spin density, $\xi\bm s^2\ll\sqrt{\lambda\eta}$, or a spin density of first order in gradients. Reference~\cite{Fang2026} makes the first: its dissipative relations, Eqs.~(11a)--(11c), are built from $u^\mu$ and $\Delta^{\mu\nu}$ alone, with $S^{\mu\nu}=O(1)$.

\subsection{The reversible terms as the transport of rescaled components}
\label{sec:fang-spin}
The reversible shear, expansion and vorticity terms of \eqref{eq:fang-rest} are those of a change of variables. Let $s^A$ be the components of the spin vector of Sec.~\ref{sec:poincare} in an orthonormal FW triad, $\dot s^A+\theta s^A=\mathcal T^A$, let $e_A$ be the dual of a Lie-transported triad, case (v) of Sec.~\ref{sec:product-rule}, and let $f$ be a positive scalar. The vector
\begin{equation}
 s'^\mu=f\,e_A{}^\mu s^A
 \label{eq:reassembled-def}
\end{equation}
has projections on comoving line elements, the vectors of the Lie-transported triad, equal to $f$ times the FW components of the spin. In the rest frame, with the triad vectors as the columns of $\bm e$, it obeys
\begin{equation}
 \dot{\bm s}'+\theta\bm s'=f\bm e\bm{\mathcal T}+(\ln f)^{\boldsymbol{\cdot}}\,\bm s'-\bm\sigma\bm s'-\tfrac13\theta\bm s'+\bm\Omega\times\bm s' ,
 \label{eq:reassembled}
\end{equation}
which for $f=1$ is \eqref{eq:reciprocal-vector}. At leading order, $\dot\varepsilon=-(\varepsilon+p)\theta$ and $\dot n=-n\theta$ give
\begin{equation}
 \dot\beta=\beta v_n^2\theta+\frac{\partial\beta}{\partial s^A}(\dot s^A+\theta s^A)+O(\partial^2),
 \label{eq:beta-rest-general}
\end{equation}
and in the counting of Ref.~\cite{Fang2026} the Maxwell relation $\partial_{s^A}\beta=-2\partial_\varepsilon(\beta\omega_A)=O(\partial)$ makes the second term $O(\partial^2)$. For $f=f_*\beta^{-2}n^{-1}$ with a constant $f_*$, therefore, $(\ln f)^{\boldsymbol{\cdot}}=(1-2v_n^2)\theta+O(\partial^2)$ and
\begin{equation}
 \dot{\bm s}'=-\bm\sigma\bm s'-\bigl(\tfrac13+2v_n^2\bigr)\theta\bm s'+\bm\Omega\times\bm s'+f\bm e\bm{\mathcal T}+O(\partial^2),
 \label{eq:reassembled-fang}
\end{equation}
with the shear, expansion and vorticity terms of \eqref{eq:fang-rest}. Read in reverse, these terms state that $\beta^2$ times the projection of the spin vector of Ref.~\cite{Fang2026} on a comoving line element changes only through the torque. The precession about the spin potential with the opposite sign, which vanishes for an equation of state isotropic in the spin, requires in addition a rotation of the triad at a rate set by the spin potential. Neither the change of triad nor the rescaling changes the spin magnitude per particle of the FW components, $(s^As^A)^{1/2}/n$, or the entropy. The vector $s'^\mu$, however, depends on the deformation that the fluid element has undergone and is not a function of the local state; it is not the spin density of the currents of Ref.~\cite{Fang2026}, for which the two theories differ by $D^{\mu\nu}$.

\FloatBarrier
\section{Shear-Driven Spin Reorientation and Its Stability}
\label{sec:bifurcation}

The aligned state loses stability above a critical shear rate and the spin reorients, as described by \Lode{} and~\LR{} of the Letter. We derive these results and the statements that accompany them, and follow the reoriented states beyond the Letter: their stresses, their stability at finite wave number, and the modulated and banded states to which the instabilities lead.

\subsection{Homogeneous simple shear}
Take $v_x=2\sigma y$, so that $\sigma_{xy}=\sigma_{yx}=\sigma$ and $\bm\Omega=(0,0,-\sigma)$, with the spin energy $\bm s^2/2\chi_s$ of Sec.~\ref{sec:generic-iso}. For uniform fields advection and precession vanish, the fluid elements are unaccelerated, and \eqref{eq:rot-output} and~\eqref{eq:cross-output} give
\begin{equation}
 \dot{\bm s}=-\Gamma_s(\bm s-\chi_s\bm\Omega)-2\xi\,\bm s\times(\bm\sigma\bm s),\qquad
 \bm s\times(\bm\sigma\bm s)=\sigma\,(-s_xs_z,\;s_ys_z,\;s_x^2-s_y^2).
 \label{eq:ode-supp}
\end{equation}
The first relation is \Lode{} of the Letter, where the time derivative is written $d\bm s/dt$. With the uniform stresses the homogeneous state is an exact solution of the full equations. The stationary states are the aligned state $\chi_s\bm\Omega$ and, for $\xi>0$ above the critical shear rate $\sigma_c$ of Sec.~\ref{sec:counting}, the two reoriented states
\begin{equation}
 \bm s=\chi_s\sigma_c\,(0,\pm\sin\vartheta,-\cos\vartheta),\qquad\cos\vartheta=\frac{\sigma_c}{\sigma},\qquad\chi_s\sigma_c=\sqrt{\frac{\lambda}{2\xi}},
\end{equation}
shown in Fig.~2 of the Letter; a third pair with $s_y=0$ would need $s_x^2<0$. The flow is invariant under the rotation by $\pi$ about $\hat{\bm z}$, which maps $(s_x,s_y,s_z)\to(-s_x,-s_y,s_z)$ and exchanges the reoriented states. The Jacobian is diagonal at the aligned state, with rates $-\Gamma_s(1+\sigma^2/\sigma_c^2)$, $\Gamma_s(\sigma^2/\sigma_c^2-1)$ and $-\Gamma_s$ for perturbations along $x$, $y$ and $z$. At the reoriented states its eigenvalues are $-2\Gamma_s$ and $-\tfrac12\Gamma_s[1\pm(1-8\tan^2\vartheta)^{1/2}]$, so perturbations about them decay monotonically for $\sigma_c<\sigma\le(9/8)^{1/2}\sigma_c$ and as damped oscillations at larger shear rates. In the invariant plane $s_x=0$ the center-manifold reduction gives, for $s_y$ in units of $\chi_s\sigma_c$, $\partial_ts_y=\Gamma_s[(\sigma^2/\sigma_c^2-1)s_y-(\sigma^2/\sigma_c^2)s_y^3]$; the cubic coefficient is negative for either sign of $\xi$, so the pitchfork bifurcation is supercritical. For $\xi<0$ the roles of $x$ and $y$ are exchanged: the threshold is $\sigma_c$ with $|\xi|$ in place of $\xi$, and the spin reorients toward the flow direction.

\subsection{General functions and other reversible terms}
\label{sec:bifurcation-general}
Let $r=\bm s^2$ and let $\chi_s(r)$, $\lambda(r)$, $\xi(r)$ be arbitrary, with $2\bm\omega=\bm s/\chi_s(r)$. The aligned state $\bm s=\chi_s\bm\Omega$ solves $|\bm s|/\chi_s(\bm s^2)=\sigma$. Reoriented states satisfy $s_x=0$, $s_z=-\lambda/(2\xi\chi_s\sigma)$ and $s_y^2+s_z^2=\lambda/2\xi$, so they lie on the spheres $\bm s^2=r_*$ with
\begin{equation}
 2\xi(r_*)\,r_*=\lambda(r_*),
\end{equation}
whose radius does not depend on the shear rate, and the aligned state is unstable if and only if $2\xi(r)\,r>\lambda(r)$ at its own value of $r$. The threshold, $2\xi\bm s^2=\lambda$ at the aligned state, involves neither $\chi_s$ nor $\Gamma_s$; with an anisotropic rotational viscosity $\lambda$ is replaced by its transverse component $\lambda_\perp$. It states that the anisotropic part of the Onsager matrix reaches half of the isotropic rotational part, and with \eqref{eq:psd-bound} it can be met only if $\lambda_\perp\le4\eta_1$. A family of reoriented states is stable if the derivative of $2\xi(r)\,r-\lambda(r)$ with respect to $r$ is positive at $r_*$. For a function $\xi(r)$ that saturates, $2\xi(r)\,r-\lambda(r)$ can have two zeros $r_1<r_2$; the aligned state is then unstable only if its $\bm s^2$ lies between $r_1$ and $r_2$, and beyond $r_2$ aligned and reoriented states coexist.

A reversible precession of the spin about the vorticity at rate $\Omega_p$, as a spin--vorticity energy $\alpha\,\bm s\cdot\bm\Omega$ would generate, changes the transverse block at the aligned state to $\bigl(\begin{smallmatrix}-\Gamma_s(1+\sigma^2/\sigma_c^2)&-\Omega_p\\ \Omega_p&\Gamma_s(\sigma^2/\sigma_c^2-1)\end{smallmatrix}\bigr)$, and the aligned state is unstable if and only if
\begin{equation}
 \sigma^4>\sigma_c^4\,(1+\Omega_p^2/\Gamma_s^2) .
\end{equation}
In regime (B) of Sec.~\ref{sec:counting}, $\Omega_p/\Gamma_s=\alpha\chi_s\sigma/\lambda$ is $O(1)$, so such a term would shift the threshold without removing it. The Thomas precession vanishes in this flow, and so does the torque of the second-order terms~\eqref{eq:H2}: without a pressure gradient and without acceleration only $\tfrac12B(\nabla\cdot\bm\Delta)^2$ remains, and its variation with respect to the spin, $-\nabla(B\,\nabla\cdot\bm\Delta)\cdot\partial\bm\Delta/\partial\bm s$, vanishes for uniform $B\,\nabla\cdot\bm\Delta$.

\subsection{General flows}
If the aligned spin $\chi_s\bm\Omega$ lies along an eigenvector $\hat{\bm e}_3$ of the shear tensor $\bm\sigma$, with eigenvalues $\sigma_{1,2,3}$, the transverse perturbations grow or decay at the rates
\begin{equation}
 -\Gamma_s\pm2\xi\chi_s|\bm\Omega|\,\bigl[(\sigma_3-\sigma_2)(\sigma_1-\sigma_3)\bigr]^{1/2}.
 \label{eq:general-strain}
\end{equation}
If $\sigma_3$ is the largest or the smallest eigenvalue the square root is imaginary and the perturbations precess and decay. If it is the intermediate one, $\hat{\bm e}_3$ is a saddle point of the rotation generated by the torque, and the state is unstable when $2|\xi|\chi_s|\bm\Omega|[(\sigma_3-\sigma_2)(\sigma_1-\sigma_3)]^{1/2}>\Gamma_s$. The instability thus requires the spin to lie along the eigenvector of $\bm\sigma$ with the intermediate eigenvalue. For the spin induced by vorticity this holds in every incompressible planar flow: for $\bm v=2\sigma(y,bx,0)$ with a constant $b$, simple shear at $b=0$, the eigenvalues of $\bm\sigma$ are $\pm\sigma(1+b)$ and $0$, the vorticity $\sigma(b-1)\hat{\bm z}$ lies along the eigenvector with eigenvalue zero, and the control parameter is $|1-b^2|\,\sigma^2/\sigma_c^2$, which vanishes for pure rotation ($b=-1$) and for irrotational shear ($b=1$) and, for $|b|\le1$, is largest for simple shear. The reflection in the flow plane maps $(s_x,s_y,s_z)\to(-s_x,-s_y,s_z)$ and protects the bifurcation. In a three-dimensional flow the vorticity is in general not an eigenvector of $\bm\sigma$; the torque then does not vanish at $\chi_s\bm\Omega$, the stationary state is displaced by $\delta\bm s=-(2\xi\chi_s^2/\Gamma_s)\,\bm\Omega\times(\bm\sigma\bm\Omega)$ for $\sigma\ll\sigma_c$, and the bifurcation becomes imperfect. For an axially symmetric shear tensor, $\sigma_1=\sigma_2=-\sigma_3/2$, with rotation about the symmetry axis, \eqref{eq:general-strain} gives the rates $-\Gamma_s\pm3i\xi\chi_s|\bm\Omega|\sigma_3$, so that there is no instability. In a non-central heavy-ion collision with Bjorken expansion along $z$ and global vorticity along $y$, the $y$ axis is the eigenvector with the intermediate eigenvalue if and only if $\tfrac94\sigma_{yy}^2<\tfrac14(\sigma_{xx}-\sigma_{zz})^2+\sigma_{xz}^2$.

For a shear rate that decays as $\sigma=\sigma_0t_0/t$ from the value $\sigma_0>\sigma_c$ at the time $t_0$, with constant coefficients, the unstable perturbation of the aligned state grows until $\sigma$ has fallen to $\sigma_c$, at $t=t_0\sigma_0/\sigma_c$, and its amplitude is multiplied by
\begin{equation}
 \exp\biggl\{\int_{t_0}^{t_0\sigma_0/\sigma_c}\Gamma_s\Bigl[\frac{\sigma^2(t)}{\sigma_c^2}-1\Bigr]dt\biggr\}
 =\exp\Bigl[\Gamma_st_0\Bigl(\frac{\sigma_0}{\sigma_c}-1\Bigr)^2\Bigr] .
 \label{eq:growth}
\end{equation}

\subsection{Stresses and flow curve}
\begin{figure}[t]
 \centering
 \includegraphics[width=\textwidth]{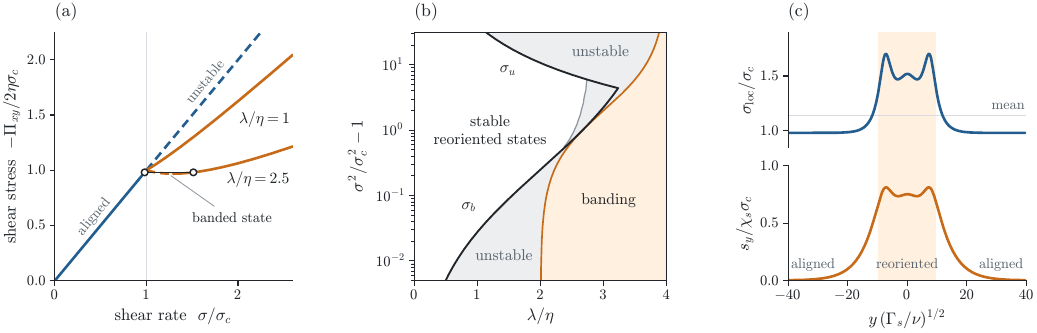}
 \caption{Stress and stability. (a) Flow curves (shear stress $-\Pi_{xy}$ versus shear rate $\sigma$) for the aligned state (blue, dashed where unstable) and the reoriented states (orange), at $\lambda/\eta=1$ and $2.5$; for $\lambda>2\eta$ the curve has a descending part (dashed). Circles joined by a line: shear rates of the two coexisting states and stress of the banded state of (c). (b) Linear stability of the homogeneous reoriented states for wave vectors in the plane of velocity gradient and vorticity, with the spin diffusivity $D_s$ equal to the kinematic viscosity $\nu=\eta/(\varepsilon+p)$. Gray: a mode grows. Black lines: long-wave boundaries $\sigma_b$ and $\sigma_u$; gray line: onset at finite wave number. Orange: the flow curve descends. (c) Stationary profiles of the local shear rate and of the transverse spin for $\lambda/\eta=2.5$ and mean shear rate $\sqrt{1.3}\,\sigma_c$ (horizontal line). The band of reoriented spin is shaded.}
 \label{fig:stress}
\end{figure}

At a stationary state $\lambda{(2\bm\omega-\bm\Omega)}=-\bm\Xi{:}\bm\sigma$, the antisymmetric stress vanishes and $\Pi_{ij}=-2\eta\sigma_{ij}+\Xi_{Aij}(\bm\Xi{:}\bm\sigma)_A/\lambda$. On the reoriented states $s_y=\pm\chi_s\sigma_c\sin\vartheta$,
\begin{equation}
 \Pi_{xy}=-2\sigma\,\eta_{\rm eff},\qquad \eta_{\rm eff}=\eta-\frac\lambda4\sin^2\vartheta,\qquad
 \Pi_{xz}=\Pi_{zx}=\mp\frac{\lambda\sigma}{4}\sin2\vartheta,\qquad \Pi_{yz}=0,
 \label{eq:tilted-stress}
\end{equation}
and the dissipated power is $4\sigma^2\eta_{\rm eff}$. On the reoriented states the bound~\eqref{eq:psd-bound} is $\lambda\le4\eta$, and on the aligned state with constant $\xi$ it is $\sigma^2\le2\sigma_c^2\sqrt{\eta/\lambda}$. The flow curve, that is, the shear stress $-\Pi_{xy}$ as a function of the shear rate $\sigma$, is $2\eta\sigma$ below $\sigma_c$ and $2\sigma[\eta-\tfrac\lambda4(1-\sigma_c^2/\sigma^2)]$ above [Fig.~\ref{fig:stress}(a)]. It is continuous, its slope drops by $\lambda$ at $\sigma_c$, and $-d\Pi_{xy}/d\sigma=2[\eta-\tfrac\lambda4(1+\sigma_c^2/\sigma^2)]$ above threshold. If $2\eta<\lambda\le4\eta$, the stress decreases with increasing shear rate in the interval $\sigma_c^2<\sigma^2<\sigma_c^2\lambda/(4\eta-\lambda)$. Between walls at $y=\pm\ell$ that move with $v_x=\pm2\sigma\ell$ and admit no spin flux, the uniform state needs only the boundary shear stress $-\Pi_{xy}$; $\Pi_{xz}$ needs none in a slab periodic in $x$ and $z$, and the heat $4\sigma^2\eta_{\rm eff}$ must leave through the walls, which requires $\eta\sigma^2\ell^2\ll\kappa T$.

\subsection{Linear stability at finite wave number}
\label{sec:stability}
Because the stresses~\eqref{eq:tilted-stress} depend on the orientation of the spin, the spin couples to the momentum equation, and the homogeneous states can lose stability at finite wave number.

We linearize the coupled equations for momentum and spin about the homogeneous states, for perturbations $\propto e^{i\bm k\cdot\bm x}$ with $\bm k=k(0,\sin\varphi,\cos\varphi)$ in the plane of the velocity gradient and the vorticity. For $k_x=0$ the shear does not advect the perturbation and these are eigenmodes. The flow is taken incompressible, the coefficients constant, and the spin equation includes diffusion, $D_s\nabla^2\bm s$ with $D_s=L_s/T\chi_s$. With $\bm u$ and $\delta\bm s$ the perturbations of the velocity and of the spin, $\varepsilon+p$ the enthalpy density and $\nu=\eta/(\varepsilon+p)$,
\begin{equation}
 (\varepsilon+p)\,\partial_tu_j=-2\sigma(\varepsilon+p)\,u_y\delta_{jx}-ik_i\,\delta\Pi_{ij},\qquad
 \partial_t\delta\bm s=\delta\bigl[-\lambda(2\bm\omega-\bm\Omega)-\bm\Xi{:}\bm\sigma\bigr]-D_sk^2\delta\bm s ,
\end{equation}
with the full linearized stresses of \eqref{eq:generic-fluxes} and~\eqref{eq:cross-output}. In units $\Gamma_s=\nu=1$ the spectrum at the reoriented states depends on $\lambda/\eta$, $\sigma/\sigma_c$ and $D_s/\nu$. It splits into a block for $(u_\perp,\delta s_x)$, with $u_\perp$ the velocity in the $(y,z)$ plane normal to $\bm k$, and a block for $(u_x,\delta s_y,\delta s_z)$; the advection term $-2\sigma u_y$ couples the first to the second in one direction only and does not affect the eigenvalues. At long wavelength each block has one hydrodynamic mode, with the growth rate $-\eta_{\rm I}(\varphi)k^2/(\varepsilon+p)$ in the block of $u_x$ and $-\eta_{\rm II}(\varphi)k^2/(\varepsilon+p)$ in the block of $u_\perp$. The effective viscosities of the two modes are
\begin{equation}
\begin{aligned}
 \eta_{\rm I}(\varphi)&=\eta-\frac\lambda4+\frac\lambda8\cot\vartheta\,\sin(2\vartheta+2\varphi),\\
 \eta_{\rm II}(\varphi)&=\eta-\frac{\lambda}{16}-\frac{\lambda}{16\cos\vartheta}\bigl[2\cos(3\vartheta+2\varphi)+\cos(3\vartheta+4\varphi)\bigr]
\end{aligned}\label{eq:eta12}
\end{equation}
on the reoriented state with $s_y>0$ (for $s_y<0$, $\varphi\to-\varphi$). For $\bm k$ along the velocity gradient $\eta_{\rm I}=\eta-\tfrac\lambda4(1+\cos^2\vartheta)$, half the slope $-d\Pi_{xy}/d\sigma$ of the flow curve; for $\bm k$ along the vorticity $\eta_{\rm I}=\eta_{\rm eff}$. Minimizing over directions,
\begin{equation}
 \min_\varphi\eta_{\rm I}=\eta-\frac\lambda4-\frac\lambda8\cot\vartheta,\qquad
 \frac{\sigma_b^2}{\sigma_c^2}=1+\frac{\lambda^2}{4(4\eta-\lambda)^2},\qquad\frac{\sigma_u^2}{\sigma_c^2}=1+\frac{(32\eta+\lambda)^2}{27\lambda^2}:
 \label{eq:Rb}
\end{equation}
every reoriented state is unstable at long wavelength for $\sigma_c<\sigma<\sigma_b$, where the susceptibility of the reorientation diverges as $\cot^2\vartheta$ while its coupling to an oblique shear vanishes only as $\tan\vartheta$, and a second long-wave instability, in the block of $u_\perp$, sets in for $\sigma>\sigma_u$ [$\min_\varphi\eta_{\rm II}=\eta(1-\lambda/4\eta)$ for $\sigma\le2\sigma_c$ and $\eta+\lambda/32-(3\sqrt3\lambda/32)\tan\vartheta$ for $\sigma\ge2\sigma_c$]. Between $\sigma_b$ and $\sigma_u$ a real eigenvalue can cross zero at finite $k$; with spin diffusion the reoriented states are stable at all wave numbers in a window of shear rates that closes at $\lambda\simeq2.7\,\eta$ for $D_s/\nu=1$ [Fig.~\ref{fig:stress}(b)], and for $D_s\to0$ an oscillatory instability at short wavelength limits the window to $\lambda<2.2\,\eta$. For $\lambda\le4\eta$ and $\sigma<\sigma_c$ the aligned state is stable at all $k$.

The linear analysis concerns perturbations with $k_x=0$; perturbations with $k_x\neq0$ are advected by the shear and have time-dependent wave vectors, and their transient growth lies outside it. Compressibility does not change the long-wave boundaries for $k_x=0$, because transverse modes decouple from sound at order $k^2$, and for $k_x=0$ the longitudinal velocity gradient exerts a torque only along $x$, the strongly damped direction. The analysis does not include the attenuation of sound in the reoriented state, and it assumes that the coefficients, which change as viscous heating raises the temperature, remain constant over the times considered.

\subsection{Nonlinear and banded states}
To follow the instabilities we integrate the nonlinear equations for fields that depend on $z'=\hat{\bm k}\cdot\bm x$ with $\hat{\bm k}=(0,\sin\varphi,\cos\varphi)$, on a periodic interval, at fixed mean shear rate. With time in units of $\Gamma_s^{-1}$, length in units of $(\nu/\Gamma_s)^{1/2}$, $\bm s$ in units of $\chi_s\sigma_c$ and velocity gradients in units of $\sigma_c$, the velocity gradient is $\partial_iv_j=2\bar\sigma\,\delta_{iy}\delta_{jx}+\hat k_i\partial_{z'} u_j$, with the mean shear rate $\bar\sigma$ and the deviation $\bm u(z',t)$ of the velocity from the mean flow, and
\begin{equation}
\begin{aligned}
 \partial_t\bm s&=-(\bm s-\bm\Omega)-\bm s\times(\bm\sigma\bm s)+\frac{D_s}{\nu}\,\partial_{z'}^2\bm s,\\
 \partial_tu_j&=-\hat k_i\partial_{z'}\Pi_{ij}-2\bar\sigma\,\frac{\sigma_c}{\Gamma_s}\,u_y\delta_{jx}-\hat k_j\partial_{z'}\delta p,\\
 \Pi_{ij}&=-2\sigma_{ij}+\frac{\lambda}{2\eta}\Bigl\{\epsilon_{Aij}\bigl[s_A-\Omega_A+(\bm s\times\bm\sigma\bm s)_A\bigr]-[(\bm s-\bm\Omega)\times\bm s]_is_j-[(\bm s-\bm\Omega)\times\bm s]_js_i\Bigr\},
\end{aligned}\label{eq:1d-model}
\end{equation}
with $\bm\sigma$ and $\bm\Omega$ computed from $\partial_iv_j$, ${2\bm\omega-\bm\Omega}={\bm s-\bm\Omega}$ in these units, and the pressure perturbation $\delta p$ enforcing $\hat{\bm k}\cdot\bm u=0$. The region allowed by~\eqref{eq:psd-bound} is $(\bm s^2)^2\le4\eta/\lambda$ in these units. Derivatives are spectral and the time integration is implicit (256 points). The linear growth rates of this code agree with \eqref{eq:eta12}.

\begin{figure}[t]
 \centering
 \includegraphics[width=\textwidth]{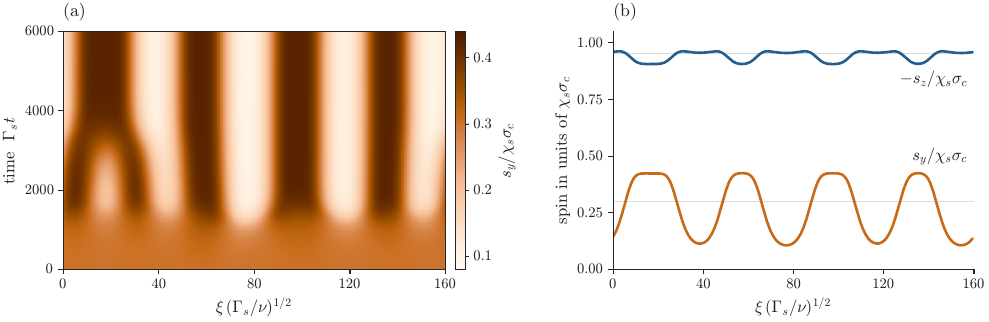}
 \caption{Modulated state near threshold from \eqref{eq:1d-model}: $\lambda/\eta=2$, $\sigma^2=1.1\,\sigma_c^2$, $\varphi=2.3$, $D_s=\nu$, $\sigma_c/\Gamma_s=5$. (a) Transverse spin component as a function of position and time. (b) Final profile of the spin components; the horizontal lines are the values in the homogeneous reoriented state.}
 \label{fig:modulated}
\end{figure}

\emph{Oblique modulation near threshold.} For $\lambda/\eta=2$, $\sigma^2=1.1\,\sigma_c^2$ (below $\sigma_b^2=1.25\,\sigma_c^2$), $\varphi=2.3$ and $\sigma_c/\Gamma_s=5$, the homogeneous reoriented state with a small random perturbation evolves into a stationary modulation of wavelength $40\,(\nu/\Gamma_s)^{1/2}$ in a box of length 160, in which $s_y/\chi_s\sigma_c$ varies between $0.11$ and $0.43$ about the homogeneous value $0.30$ and the local shear rate by about $\pm10\%$; the state satisfies the bound~\eqref{eq:psd-bound} throughout (Fig.~\ref{fig:modulated}).

\begin{figure}[t]
 \centering
 \includegraphics[width=\textwidth]{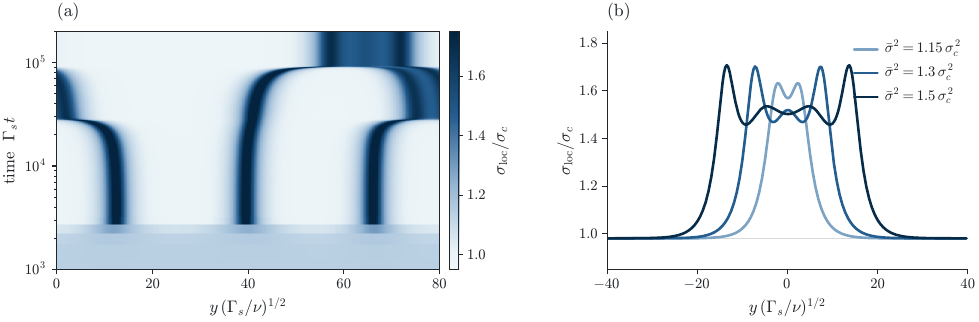}
 \caption{Shear banding from \eqref{eq:1d-model} for $\lambda/\eta=2.5$, $D_s=\nu$ and wave vector along the velocity gradient, in a periodic box of length $L=80\,(\nu/\Gamma_s)^{1/2}$. (a) Local shear rate as a function of position and time (logarithmic scale) for $\bar\sigma=\sqrt{1.3}\,\sigma_c$, starting from the homogeneous reoriented state with a small random perturbation: three bands form and coarsen to one. (b) Stationary profiles of the local shear rate for three mean shear rates (darker for larger $\bar\sigma$), shifted to a common center; the horizontal line is the shear rate of the aligned state that coexists with the band.}
 \label{fig:banding}
\end{figure}

\emph{Shear banding.} A flow curve with a descending part leads to shear banding~\cite{Olmsted2008}, the coexistence of domains with different shear rates at a common shear stress. For $\lambda/\eta=2.5$ the flow curve descends in the interval $\sigma_c^2<\sigma^2<5\sigma_c^2/3$. Starting from the homogeneous reoriented state at the mean shear rate $\bar\sigma=\sqrt{1.3}\,\sigma_c$ with $\bm k$ along the velocity gradient, three bands of high shear rate form at $t\simeq3\times10^3\,\Gamma_s^{-1}$ and then coarsen: two of them merge at $t\simeq2.9\times10^4\,\Gamma_s^{-1}$ and the remaining two at $t\simeq9.1\times10^4\,\Gamma_s^{-1}$ [Fig.~\ref{fig:banding}(a)]. The final state is stationary (all time derivatives are below $10^{-12}$). It consists of one band in which the spin is reoriented, with a plateau at the local shear rate $1.52\,\sigma_c$ and an oscillatory overshoot at each interface, surrounded by fluid at $0.981\,\sigma_c$ whose spin has returned to the vorticity direction [Fig.~\ref{fig:stress}(a),(c)]. The shear stress is uniform, $-\Pi_{xy}=0.9806\times2\eta\sigma_c$; this is the stress of the aligned state at $\sigma^2=0.962\,\sigma_c^2$ and of the homogeneous reoriented state at $\sigma^2=2.30\,\sigma_c^2$, and it lies between the minimum, $0.968$, and the maximum, $1$, of the stress on the multivalued part of the flow curve. Runs at $\bar\sigma^2=1.15\,\sigma_c^2$ and $1.5\,\sigma_c^2$ started from this state relax to one band with the same stress within $3\times10^{-4}$; the width of the band grows linearly with the mean shear rate, with the slope $L/(\sigma_{\rm high}-\sigma_{\rm low})$ of the lever rule to within $3\%$ [Fig.~\ref{fig:banding}(b)]. The banded state is thus a coexistence of the aligned and the reoriented state at a selected stress. Throughout these runs $|\bm s|\le1.05\,\chi_s\sigma_c$, inside the region $|\bm s|\le1.125\,\chi_s\sigma_c$ allowed by \eqref{eq:psd-bound}. For $\lambda/\eta=3$ and $\bar\sigma=\sqrt2\,\sigma_c$ with constant $\xi$ the high-shear bands become thin and reach $|\bm s|=1.12\,\chi_s\sigma_c$, outside the allowed region $|\bm s|\le1.075\,\chi_s\sigma_c$; the constant-$\xi$ model is then not adequate, and a function $\xi(\bm s^2)$ that respects \eqref{eq:psd-bound} for all $\bm s$ is required.

\FloatBarrier
\section*{Acknowledgments}
We thank Jin Hu and Yi Yin for fruitful discussions, and Giorgio Torrieri for insightful questions. Large language models (OpenAI GPT and Anthropic Claude) assisted with symbolic cross-checks, numerical code, figure preparation, and text editing. The authors take full responsibility for the content of this work. This work was supported in part by the National Natural Science Foundation of China (NSFC) under Grant Nos. 12235016 and 12221005.

\bibliographystyle{apsrev4-2}
\bibliography{references}